\documentclass[preprint,trackchanges]{aastex63}
\usepackage{amssymb}
\usepackage{ulem}

\newcommand{\FeH}{$\mbox{[Fe/H]}$} 
\newcommand{\AB}[2]{$\mbox{[#1/#2]}$} 

\newcommand{\FeHeq}[1]{$\mbox{[Fe/H]}={#1}$}    
\newcommand{\FeHsim}[1]{$\mbox{[Fe/H]}\sim{#1}$}  
\newcommand{\FeHlt}[1]{$\mbox{[Fe/H]}<{#1}$}    
\newcommand{\FeHle}[1]{$\mbox{[Fe/H]}\le{#1}$}   
\newcommand{\FeHgt}[1]{$\mbox{[Fe/H]}>{#1}$}    
\newcommand{\ABeq}[3]{$\mbox{[#1/#2]}={#3}$}    
\newcommand{\ABlt}[3]{$\mbox{[#1/#2]}<{#3}$}    
\newcommand{\ABgt}[3]{$\mbox{[#1/#2]}>{#3}$}    
\newcommand{\ABge}[3]{$\mbox{[#1/#2]}\ge{#3}$}   
\newcommand{\ABsim}[3]{$\mbox{[#1/#2]}\sim {#3}$} 
\newcommand{\ABlesssim}[3]{$\mbox{[#1/#2]}\lesssim {#3}$} 
\newcommand{\tefft}{$T_{\mbox{\scriptsize eff}}$} 
\newcommand{\teffm}{T_{\mbox{\scriptsize eff}}}  

\newcommand{\logg}{\ensuremath{\log g}}
\newcommand{\logL}{\ensuremath{\log L}}

\newcommand{\mlp}{\ensuremath{\alpha_{\mathrm{MLT}}}}

\newcommand{\Tefft}{\ensuremath{T_{\mathrm{eff}}}}

\received{}
\revised{}
\accepted{}
\submitjournal{ApJ}

\shorttitle{LAMOST/Subaru VMP Stars. II. Abundances}
\shortauthors{Li et al.}

\begin{document}

\title{Four-hundred Very Metal-Poor Stars Studied with LAMOST and Subaru. II. Elemental abundances}

\correspondingauthor{Wako Aoki, Haining Li, Gang Zhao}
\email{aoki.wako@nao.ac.jp, lhn@nao.cas.cn, gzhao@nao.cas.cn}

\author[0000-0002-0786-7307]{Haining Li}
\affiliation{Key Lab of Optical Astronomy, National Astronomical
  Observatories, Chinese Academy of Sciences (CAS) \\
A20 Datun Road, Chaoyang, Beijing 100101, China}
  
\author[0000-0002-8975-6829]{Wako Aoki}
\affiliation{National Astronomical Observatory of Japan \\ 2-21-1 Osawa,
  Mitaka, Tokyo 181-8588, Japan
}
\affiliation{Department of Astronomical Science, School of Physical Sciences, The Graduate University of Advanced Studies (SOKENDAI) \\
 2-21-1 Osawa, Mitaka,
Tokyo 181-8588, Japan}


\author{Tadafumi Matsuno}
\affiliation{Kapteyn Astronomical Institute, University of Groningen  \\ Landleven 12, 9747 AD Groningen, The Netherlands}
\affiliation{Department of Astronomical Science, School of Physical Sciences, The Graduate University of Advanced Studies (SOKENDAI) \\
 2-21-1 Osawa, Mitaka,
Tokyo 181-8588, Japan}\affiliation{National Astronomical Observatory of Japan \\ 2-21-1 Osawa,
  Mitaka, Tokyo 181-8588, Japan
}

\author{Qianfan Xing}
\affiliation{Key Lab of Optical Astronomy, National Astronomical
  Observatories, Chinese Academy of Sciences \\
A20 Datun Road, Chaoyang, Beijing 100101, China}

\author{Takuma Suda}
\affiliation{Department of Liberal Arts, Tokyo University of Technology, Nishi Kamata 5-23-22, Ota-ku, Tokyo 144-8535, Japan
}

\author[0000-0001-8537-3153]{Nozomu Tominaga}
\affiliation{National Astronomical Observatory of Japan \\ 2-21-1 Osawa,
  Mitaka, Tokyo 181-8588, Japan
}
\affiliation{Department of Astronomical Science, School of Physical Sciences, The Graduate University of Advanced Studies (SOKENDAI) \\
 2-21-1 Osawa, Mitaka,
Tokyo 181-8588, Japan}
\affiliation{Department of Physics, Faculty of Science and Engineering, Konan University \\ 8-9-1 Okamoto, Kobe, Hyogo 658-8501, Japan
}
\affiliation{Kavli Institute for the Physics and Mathematics of the
Universe (WPI), The University of Tokyo, 5-1-5 Kashiwanoha, Kashiwa, Chiba
277-8583, Japan
}

\author{Yuqin Chen}
\affiliation{Key Lab of Optical Astronomy, National Astronomical
  Observatories, Chinese Academy of Sciences \\
A20 Datun Road, Chaoyang, Beijing 100101, China}

\author{Satoshi Honda}
\affiliation{Nishi-Harima Astronomical Observatory, Center for Astronomy, University of Hyogo \\ 407-2,
Nishigaichi, Sayo-cho, Sayo, Hyogo 679-5313, Japan}

\author[0000-0003-4656-0241]{Miho N. Ishigaki}
\affiliation{National Astronomical Observatory of Japan \\ 2-21-1 Osawa,
  Mitaka, Tokyo 181-8588, Japan
}

\author{Jianrong Shi}
\affiliation{Key Lab of Optical Astronomy, National Astronomical
  Observatories, Chinese Academy of Sciences \\
A20 Datun Road, Chaoyang, Beijing 100101, China}

\author{Jingkun Zhao}
\affiliation{Key Lab of Optical Astronomy, National Astronomical
  Observatories, Chinese Academy of Sciences \\
A20 Datun Road, Chaoyang, Beijing 100101, China}

\author[0000-0002-8980-945X]{Gang Zhao}
\affiliation{Key Lab of Optical Astronomy, National Astronomical
  Observatories, Chinese Academy of Sciences \\
A20 Datun Road, Chaoyang, Beijing 100101, China}
\altaffiliation{Schoold of Astronomy and Space Science, University of Chinese Academy of 
Sciences \\
No.19(A) Yuquan Road, Shijingshan District, Beijing, 100049, China}

\begin{abstract}
We present homogeneous abundance analysis of over 20 elements for 385 very metal-poor (VMP) stars based on the LAMOST survey 
and follow-up observations with the Subaru Telescope. It is the largest high-resolution VMP sample (including 363 new objects) studied by a single program, 
and the first attempt to accurately determine evolutionary stages for such a large sample based on {\it Gaia} parallaxes. 
The sample covers a wide metallicity range from \ABlesssim{Fe}{H}{-1.7} down to \FeHsim{-4.3}, including over 110 objects with \FeHle{-3.0}. 
The expanded coverage in evolutionary status makes it possible to define abundance trends respectively for giants and turnoff stars. 
The newly obtained abundance data confirm most abundance trends found by previous studies, but also provide useful update and new sample of outliers. 
The Li plateau is seen in $-2.5<$ \FeHlt{-1.7} in our sample, 
whereas the average Li abundance is clearly lower at lower metallicity. 
Mg, Si, and Ca are overabundant with respect to Fe, showing decreasing trend with increasing metallicity. 
Comparisons with chemical evolution models indicate that the over-abundance of Ti, Sc, and Co are not well reproduced by current theoretical predictions. 
Correlations are seen between Sc and $\alpha$-elements, while Zn shows a detectable correlation only with Ti but not with other $\alpha$-elements. 
The fraction of carbon-enhanced stars ([C/Fe]$>0.7$) is in the range of $20 - 30\%$ for turnoff stars depending on the treatment of objects for which C abundance is not determined, 
which is much higher than that in giants ($\sim 8\%$). 
Twelve Mg-poor stars (\ABlt{Mg}{Fe}{0.0}) have been identified in a wide metallicity range from \FeHsim{-3.8} through {$-1.7$}. 
Twelve Eu-rich stars ([Eu/Fe]$>1.0$) have been discovered in $-3.4<$[Fe/H]$<-2.0$, enlarging the sample of r-process-enhanced stars with relatively high metallicity. 
\end{abstract}




\section{Introduction}\label{sec:intro}

Stars lacking metals have long been regarded as fossil records of the early evolution of the Galaxy
and preserve important clues to the first generation of stars in the universe. 
Low-metallicity stars with \FeHlt{-2.0}, $-3.0$, and $-4.0$
\footnote{
The standard notations 
[X/Y]$=\log(N_{\rm X}/N_{\rm Y})-\log(N_{\rm X}/N_{\rm Y})_{\odot}$ and $\log A({\rm X})=\log(N_{\rm X}/N_{\rm H})+12$ for elements X and Y are adopted in this work.}
are referred to as very metal-poor (VMP), extremely metal-poor (EMP),
and ultra metal-poor (UMP) stars, respectively \citep{Beers&Christlieb2005ARAA}. 
Detailed chemical abundance patterns of these objects can be compared to 
theoretical models and constrain the nucleosynthesis of early generations of supernovae \citep{Nomoto2013ARAA}, 
while their observed abundance trends along metallicities provide essential information  
about the chemical history of the Milky Way \citep{Frebel&Norris2015ARAA}.
Extending the overall abundance analysis to a large sample of VMP and EMP stars allow us to examine nucleosynthetic yields to constrain the detailed physics of these events, 
such as progenitor stellar masses, rotation rates, masses of the compact remnant (neutron stars or black holes), explosion energies, mixing efficiencies of ejecta, etc.
\citep{Heger&Woosley2010ApJ,Limongi2012ApJS,Tominaga2014ApJ,Limongi2018ApJS,Wanajo2018ApJ,Jones2019ApJ}.
These yields are important for understanding the chemical enrichment and initial condition of the early Milky Way.

In the past two decades, abundance patterns of VMP/EMP stars have been studied extensively 
through high-resolution spectroscopic observations for several hundred VMP stars, 
including the samples of First Stars \citep{Cayrel2004AA,Bonifacio2009AA} (hereafter FS), 
the Most Metal-poor Stars \citep{Norris2013ApJa,Yong2013ApJa} (hereafter Y13), 
and other follow-up studies on metal-poor stars selected from survey projects 
(e.g., \citealt{Barklem2005AA,Aoki2013ApJ,Cohen2013ApJ,Roederer2014AJ,Jacobson2015ApJ}).
Such analyses have led to better understanding about the abundance trends of VMP stars, 
and also obtained a number of important discoveries such as 
first discoveries of hyper metal-poor stars with metallicities below \FeHsim{-5.0} \citep{Christlieb2002Nature,Frebel2005Nature}, objects with unprecedentedly low iron abundance of \FeHlt{-7.0} \citep{Keller2014Nature} or total metal content \citep{Caffau2011Nature},
and stars with extreme abundance patterns. 

Meanwhile, larger number of VMP/EMP stars have made us realize that there exists a large scatter 
in the observed elemental abundance ratios among these old stars, 
indicating rather stochastic chemical enrichment at the early stage. 
An example is the large fraction of carbon-enhanced metal-poor (CEMP) stars 
that do not present enhancement in heavy elements, i.e., CEMP-no stars \citep{Aoki2002ApJ}. 
Dominating the most metal-poor stars, the surface abundances of the majority of CEMP-no stars are believed to reflect the abundances of natal gas where they were born. 
However, the origin of the carbon-excess in these stars are still in debate, 
with proposed scenarios including rotating massive stars \citep{Maeder2015AA},
faint supernova \citep{Umeda&Nomoto2005ApJ}, and inhomogeneous metal-mixing \citep{Hartwig2019ApJ}.
Enhancement of carbon by binary mass transfer from the former AGB stars in which s-process is not efficient is also proposed \citep{Suda2004ApJ}. 
More recent observations have revealed that some CEMP-no stars can also be enhanced 
in intermediate-mass elements such as Na, Mg, Al, Si, etc. (e.g., \citealt{Bonifacio2018AA,Aoki2018PASJ}), 
which could be a useful constraint on the 
 mechanisms that produce the C enhancement in CEMP-no stars.
Another sign of stochastic chemical enrichment is the discovery of VMP/EMP stars 
showing clearly lower \AB{$\alpha$}{Fe} ratios compared to the typical halo stars 
\citep{Cohen2013ApJ,Caffau2013AA,Bonifacio2018AA}. 
Some objects could have been enriched by more than one first generation of supernova  \citep{Hartwig2019MNRAS,Salvadori2019MNRAS}.
A possible signature of supernovae by very massive stars is also suggested \citep{Aoki2014Science}.

The above mentioned remarkable abundance patterns observed in VMP/EMP stars 
are usually considered as typical features of ultra-faint dwarf galaxies and some classical dwarf galaxies 
\citep{Tolstoy2009ARAA,Venn2012ApJ,Salvadori2015MNRAS}. 
Increasing evidence for accretion of small stellar systems in the formation of the Milky Way halo has indeed been obtained; 
stellar streams such as Sagittarius stream \citep{Ibata1994Nature} and kinematic substructures including Gaia-Enceladus, Sequoia, Thamnos, 
and the Helmi streams are considered to be signatures of past galaxy accretions \citep[e.g.,][]{Helmi1999Nature,Helmi2018Nature,Koppelman2019AA}.
However, most of these systems have mean metallicities with \FeHgt{-2.0}\citep[e.g.,][]{Naidu2020ApJ,Zhao&Chen2021SCPMA,Ruiz-Lara2022arXiv}, 
and the presence of the aforementioned kinematic substructures is less prominent at metallicities below \FeHsim{-2.0} \citep[see e.g., Fig. 2 of ][]{Myeong2018ApJ}.
Though there have been a few studies on (limited number of stars in) substructures with lower metallicities \citep[e.g.,][]{Roederer2019ApJ,Yuan2020ApJ,Aguado2021MNRAS},
the assembly history of VMP/EMP stars is still far less clear compared to stars with \FeHgt{-2.0} 
Interestingly, \citet{Sestito2019MNRAS} has found out that some ultra metal-poor stars 
orbit close to or within the plane of the Galactic disk, 
which has been later on confirmed with a much larger VMP star sample \citep{Sestito2020MNRAS}, 
and is reproduced by high-resolution cosmological simulations \citep{Sestito2021MNRAS}. 
Such observations raise challenge to the conventional models 
where metal-poor stars should be much older than the Galactic plane. 
Hereby, after the launching of {\it Gaia} mission \citep{Gaia2016AA}, combination of spectroscopic observations  
and accurate parallax and proper motions provided by {\it Gaia} have made it possible 
to investigate the chemo-dynamical structure and evolution of the Milky Way in more detail, 
through multi-dimensional database of VMP/EMP stars. 

For previous studies that have focused on VMP/EMP stars, 
the number of program stars studied by individual work is usually from about a few dozens to at most a few hundred, 
which is still much smaller than the sample of stars with higher metallicity (i.e., disk stars) 
measured by recent large spectroscopic surveys (e.g., Gaia/ESO survey: \citealt{Gilmore2012Msngr}, GALAH: \citealt{DeSilva2015MNRAS}). 
Detailed chemical abundances of VMP stars have been provided by more recent efforts 
such as the $R-$Process Alliance (RPA, \citealt{Holmbeck2020ApJS}) and high-resolution follow-up observations of the Pristine survey (e.g., \citealt{Venn2020MNRAS})  
and the SkyMapper survey (e.g., \citealt{Jacobson2015ApJ, Yong2021MNRAS}). 
However, the samples of VMP stars in these studies include a few dozen to over one hundred VMP stars, e.g., about 150 VMP stars in the RPA database.
Hence, a large sample of homogeneously observed and analyzed VMP stars are very much in need. 

Decades of studies have allowed us to recognize that the most metal-poor stars are rare objects to find, 
and thus studies of chemical abundances of VMP and EMP stars have to take advantages from wide field spectroscopic surveys.  
Such efforts include the early objective prism surveys such as the HK Survey \citep{Beers1985AJ,Beers1992AJ} and the Hamburg/ESO Survey \citep{Christlieb2008AA}, 
low-resolution spectroscopic surveys (e.g., SEGUE,
the Sloan Extension for Galactic Understanding and Exploration; \citealt{Yanny2009AJ}),
and photometric survey projects including the SkyMapper survey  \citep{Keller2007PASA} and the Pristine survey \citep{Starkenburg2017MNRAS}.
However, existing candidate samples are either limited in their number of stars,
or comprise stars that are too faint for efficient high-resolution follow-up spectroscopy with 4-10 meter telescopes.
As a result, the number of EMP/UMP stars with high-resolution follow-up is still
not sufficiently large for statistical investigations of the most metal-poor stars.

LAMOST (Large sky Area Multi-Object fiber Spectroscopic Telescope)
\footnote{\url{http://www.lamost.org/public/}}
has completed its first five-year low-resolution spectroscopic survey from 2012 through 2017 (LAMOST-I),
and released more than seven million spectra of stars in the Milky Way in its fifth data release (DR5).
This huge database of spectra provides a great opportunity
to explore the various stellar components of the Galaxy \citep{Zhao2006ChJAA,Zhao2012RAA,Liu2015RAA},
including searches for VMP stars over a large area of sky.
One advantage of the LAMOST sample of stars is that they cover a range in apparent magnitude suitable
for efficient follow-up high-resolution spectroscopic observations with $4-10$ meter telescopes.
In addition, the wavelength coverage (3700 $-$ 9100\,{\AA}) and resolving power ($R \sim 1800$) of LAMOST spectra
allow for quite reliable estimation of the stellar parameters including effective temperatures and metallicities. 
The LAMOST database has been provided a great opportunity to search for large number of 
VMP stars \citep{Aguado2017AA,Li2015ApJ,Li2018ApJS}.

To further extend previous efforts and obtain a large homogeneous sample of VMP stars, 
a joint project was initiated to obtain high-resolution spectra for over 400 VMP star candidates 
selected from the LAMOST survey, using the High Dispersion Spectrograph (HDS) equipped at the Subaru Telescope. 
The ``snapshot'' spectra with $R=36,000$ have been obtained for the main sample, while for 
a few interesting objects with \FeHlt{-4.0} or with peculiar elemental abundances,
spectra with higher SNR and/or resolving power ($R=$60,000) were then obtained for more detailed analysis, 
e.g., for two UMP stars \citep{Li2015PASJ}, a bright EMP r-II star \citep{Li2015RAA}, etc.

Detailed description about target selection and observations of the LAMOST/Subaru 
VMP sample is provided in Aoki et al. (hereafter Paper I),
in which radial velocities and interstellar reddening are also reported.  
This is the second paper of the series, 
presenting a homogeneous chemical abundance analysis for this LAMOST/Subaru VMP star sample.
The outline of the paper is as follows: analysis of stellar parameters 
as well as elemental abundances are described in \S~\ref{sec:method}, 
interpretations about the abundance results including the observed abundance trend, 
and comparison with theoretical models are presented in \S~\ref{sec:discussion}, 
which also includes discussions on various categories of chemically peculiar objects. 
A brief summary is given in \S~\ref{sec:summary}.

\section{Observations and analysis of program stars} \label{sec:method}

\begin{figure*}
\hspace{-2.0cm}
\plotone{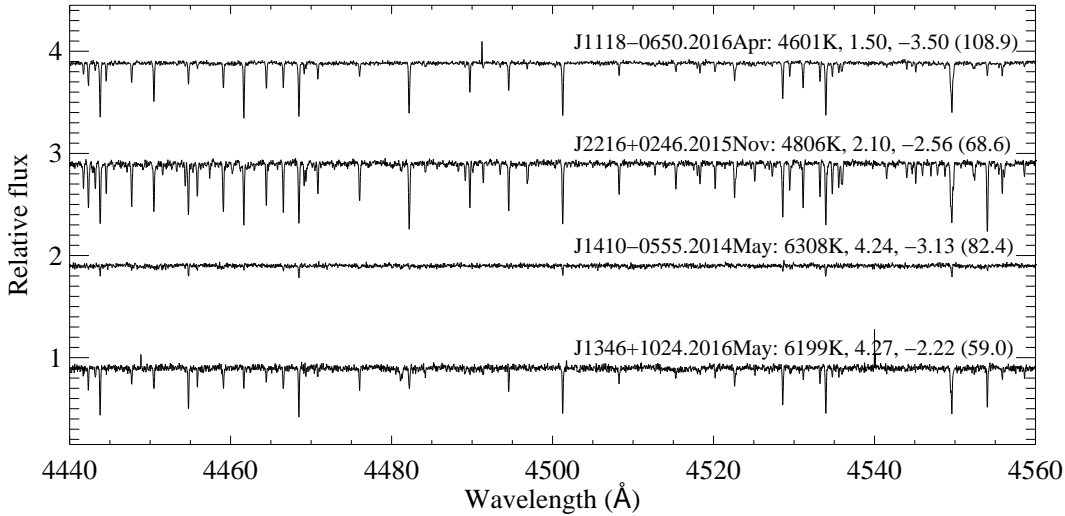}
\caption{Four normalized Subaru/HDS spectra nearby 4500\,{\AA}, 
from top to bottom respectively for EMP red giant, VMP red giant, EMP turnoff, and VMP turnoff stars. 
Stellar parameters (\Tefft, \logg, and \FeH) for each object have been labelled with the object name, 
and corresponding spectral SNR has also been shown in parenthesis. 
\label{fig:HDS_spec}}
\end{figure*}

The program stars were first selected from LAMOST DR1 through DR5.
More than 15,000 candidates of VMP stars have been initially selected 
based on metallicities derived from LAMOST low-resolution spectra
out of over 4.9 million spectra with signal-to-noise ratio (SNR) higher than 15 in the $g$-band,
using two independent methods, i.e., template matching of observed flux and line index matching 
(readers may refer to \citet{Li2018ApJS} for more details). 
To focus on relatively bright (e.g., $g < 13.0$) VMP targets 
or most probable EMP candidates (\FeHlt{-3.0} derived in both methods), 
over 400 of the above VMP candidates were selected for high-resolution followed-up observation 
using the Subaru/HDS. 
The obtained ``snapshot'' spectra with $R=$36,000 cover a wavelength coverage of 4000$-$6800\,{\AA}, 
which were taken with exposure times of 10$-$20 minutes in most cases during seven observing runs 
from May 2014 to August 2017. Readers can find more details about the sample and observations in Paper I.
The original sample includes 420 objects (with 445 spectra).
In order to obtain sufficient spectral quality to derive reliable parameters and elemental abundances, 
spectra with SNR higher than 10 per pixel ($\sim$25 per resolution element) around 4500~{\AA}, 
or those in which more than 10 \ion{Fe}{1} lines are detected, 
have been selected for the following analysis. 
This results in 411 spectra (385 unique stars) for estimation of stellar parameters and elemental abundances. 
Figure~\ref{fig:HDS_spec} shows sample Subaru/HDS spectra for four program stars, 
respectively representing VMP and EMP red giants, and also VMP and EMP turnoff stars.

\subsection{Equivalent width measurements}\label{subsec:EW_measurements}

The equivalent widths (EWs) of atomic lines were then measured for all 411 selected spectra, 
based on a linelist compiled from literature (\citealt{Aoki2013AJ}, and \citealt{Mashonkina2010AA}), 
fitting a Gaussian profile 
for corresponding single and unblended atomic absorption lines using an IDL code Tame \citep{Kang&Lee2012MNRAS}. 


There have been 26 targets observed more than once in our program. 
Their spectra could be useful to check the consistency of the reduced spectra taken under different condition, 
stability of our EW measurements, and the measurement uncertainty in EWs due to noise. 
The systematic offset between multiple EW measurements is negligible (no larger than 0.03\,m{\AA}), 
demonstrating the consistency of the reduced spectra and our EW measurements. 
The scatter is usually smaller than 0.20\,m{\AA}, which is comparable to the typical uncertainty of EW measurements.

\begin{figure*}
\plotone{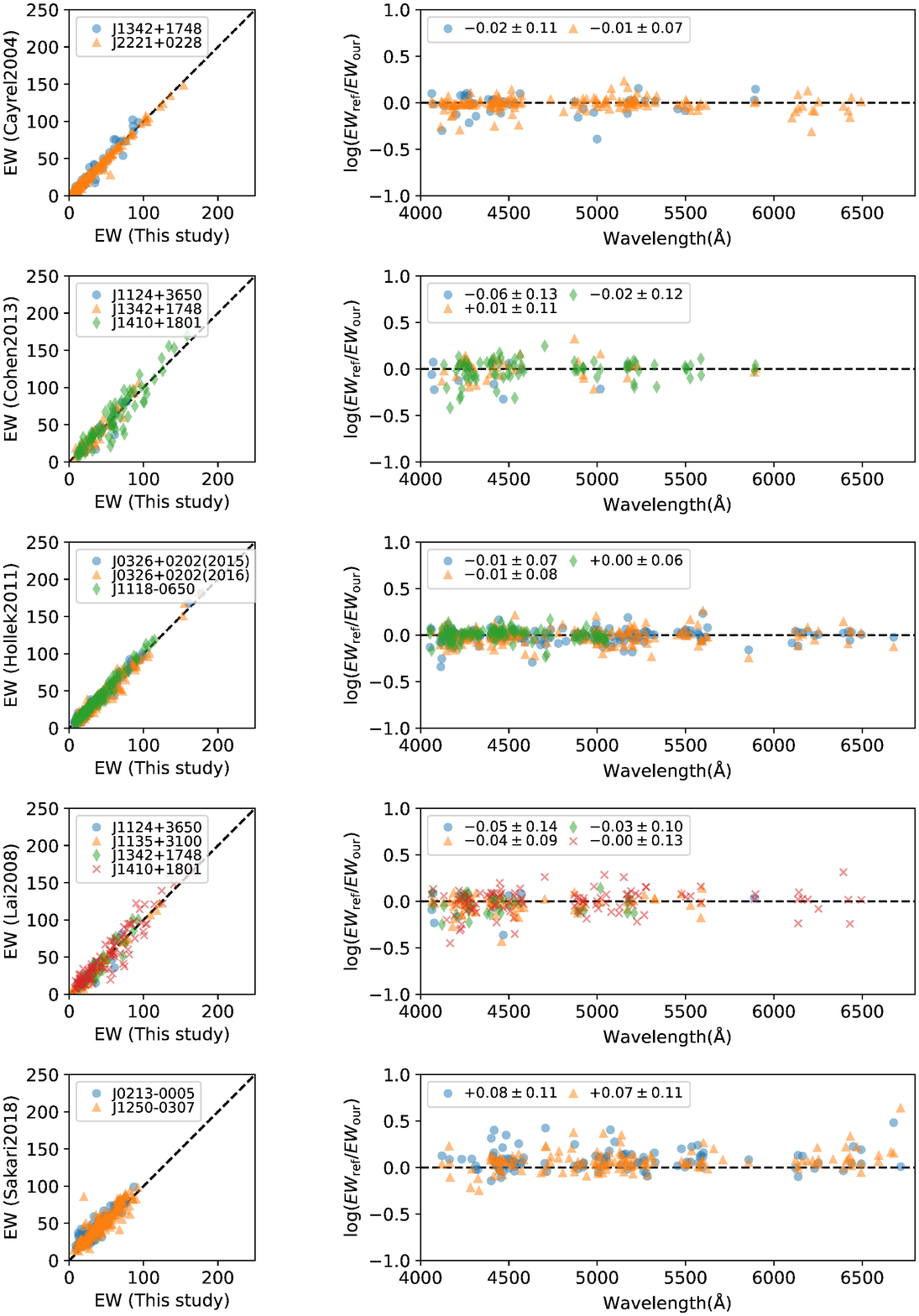}
\caption{Comparison of equivalent widths between our study and literature. The literature was selected to have more than one star in common with our study. 
The average logarithmic difference of equivalent widths and the standard deviation are presented in the right panels. \label{fig:ewlite}}
\end{figure*}

\begin{deluxetable}{lclll}
\tablecaption{Literatures for the comparison purpose 
\label{tab:literature_summary}}
\tablehead{   & \colhead{EW\tablenotemark{a}} & \colhead{$T_{\rm eff}$\tablenotemark{b}} & \colhead{$\log g$\tablenotemark{c}} &  \colhead{Figures\tablenotemark{d}} }
\startdata
\citet{Cayrel2004AA}     &  G  & phot. & ion. & \ref{fig:ewlite} \& \ref{fig:abcompare1}  \\     
\citet{Honda2004ApJ,Honda2004ApJS}      &  G  & phot. & ion. &  \ref{fig:abcompare2} \\    
\citet{Lai2008ApJ}        &  V  & phot. & iso. & \ref{fig:ewlite} \& \ref{fig:abcompare1} \\
\citet{Hollek2011ApJ}     &  V  & ex. & ion. & \ref{fig:ewlite} \& \ref{fig:abcompare2} \\
\citet{Cohen2013ApJ}      &  G  & phot. & iso.+ion. & \ref{fig:ewlite} \& \ref{fig:abcompare1} \\
\citet{Li2015PASJ}         &  G  & ex. & ion. &  \ref{fig:abcompare2} \\
\citet{Matsuno2017PASJ}    &  G  & Balmer & Balmer & \ref{fig:abcompare2}\\
\citet{Sakari2018ApJ}     &  G  & ex. & ion. & \ref{fig:ewlite} \& \ref{fig:abcompare2}   \\     
\citet{Li2018ApJL}         &  G  & ex. & ion. & \ref{fig:abcompare3}  
\enddata
\tablenotetext{}{A summary of the methods adopted in literatures.}
\tablenotetext{a}{G: Gaussian profile is assumed for equivalent widths measurements. V: Viogt profiles are assumed for a part of the lines.}
\tablenotetext{b}{The method of temperature determination. phot: photometric temperature, 
ex.: temperature from exicitation balance of neutral iron lines, Balmer: temperature from Balmer line profiles.}
\tablenotetext{c}{The method of $\log g$ determination. ion.: $\log g$ from ionization balance between neutral and singly-ionized iron, 
iso.: $\log g$ estimated using relations between $T_{\rm eff}$ and $\log g$ from isochrones, Balmer: $\log g$ from Balmer line profiles.}
\tablenotetext{d}{Reference to figures that compare equivalent widths, and stellar parameters and abundances.}
\tablenotetext{e}{$\langle \mathrm{3D}\rangle$ NLTE correction by \citet{Amarsi2016MNRAS} was applied.}
\end{deluxetable}

Figures \ref{fig:ewlite} shows comparisons of EWs measured in the present work with literature that has more than two stars in common with our study (Table~\ref{tab:literature_summary}). 
Note that some studies adopt Voigt profile fitting, while others assume Gaussian profiles for all the lines. 
This information is also included in Table~\ref{tab:literature_summary}.
In general, there is a good agreement between our measurements and those from literature. 

Two stars that are in common with \citet{Sakari2018ApJ} show small but significant offset in EWs 
($\frac{|\langle \Delta\log \mathrm{EW}\rangle|}{\sigma(\Delta\log \mathrm{EW})/\sqrt{N}}>8$) for both stars.
The difference is seen over the entire wavelength range. 
We note that we did not find any sign of problems on sky subtraction during the visual inspection of the reduced spectra for the two objects.  
We also note that our measured equivalent widths are in agreement with other studies ($\frac{|\langle \Delta\log \mathrm{EW}\rangle|}{\sigma(\Delta\log \mathrm{EW})/\sqrt{N}}<5$).

\subsection{Determination of stellar parameters}\label{subsec:stellar_parameter}

In Paper I, we describe high-resolution spectroscopic observations of over 400 program stars 
obtained using Subaru/HDS, including sample selection, radial velocity measurements, and discussion on binarity. 
Our sample consists of 363 newly discovered and analyzed metal-poor stars, 
together with 22 objects whose abundances have already been published in previous studies. 

\begin{figure*}
\epsscale{1.0}
\plotone{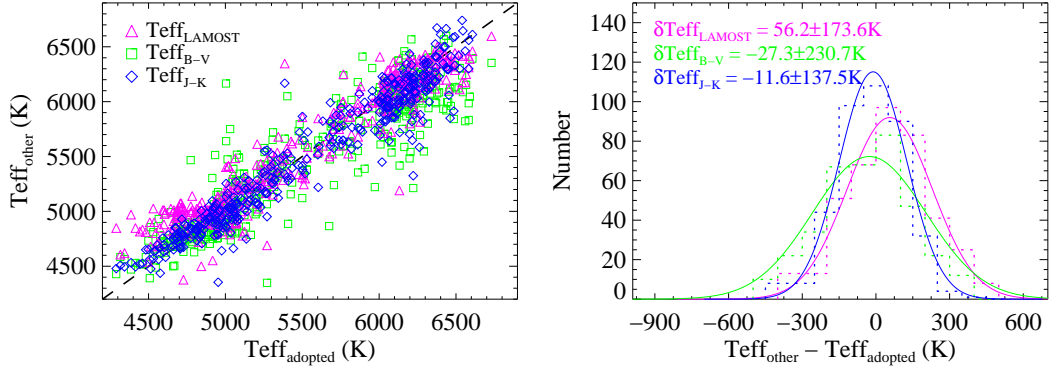}
\caption{Left: Comparison between the adopted {\tefft} and those derived from LAMOST spectra (pink triangles), 
and that based on $(J-K)_0$ (blue diamonds) and $(B-V)_0$ (green squares) .
The one-to-one line is plotted for reference.
Right: Distribution of the {\tefft} difference, together with a best-fit Gaussian
to indicate the typical difference and scatter. Same colors to the left panel are used for the three sets of {\tefft} measurements.
\label{fig:Teff_comparison}}
\end{figure*}

Since our sample covers a relatively wide metallicity range and evolutionary status, 
we have adopted the calibration for effective temperatures from \citet{Alonso1996AAS} and \citet{Alonso1999AAS}, 
based on the dereddened color $(V-K)_0$, obtained from $V$ magnitude coming from APASS, 
$K$ magnitude transformed from 2MASS $K_s$, and the extinction estimated from Green (2018).  
These data are given in Paper I.
The {\tefft} estimated from different colors are compared in Figure~\ref{fig:Teff_comparison}. 
We find good correlations between the adopted {\tefft} and those from colors with offset about 50~K or less. 
The scatter is large for the {\tefft} from $B-V$, which is most sensitive to the metallicity and molecular features.

It is known that the Fe abundances derived from Fe\,I lines can suffer relatively large NLTE 
(e.g., \citealt{Amarsi2016MNRAS}), 
and that the number of Fe\,II lines could be very limited in spectra of VMP stars, 
which makes it difficult to derive the surface gravity by forcing the Fe I and Fe II lines 
to result in consistent Fe abundance. 
Therefore, we adopt the surface gravity derived using parallax data for our sample when reliable data are available. 

For 315 objects, the {\logg} are derived from reliable parallax (err\_parallax/parallax $<$ 0.2) 
measurements by {\it Gaia} DR2 \citep{Gaia2018AA}, based on the following equation:
\begin{equation}
\logg = \logg_\odot + \log\frac{M}{M_\odot} + 4\log \frac{\teffm}{T_{\rm eff \odot}} + 0.4(M_{bol} - M_{bol_\odot})
\label{equa:logg}
\end{equation}
where $M_{bol}$ refers to the bolometric magnitude, and a fixed stellar mass of $0.8\,M_\odot$ has been adopted for our calculation. 
Quantitative tests have shown that adopting different stellar mass for each object 
would only result in a 0.01 difference in {\logg} on average.

\begin{figure*}
\plotone{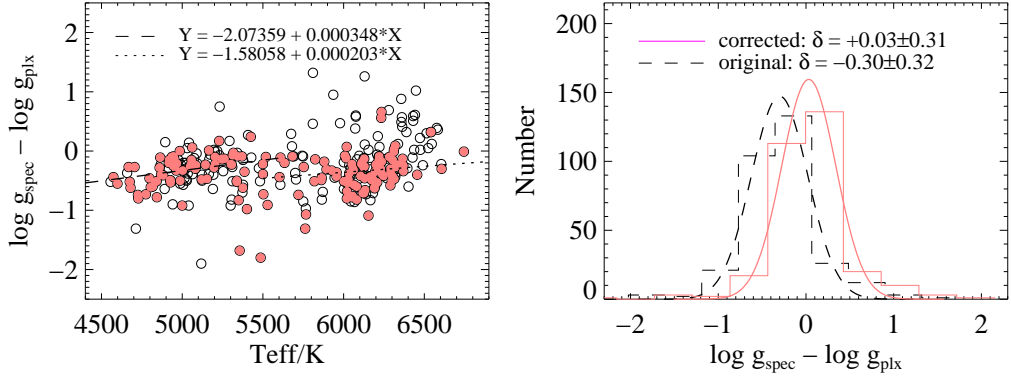}
\caption{Left: The difference between the spectroscopic $\logg_{\rm spec}$ 
and parallax-derived $\logg_{\rm plx}$ along {\tefft} for all 315 objects which have reliable parallax measurements.  
Open circles indicate objects with lower spectral quality (S/N$<50$), 
while filled circles for objects with S/N$>50$. 
The {\logg} corrections are derived by linear fitting to the data in two {\tefft} ranges (dashed and dotted lines). 
Right: Distribution of the difference between $\logg_{\rm plx}$ vs $\logg_{\rm spec}$ (dashed) 
and the corrected {\logg} (solid) for all 315 objects which have reliable parallax measurements.
Gaussian profile has been fit to estimate the typical difference and scatter.
\label{fig:logg_correction}}
\end{figure*}

For the rest of the sample for which Gaia DR2 have not provided (reliable) parallax measurements, 
the {\logg} were first determined using the spectroscopic method, i.e., forcing the \ion{Fe}{1} and \ion{Fe}{2} lines 
to derive similar iron abundances, and then the spectroscopic {\logg} ($\logg_{\mathrm{spec}}$) were corrected to the parallax-derived scale adopting the following procedure. 
Using the 315 objects which have reliable parallax-derived {\logg}, 
the correction is determined by linear fittings of the difference between their $\logg_{\mathrm{spec}}$ 
and the parallax-derived $\logg_{\mathrm{plx}}$ as a function of {\tefft}. 
As shown in the left panel of Fig.~\ref{fig:logg_correction}, 
the correction is separately fitted by cooler (giant) stars and warmer (turnoff) stars, 
respectively covering $\teffm < 5700$\,K (dashed line) and $\teffm > 5300$\,K (dotted line). 
For objects located in the overlapping region 5300\,K $< \teffm < 5700$\,K, 
an average of the corrected {\logg} has been adopted. 
We note that only spectra with SNR higher than 50 (filled circles) are used in the fitting to derive the correction, 
which ensures reliable measurements of iron lines (especially \ion{Fe}{2}) and thus robust $\logg_{\mathrm{spec}}$.
It can be seen from the right panel of Fig.~\ref{fig:logg_correction} that 
the systematic offset between $\logg_{\mathrm{spec}}$ and $\logg_{\mathrm{plx}}$ has been removed after the correction,  
whereas the dispersion does not change significantly.

\begin{figure*}
\plotone{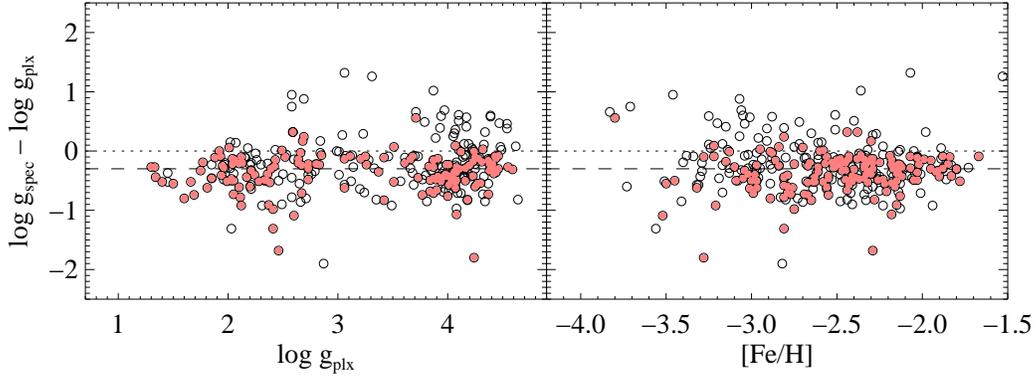}
\caption{Variation of the difference between the parallax-derived and spectroscopic {\logg} 
along different {\logg} (left) and metallicity (right) for all program stars with reliable parallax measurements. 
Symbols are the same as in Figure~\ref{fig:logg_correction}. 
The dotted and dashed lines refer to a zero difference 
and the average difference, respectively, between the two sets of {\logg} for our sample. \label{fig:dlogg_logg_FeH}}
\end{figure*}

The dependence of the {\logg} difference (before correction) has been examined 
against the other two atmospheric parameters,  i.e., {\logg} and {\FeH}, 
as presented in Figure~\ref{fig:dlogg_logg_FeH}. 
No distinct correlation with either parameter is found, confirming that 
the adopted {\logg} correction is applicable to objects with various evolutionary status and metallicities. 
It is noticed that slightly larger difference between the two sets of {\logg} is found 
for EMP turnoff stars with relatively lower S/N spectra (open circles).
This is an expected result, as the number of Fe II lines significantly decreases 
for warmer objects or those with lower metallicities, 
which leads to larger uncertainties of the spectroscopically derived $\logg_{\mathrm{spec}}$.

\begin{figure*}
\epsscale{0.7}
\plotone{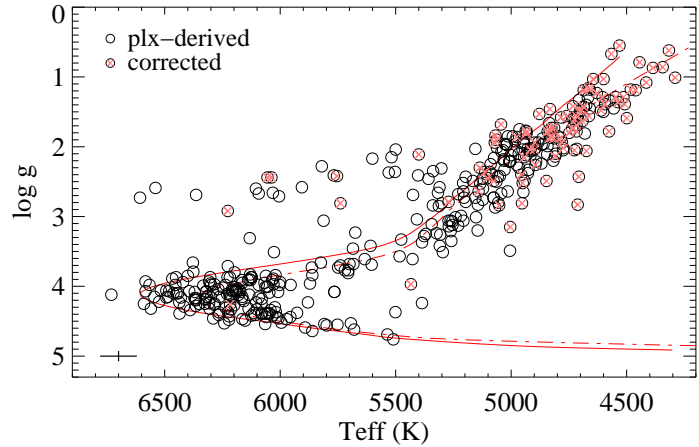}
\caption{The distribution of the program stars in the {\Tefft} vs. {\logg} diagram. 
Open circles refer to the 315 objects with {\logg} derived from the Gaia parallax, 
while circle+crosses represent targets with {\logg} derived from the empirical corrections 
as described in the text. The solid and dashed-dotted lines respectively stand for 
the Y$^{2}$ isochrone for \FeHeq{-3.0} and $-2.0$. \label{fig:HRD}}
\end{figure*}

After deriving the empirical {\logg} correction, it has been adopted to the $\logg_{\mathrm{spec}}$ 
measured from the 70 objects that do not have $\logg_{\mathrm{plx}}$, 
so as to obtain their final {\logg} values.
The distribution of the final adopted {\tefft} and {\logg} of the whole sample are shown in Figure~\ref{fig:HRD}.
It can been seen that the majority of the 70 objects for which the {\logg} correction 
has been adopted are located on the giant branch, i.e., they are more distant objects and thus 
do not have (reliable) parallax measurements compared with relatively nearby turnoff stars. 
In general, adopting the final stellar parameters, stars align on the theoretical isochrone 
(the solid and dashed lines in Figure~\ref{fig:HRD}), confirming robust estimation of parameters 
and the effective correction on {\logg}.

Since the number of measured \ion{Fe}{1} lines of the analyzed sample stars (10--170 lines) is significantly larger than that of Fe II (0--20 lines), 
the average abundance obtained from \ion{Fe}{1} lines is adopted as the final value of the metallicity.  
The resulted metallicities confirm that over 93\% of the measured objects are with \FeHlt{-2.0}.
The remaining 7\% objects are still metal-poor, e.g., 
the median and average metallicities of these objects are $-1.86$ and $-1.79$, respectively,  
and there is only one exception with \FeHgt{-1.5}.
Most of these less metal-poor (7\%) objects are turnoff stars, for which estimation based on low-resolution spectra 
tends to underestimate metallicities compared to giants. 
The overall result demonstrates the reliability in the estimation of metallicities based on LAMOST low-resolution spectra, 
which enables a quite efficient selection of targets for the follow-up high-resolution spectroscopy with the Subaru Telescope.
The microturbulence velocities of all program stars have been derived using the traditional spectroscopic method, 
i.e., by forcing \ion{Fe}{1} lines with different equivalent widths to result in similar iron abundances. 
With the above stellar parameters derived, the luminosity is then  estimated for each objects based on: 
$\logL = \logL_{\odot} - (\logg - \logg_{\odot}) + (\log M - \log M_{\odot}) + 4(\log \teffm - \log T_{\rm eff\odot})$,  
where the final {\logg} (parallax-derived plus spectroscopic-corrected) values have been adopted (Table~\ref{tab:stellar_param}).

\begin{longrotatetable}
\begin{deluxetable*}{rllllccrccccccccrcccc}
\tablecolumns{21}
\tablewidth{0pt}
\tabletypesize{\scriptsize}
\tablecaption{Basic information and stellar Parameters of the Program Stars.\label{tab:stellar_param}}
\tablehead{
\colhead{}&\multicolumn{4}{c}{Observational info.} & \colhead{} &
\multicolumn{11}{c}{Adopted parameters (Subaru measurement)}&\colhead{}&\multicolumn3c{LAMOST measurement}\\
\cline{2-5}\cline{7-17}\cline{19-21}
\colhead{ID}& \colhead{R.A.}& \colhead{Decl.}& \colhead{RV}& \colhead{SNR$^{a}$} & \colhead{} & \colhead{\Tefft}& \colhead{$\sigma$\,{\tefft}}& \colhead{\logg}& \colhead{$\sigma$\,{\logg}}& \colhead{\FeH}& \colhead{$\sigma$\,{\FeH}}& \colhead{$\xi$}&$\sigma$\,{$\xi$}&
\colhead{$\log (L/L_{\odot})$} & \colhead{$\sigma$\,$\log (L/L_{\odot}$)} & \colhead{$\mathrm{f}_{\logg}^{b}$}& \colhead{}& \colhead{\Tefft}& \colhead{{\logg}}& \colhead{\FeH}\\
\colhead{}& \colhead{deg.}& \colhead{deg.} &\colhead{$\rm km~s^{-1}$} & \colhead{} &\colhead{} & \colhead{K}& \colhead{K}& \colhead{}& \colhead{}& \colhead{}& \colhead{}& \colhead{$\rm km~s^{-1}$}& \colhead{$\rm km~s^{-1}$}& 
\colhead{}& \colhead{}& \colhead{}& \colhead{}& \colhead{(K)}& \colhead{}& \colhead{}}
\startdata
 J0002+0343 &0.645994&  3.727085&  -91.4& 23.7&& 6207 & 110 & 4.16 & 0.05 & $-$3.02 & 0.24 & 1.19 & 0.30 &  0.307 &  0.060 & plx  && 6361& 4.30& $-$3.4\\
 J0003+1556 &0.795383& 15.933828& -185.5& 32.2&& 5302 &  45 & 2.31 & 0.08 & $-$2.23 & 0.11 & 2.11 & 0.12 &  1.883 &  0.081 & plx  && 5289& 1.09& $-$2.6\\
 J0006+0123 &1.657593&  1.395354&  -65.9& 44.1&& 5675 &  53 & 3.23 & 0.04 & $-$2.27 & 0.11 & 1.52 & 0.08 &  1.081 &  0.043 & plx  && 5500& 2.52& $-$2.6\\
 J0006+1057 &1.571676& 10.961624& -312.4& 20.9&& 4691 &  38 & 1.63 & 0.14 & $-$3.15 & 0.14 & 2.51 & 0.11 &  2.350 &  0.140 & spec && 5012& 1.72& $-$3.2\\
 J0013+2350 &3.463742& 23.846972& -259.2& 41.4&& 6093 & 120 & 3.98 & 0.04 & $-$2.26 & 0.09 & 1.49 & 0.10 &  0.455 &  0.054 & plx  && 6238& 3.91& $-$2.2\\
 J0022+4254 &5.510848& 42.916500& -183.3& 23.2&& 6566 & 133 & 3.98 & 0.06 & $-$2.99 & 0.25 & 0.96 & 0.30 &  0.585 &  0.072 & plx  && 6250& 4.04& $-$3.4\\
 J0023+2023 &5.795711& 20.392651& -126.0& 37.3&& 5197 &  68 & 2.79 & 0.05 & $-$3.29 & 0.09 & 1.53 & 0.11 &  1.368 &  0.054 & plx  && 5250& 2.55& $-$3.1\\
 J0023+3558 &5.916771& 35.967957& -303.0& 38.0&& 5145 &  82 & 3.20 & 0.06 & $-$2.56 & 0.09 & 1.35 & 0.08 &  0.941 &  0.065 & plx  && 4960& 3.36& $-$3.1\\
 J0025+2305 &6.432368& 23.093561& -229.6& 48.3&& 6308 & 144 & 3.93 & 0.04 & $-$2.81 & 0.10 & 1.99 & 0.17 &  0.565 &  0.059 & plx  && 6152& 3.26& $-$3.4
\enddata
\tablenotetext{a}{The SNR is estimated from the normalized Subaru/HDS spectra around 4500\,{\AA}.}
\tablenotetext{b}{Flag of estimation of {\logg}, where ``plx'' refers to program stars whose {\logg} have been 
estimated based on the reliable parallax, while ``spec'' refers to those based on empirical correction from 
the spectroscopic estimation. See the main text in \S~\ref{sec:method} for details.}
\tablecomments{The table is published in its entirety in the machine-readable format.
      A portion is shown here for guidance regarding its form and content.}
\end{deluxetable*}
\end{longrotatetable}

We have estimated the uncertainty of stellar parameters for individual spectra, 
which are presented in Table~\ref{tab:stellar_param} together with the final adopted stellar parameters. 
The uncertainty of the {\tefft} has been derived from propagation of errors 
which mainly include photometric uncertainties in $V$ and $K$ magnitudes, and uncertainty in extinction estimations.
The uncertainty of the {\logg} could be divided into two categories: for those with parallax-derived {\logg}, 
the uncertainty has been estimated from error propagation of Equation~\ref{equa:logg}, 
which mainly comes from the uncertainty of parallax measurements and photometry, 
and also includes uncertainties of {\tefft}. 
The typical uncertainty of parallax-derived {\logg} is about 0.07\,dex. 
Note that we have adopted a fixed stellar mass of 0.8\,M$_{\odot}$, 
which may result in another $\sim$0.04\,dex uncertainty in the derived {\logg} if a 10\% uncertainty in the stellar mass is adopted.
For objects whose {\logg} is derived from the corrected spectroscopic estimation, 
the uncertainty (on average 0.19\,dex) comes from the scatter of abundances derived from \ion{Fe}{2} lines (especially in cases with less than 8 \ion{Fe}{2} lines), 
the non-negligible NLTE effect on abundances derived from \ion{Fe}{1} lines for VMP stars, 
and the uncertainty from the empirical correction (shown in Figure~\ref{fig:logg_correction}) of the adopted {\logg}.  

\subsection{Elemental abundances}

With EWs and stellar parameters obtained above, elemental abundances were then derived for the program stars. 
For abundance analysis of all species, the 1D plane-parallel, 
hydrostatic model atmospheres of the ATLAS NEWODF grid of \citet{Castelli&Kurucz2003IAUS} has been adopted, 
assuming a mixing-length parameter of $\mlp=1.25$, no convective overshooting, and local thermodynamic equilibrium (LTE). 
This grid of model atmospheres is also applied to the analysis of carbon-enhanced metal-poor stars in our sample. 
The impact of carbon-excess on 1D model atmospheres is not significant in main sequence turnoff stars 
and relatively warm red giants where molecular features are not strong, whereas it could be large in 3D models \citep{Gallagher2017}. 
\citet{Christlieb2004ApJ} investigate the effect of non-solar abundance ratios on model atmospheres 
for their study of the hyper metal-poor star HE~0107--5240 with large carbon-excess. 
They discuss that the effect of abundance changes is not very significant for stars with extremely low metallicity, 
because of the dominant role of hydrogen as opacity source and electron donor \citep[see also ][]{Collet2006ApJ}. 
Our sample includes a few cool giants with relatively high metallicity and carbon abundances, 
for which analysis using model atmospheres with carbon excess might be useful to derive more accurate abundances in future work. 
We use an updated version of the abundance analysis code MOOG \citep{Sneden1973ApJ}, 
which treats continuous scattering as a source function which sums both absorption and scattering \citep{Sobeck2011AJ}.

The lithium and carbon abundances of the program stars were respectively derived
by matching the synthetic spectra of \ion{Li}{1}\,6707.8\,{\AA} doublet  
and the CH\,A-X band around $4300-4330$\,{\AA} to the observed ones.
The Li subordinate line at 6103.6\,{\AA} was used in cases of extremely Li-enhanced objects.

\begin{figure*}
\hspace{-0.9cm}\epsscale{1.1}
\plotone{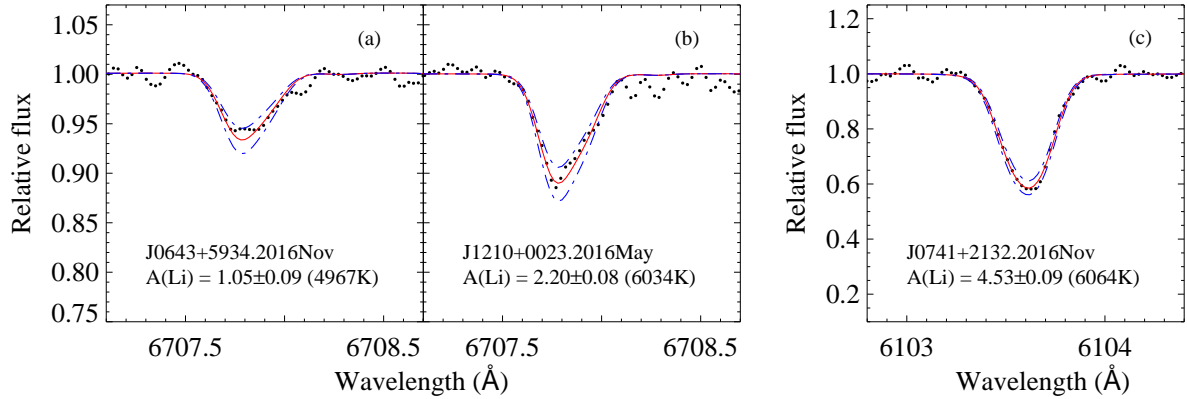}
\caption{Examples of spectral fitting of Li 6708\,{\AA} line for a giant (a) and for a turnoff star (b) 
which present normal lithium abundances, and of Li 6103\,{\AA} line for a Li-rich star (c) whose Li 6708\,{\AA} line becomes saturated under LTE models. 
The observed spectrum is shown with dots, and the best fit and $1\sigma$ uncertainty has been respectively shown in solid and dashed lines.
\label{fig:Li_fitting}}
\end{figure*}

\begin{figure*}
\hspace{-1.5cm}\epsscale{1.1}
\plotone{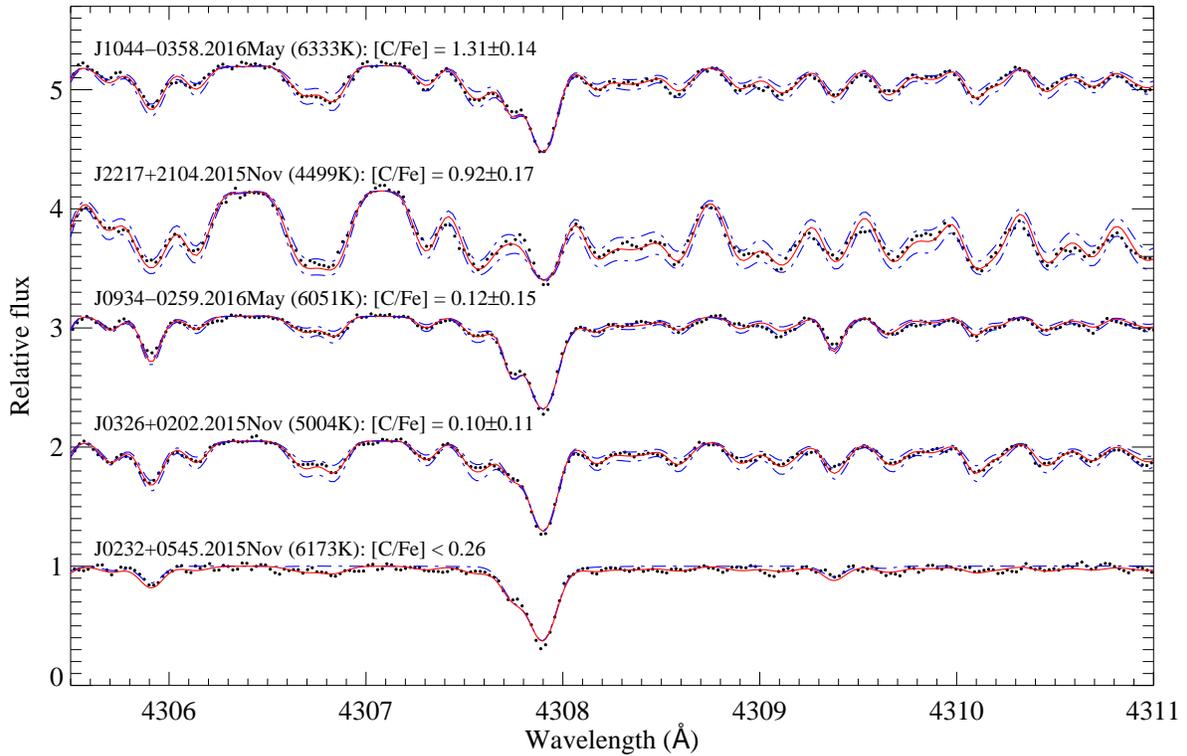}
\caption{Examples of spectral fitting of the CH-band around 4310\,{\AA}. 
The top four spectra respectively represent CEMP (turnoff and giant) and non-CEMP (turnoff and giant). 
Definition of symbols are the same as Figure~\ref{fig:Li_fitting}. 
The bottom spectrum shows the upper limit (solid line), while the dashed line stands for the case with no carbon. 
\label{fig:CH_fitting}}
\end{figure*}

For 19 other elements (Na, Mg, Si, Ca, Sc, Ti, V, Cr, Mn, Fe, Co, Ni, Zn, Sr, Y, Zr, Ba, La, and Eu), 
the abundances were computed from the measured equivalent widths using model atmospheres (see below) with the derived stellar parameters.
If the elemental abundance was derived from a single line, 
or if the derived abundances deviated by more than 3$\,\sigma$ from the average
values computed for an atomic species from multiple lines,
the abundances were examined with spectral synthesis.
For most of the used lines, synthetic spectra with abundances
derived from the measured equivalent widths match the observed spectra well.
However, for lines that suffer from blending or problems in setting the continuum level,
the observed spectral line was not properly reproduced by synthetic spectra with the abundance derived from equivalent widths. 
In such cases, the abundance derived from spectral synthesis was adopted. 
Hyper-fine structure and isotopic splitting have been considered for Sr, Ba, La, and Eu 
\citep{Borghs1983HyInt,McWilliam1995AJ,Lawler2001ApJ_La,Lawler2001ApJ_Eu}. 
Note that the above mentioned analysis is expected to subject to systematic uncertainties 
from NLTE effects. The NLTE effects on spectral features will be studied by 
more detailed analysis for a sub-sample with relatively high-SNR in a separate work of the series. 

The derived abundances (upper-limits) of the program stars are listed in Table~\ref{tab:abun_error},
which also includes the number of lines that have been used for deriving the abundance, $N$,
together with the abundance error as described in the following text. 
The photospheric solar abundances of \citet{Asplund2009ARAA} are adopted when calculating the \AB{X}{Fe} abundance ratios. 

\subsection{Abundances uncertainties}\label{subsec:abun_error}

The errors of the derived abundances of the sample were estimated from two aspects: 
the uncertainties in measurements and in the determination of stellar parameters.
The error due to the uncertainty of measurement was estimated from the standard deviation $\sigma_{\rm X}$ of abundances 
derived from individual lines and the number of lines $N_{\rm X}$, as $\sigma_{\rm X}/\sqrt{N_{\rm X}}$. 
This estimate provides a kind of random error also including the uncertainties of {\it gf} values of spectral lines. 
When the number of adopted atomic lines is smaller than four, 
the larger of $\sigma_{\rm X}/\sqrt{N_{X}}$ and $\sigma({\rm FeI})/\sqrt{N_{X}}$ was then adopted. 
The columns of $\sigma \rm (N)$ and N in Table~\ref{tab:abun_error} 
standard deviation of abundances derived from individual lines
and the number of adopted lines for corresponding species. 

The abundance errors due to uncertainties of the stellar parameters were estimated from changes of abundances 
by individually varying {\tefft}, {\logg}, and {\FeH} according to $1\sigma$ parameter uncertainties. 
In Table~\ref{tab:abun_error}, columns $\sigma${\tefft}, $\sigma${\logg}, $\sigma${\FeH}, and $\sigma \xi$
summarize corresponding quantities in abundance [X/Fe] uncertainties for the program stars, 
and the total error derived by adding individual errors in quadrature is given in the column of ``Total''.
We note that the effect of $\sigma T_{\rm eff}$ more or less cancels in [X/H] and [Fe/H] in many cases, 
resulting in smaller errors in [X/Fe] than that in [Fe/H].

\begin{longrotatetable}
\begin{deluxetable*}{rrrrrrrrrrcrrrrrrrrr}
\tablecaption{Abundances and abundance error estimations for six sample stars}\label{tab:abun_error}
\tablewidth{500pt}
\tabletypesize{\scriptsize}
\tablehead{
}
\startdata
\colhead{} & \multicolumn9c{J1346+1024 (plx\_logg, TO VMP)}& \colhead{} &\multicolumn9c{J0934-0108 (plx\_logg, RGB VMP)}\\
\cline{1-10}\cline{12-20}
Ion  & log$\epsilon$(X)& [X/Fe] & $\sigma$\,{\tefft}& $\sigma$\,{\logg}& $\sigma$\,{\FeH}& $\sigma$\,$\xi$& $\sigma$(N)& Total& N &
     & log$\epsilon$(X)& [X/Fe] & $\sigma$\,{\tefft}& $\sigma$\,{\logg}& $\sigma$\,{\FeH}& $\sigma$\,$\xi$& $\sigma$(N)& Total& N \\
     &                 &        &       77\,K       &       0.02       &        0.10     &   0.12\,km~s$^{-1}$   &            &      &   &
     &                 &        &       44\,K       &       0.04       &        0.09     &   0.05\,km~s$^{-1}$   &            &      &   \\
\cline{1-10}\cline{12-20}
   LiI&    2.21&\nodata&$-$0.01&$-$0.00&$-$0.00&   0.01&   0.10&   0.10&    1&&   1.00&\nodata&   0.00&   0.00&   0.00&   0.01&   0.09&   0.09&    1\\
     C&    6.51&   0.30&   0.09&$-$0.00&   0.02&   0.01&   0.10&   0.14&    1&&   6.17&   0.08&   0.04&$-$0.01&   0.02&   0.01&   0.09&   0.10&    1\\
   NaI&    3.91&$-$0.11&   0.01&$-$0.01&$-$0.00&$-$0.03&   0.07&   0.08&    2&&   3.68&$-$0.22&   0.01&$-$0.01&   0.00&$-$0.01&   0.06&   0.07&    2\\
   MgI&    5.80&   0.42&$-$0.01&$-$0.00&$-$0.00&$-$0.00&   0.10&   0.10&    8&&   5.50&   0.24&$-$0.02&   0.00&   0.00&$-$0.00&   0.09&   0.09&    5\\
   SiI& \nodata&\nodata&\nodata&\nodata&\nodata&\nodata&\nodata&\nodata&    0&&   5.71&   0.54&   0.00&   0.00&   0.00&$-$0.01&   0.09&   0.09&    1\\
   CaI&    4.53&   0.41&$-$0.01&$-$0.00&   0.01&   0.01&   0.09&   0.09&   23&&   4.27&   0.27&$-$0.02&   0.00&   0.00&   0.00&   0.08&   0.08&   24\\
  ScII&    1.17&   0.24&$-$0.02&   0.01&   0.01&   0.00&   0.14&   0.14&   10&&   1.00&   0.19&$-$0.03&   0.02&   0.01&   0.00&   0.11&   0.12&   11\\
  TiII&    3.31&   0.58&$-$0.03&   0.01&$-$0.00&$-$0.01&   0.13&   0.13&   28&&   2.98&   0.37&$-$0.03&   0.01&   0.01&$-$0.01&   0.10&   0.11&   33\\
    VI& \nodata&\nodata&\nodata&\nodata&\nodata&\nodata&\nodata&\nodata&    0&&   1.59&$-$0.00&   0.00&   0.00&   0.00&   0.01&   0.09&   0.09&    1\\
   CrI&    3.26&$-$0.16&   0.01&$-$0.00&   0.01&$-$0.01&   0.09&   0.09&    6&&   3.05&$-$0.25&   0.00&$-$0.01&   0.00&$-$0.01&   0.05&   0.05&   10\\
   MnI& \nodata&\nodata&\nodata&\nodata&\nodata&\nodata&\nodata&\nodata&    0&&   2.64&$-$0.45&$-$0.01&   0.00&   0.00&   0.01&   0.05&   0.05&    3\\
   FeI&    5.28&   0.00&$-$0.00&$-$0.00&$-$0.00&$-$0.00&   0.10&   0.10&  107&&   5.16&$-$0.00&   0.00&   0.00&   0.00&   0.00&   0.09&   0.09&  159\\
  FeII&    5.47&   0.19&$-$0.05&   0.01&$-$0.00&$-$0.00&   0.13&   0.14&   11&&   5.25&   0.09&$-$0.05&   0.01&   0.00&$-$0.00&   0.10&   0.11&   17\\
   CoI&    2.99&   0.22&   0.01&$-$0.00&$-$0.00&   0.00&   0.07&   0.07&    2&&   3.01&   0.36&   0.00&$-$0.01&   0.00&$-$0.02&   0.05&   0.06&    3\\
   NiI&    4.15&   0.15&$-$0.01&$-$0.00&   0.01&   0.01&   0.23&   0.23&    3&&   3.88&   0.00&$-$0.01&   0.00&   0.00&   0.01&   0.09&   0.09&   14\\
   ZnI& \nodata&\nodata&\nodata&\nodata&\nodata&\nodata&\nodata&\nodata&    0&&   2.41&   0.19&$-$0.04&   0.01&   0.00&   0.01&   0.06&   0.08&    2\\
  SrII&    1.08&   0.43&$-$0.00&$-$0.00&$-$0.00&$-$0.07&   0.16&   0.17&    2&&   0.46&$-$0.07&$-$0.01&   0.00&   0.00&$-$0.03&   0.13&   0.13&    2\\
   YII&    0.23&   0.24&$-$0.03&$-$0.00&$-$0.00&   0.00&   0.16&   0.16&    2&&$-$0.51&$-$0.38&$-$0.03&   0.01&   0.01&   0.00&   0.14&   0.14&    2\\
  ZrII& \nodata&\nodata&\nodata&\nodata&\nodata&\nodata&\nodata&\nodata&    0&&   0.24&   0.00&$-$0.02&   0.02&   0.01&   0.01&   0.06&   0.07&    2\\
  BaII& $-$0.02&   0.02&$-$0.01&$-$0.00&$-$0.00&$-$0.03&   0.08&   0.09&    3&&$-$0.37&$-$0.21&$-$0.01&   0.01&   0.01&$-$0.01&   0.04&   0.04&    4\\
  LaII& \nodata&\nodata&\nodata&\nodata&\nodata&\nodata&\nodata&\nodata&    0&&$-$0.90&   0.34&$-$0.02&   0.01&   0.01&   0.01&   0.26&   0.26&    3\\
  EuII& \nodata&\nodata&\nodata&\nodata&\nodata&\nodata&\nodata&\nodata&    0&&$-$1.35&   0.47&$-$0.02&   0.02&   0.01&   0.01&   0.06&   0.07&    2\\
\hline
\hline
\colhead{} & \multicolumn9c{J1410-0555 (plx\_logg, TO EMP)}& \colhead{} &\multicolumn9c{J1733+2633 (plx\_logg, RGB EMP)}\\
\cline{1-10}\cline{12-20}
Ion  & log$\epsilon$(X)& [X/Fe] & $\sigma$\,{\tefft}& $\sigma$\,{\logg}& $\sigma$\,{\FeH}& $\sigma$\,$\xi$& $\sigma$(N)& Total& N &
     & log$\epsilon$(X)& [X/Fe] & $\sigma$\,{\tefft}& $\sigma$\,{\logg}& $\sigma$\,{\FeH}& $\sigma$\,$\xi$& $\sigma$(N)& Total& N \\
     &                 &        &       80\,K       &       0.02       &        0.14     &   0.20\,km~s$^{-1}$   &            &      &   &
     &                 &        &       45\,K       &       0.04       &        0.11     &   0.10\,km~s$^{-1}$   &            &      &   \\
\cline{1-10}\cline{12-20}
   LiI&    2.13&\nodata&$-$0.01&   0.00&   0.00&   0.02&   0.14&   0.14&    1&&    0.96& \nodata&   0.00&   0.00&   0.00&   0.02&   0.11&   0.11&    1\\
     C&    6.97&   1.67&   0.07&   0.00&   0.02&   0.02&   0.14&   0.16&    1&&    5.58&    0.43&   0.04&$-$0.01&   0.02&   0.02&   0.11&   0.12&    1\\
   NaI&    3.45&   0.34&   0.00&   0.00&   0.01&$-$0.02&   0.10&   0.10&    2&&    2.60& $-$0.36&$-$0.01&   0.00&   0.00&   0.00&   0.08&   0.08&    2\\
   MgI&    5.38&   0.91&$-$0.01&   0.00&   0.00&   0.00&   0.09&   0.09&    6&&    4.93&    0.61&$-$0.01&$-$0.01&   0.00&   0.00&   0.09&   0.09&    6\\
   SiI& \nodata&\nodata&\nodata&\nodata&\nodata&\nodata&\nodata&\nodata&    0&&    5.00&    0.77&   0.00&   0.00&   0.01&   0.02&   0.11&   0.11&    1\\
   CaI&    3.63&   0.42&$-$0.02&   0.00&   0.00&   0.01&   0.19&   0.19&    9&&    3.42&    0.36&$-$0.02&   0.00&   0.00&   0.01&   0.12&   0.12&   13\\
  ScII&    0.46&   0.44&$-$0.02&   0.01&   0.00&   0.01&   0.23&   0.23&    2&&    0.24&    0.37&$-$0.03&   0.01&   0.00&   0.00&   0.08&   0.09&    9\\
  TiII&    2.12&   0.30&$-$0.03&   0.00&   0.00&   0.01&   0.14&   0.14&   10&&    2.06&    0.39&$-$0.02&   0.01&   0.01&   0.01&   0.11&   0.11&   24\\
    VI& \nodata&\nodata&\nodata&\nodata&\nodata&\nodata&\nodata&\nodata&    0&& \nodata& \nodata&\nodata&\nodata&\nodata&\nodata&\nodata&\nodata&    0\\
   CrI&    2.28&$-$0.23&   0.00&   0.00&   0.00&   0.01&   0.12&   0.12&    5&&    1.96& $-$0.40&   0.01&   0.00&   0.01&   0.00&   0.08&   0.08&    5\\
   MnI& \nodata&\nodata&\nodata&\nodata&\nodata&\nodata&\nodata&\nodata&    0&& \nodata& \nodata&\nodata&\nodata&\nodata&\nodata&\nodata&\nodata&    0\\
   FeI&    4.37&   0.00&   0.00&   0.00&   0.00&   0.00&   0.14&   0.14&   42&&    4.22& $-$0.00&   0.00&   0.00&   0.00&   0.00&   0.11&   0.11&   95\\
  FeII&    4.39&   0.02&$-$0.05&   0.00&   0.00&   0.01&   0.18&   0.19&    4&&    4.25&    0.03&$-$0.04&   0.02&   0.01&   0.02&   0.11&   0.12&   12\\
   CoI&    2.53&   0.67&   0.01&   0.00&   0.01&   0.02&   0.10&   0.10&    2&&    2.01&    0.30&   0.01&$-$0.01&   0.00&   0.01&   0.09&   0.09&    2\\
   NiI& \nodata&\nodata&\nodata&\nodata&\nodata&\nodata&\nodata&\nodata&    0&&    2.88& $-$0.06&   0.00&   0.00&   0.00&   0.01&   0.11&   0.11&    1\\
   ZnI& \nodata&\nodata&\nodata&\nodata&\nodata&\nodata&\nodata&\nodata&    0&& \nodata& \nodata&\nodata&\nodata&\nodata&\nodata&\nodata&\nodata&    0\\
  SrII& $-$0.11&   0.15&$-$0.01&   0.01&   0.00&$-$0.03&   0.10&   0.10&    2&& $-$1.12& $-$0.71&$-$0.01&   0.01&   0.00&$-$0.01&   0.08&   0.08&    2\\
   YII& \nodata&\nodata&\nodata&\nodata&\nodata&\nodata&\nodata&\nodata&    0&& \nodata& \nodata&\nodata&\nodata&\nodata&\nodata&\nodata&\nodata&    0\\
  ZrII& \nodata&\nodata&\nodata&\nodata&\nodata&\nodata&\nodata&\nodata&    0&& \nodata& \nodata&\nodata&\nodata&\nodata&\nodata&\nodata&\nodata&    0\\
  BaII& $-$1.05&$-$0.10&$-$0.01&   0.01&   0.01&   0.02&   0.10&   0.10&    2&& $-$2.22& $-$1.12&$-$0.01&   0.01&   0.01&   0.02&   0.08&   0.08&    2\\
  LaII& \nodata&\nodata&\nodata&\nodata&\nodata&\nodata&\nodata&\nodata&    0&& \nodata& \nodata&\nodata&\nodata&\nodata&\nodata&\nodata&\nodata&    0\\
  EuII& \nodata&\nodata&\nodata&\nodata&\nodata&\nodata&\nodata&\nodata&    0&& \nodata& \nodata&\nodata&\nodata&\nodata&\nodata&\nodata&\nodata&    0\\
\hline
\hline
\colhead{} & \multicolumn9c{J1432+3755 (corrected\_logg, RGB EMP)}& \colhead{} &\multicolumn9c{J1102+0102 (corrected\_logg, RGB VMP)}\\
\cline{1-10}\cline{12-20}
Ion  & log$\epsilon$(X)& [X/Fe] & $\sigma$\,{\tefft}& $\sigma$\,{\logg}& $\sigma$\,{\FeH}& $\sigma$\,$\xi$& $\sigma$(N)& Total& N &
     & log$\epsilon$(X)& [X/Fe] & $\sigma$\,{\tefft}& $\sigma$\,{\logg}& $\sigma$\,{\FeH}& $\sigma$\,$\xi$& $\sigma$(N)& Total& N \\
     &                 &        &       24\,K       &       0.05       &        0.14     &   0.10\,km~s$^{-1}$   &            &      &   &
     &                 &        &       45\,K       &       0.07       &        0.09     &   0.05\,km~s$^{-1}$   &            &      &   \\
\cline{1-10}\cline{12-20}
   LiI&\nodata&\nodata&\nodata&\nodata&\nodata&\nodata&\nodata&\nodata&    0&&   0.99&\nodata&   0.01&   0.00&   0.00&   0.01&   0.09&   0.09&    1\\
     C&   4.95&$-$0.25&   0.04&$-$0.10&   0.02&   0.01&   0.14&   0.18&    1&&   6.08&$-$0.03&   0.04&$-$0.09&   0.03&   0.01&   0.09&   0.14&    1\\
   NaI&   3.12&   0.11&$-$0.00&$-$0.01&$-$0.01&$-$0.05&   0.10&   0.11&    2&&   4.20&   0.28&   0.02&$-$0.02&   0.01&$-$0.00&   0.09&   0.09&    2\\
   MgI&   4.86&   0.49&$-$0.01&$-$0.01&$-$0.01&$-$0.01&   0.12&   0.12&    8&&   5.48&   0.20&$-$0.02&$-$0.01&$-$0.01&$-$0.00&   0.09&   0.09&    6\\
   SiI&   4.83&   0.55&$-$0.01&$-$0.01&$-$0.01&$-$0.01&   0.14&   0.14&    1&&   5.43&   0.24&   0.00&   0.00&   0.00&$-$0.01&   0.09&   0.09&    1\\
   CaI&   3.43&   0.32&$-$0.01&$-$0.01&   0.00&$-$0.00&   0.09&   0.09&   17&&   4.22&   0.20&$-$0.02&   0.00&   0.00&   0.00&   0.10&   0.10&   23\\
  ScII&   0.00&   0.08&$-$0.02&   0.01&   0.00&$-$0.01&   0.10&   0.10&   12&&   0.98&   0.15&$-$0.03&   0.02&   0.01&   0.00&   0.12&   0.13&   13\\
  TiII&   2.06&   0.34&$-$0.02&   0.01&   0.00&$-$0.02&   0.11&   0.11&   32&&   3.07&   0.44&$-$0.04&   0.02&   0.01&$-$0.01&   0.14&   0.15&   33\\
    VI&   0.64&$-$0.06&$-$0.00&   0.00&   0.00&   0.01&   0.14&   0.14&    1&&   1.47&$-$0.14&   0.01&   0.00&   0.00&   0.01&   0.09&   0.09&    1\\
   CrI&   2.11&$-$0.30&$-$0.00&$-$0.01&$-$0.01&$-$0.01&   0.17&   0.17&    9&&   3.06&$-$0.26&   0.01&$-$0.01&$-$0.01&$-$0.01&   0.09&   0.09&   10\\
   MnI&   1.79&$-$0.41&$-$0.00&   0.00&   0.00&   0.01&   0.11&   0.11&    2&&   2.57&$-$0.54&   0.00&   0.00&   0.00&   0.01&   0.05&   0.05&    3\\
   FeI&   4.27&   0.00&   0.00&   0.00&   0.00&   0.00&   0.14&   0.14&  149&&   5.18&$-$0.00&   0.00&   0.00&   0.00&   0.00&   0.09&   0.09&  147\\
  FeII&   4.41&   0.14&$-$0.03&   0.02&   0.01&   0.01&   0.08&   0.09&   15&&   5.33&   0.15&$-$0.06&   0.02&   0.01&$-$0.01&   0.07&   0.10&   15\\
   CoI&   2.03&   0.27&   0.01&   0.00&   0.00&$-$0.01&   0.10&   0.10&    3&&   2.91&   0.24&   0.02&   0.00&   0.00&$-$0.02&   0.20&   0.20&    3\\
   NiI&   3.05&   0.06&$-$0.01&   0.00&   0.00&   0.01&   0.12&   0.12&   10&&   3.82&$-$0.08&$-$0.01&   0.00&   0.00&   0.01&   0.12&   0.12&   15\\
   ZnI&   1.51&   0.18&$-$0.02&   0.01&   0.00&   0.01&   0.22&   0.22&    2&&   2.43&   0.19&$-$0.04&   0.02&   0.01&   0.01&   0.06&   0.08&    2\\
  SrII&$-$0.41&$-$0.05&$-$0.01&   0.01&$-$0.01&$-$0.08&   0.10&   0.13&    2&&   0.69&   0.14&$-$0.01&   0.00&   0.01&$-$0.01&   0.10&   0.10&    2\\
   YII&$-$1.44&$-$0.42&$-$0.02&   0.01&   0.00&$-$0.00&   0.11&   0.11&    3&&$-$0.31&$-$0.20&$-$0.03&   0.03&   0.01&   0.01&   0.18&   0.19&    4\\
  ZrII&$-$0.39&   0.26&$-$0.02&   0.01&   0.00&$-$0.00&   0.14&   0.14&    2&&   0.50&   0.24&$-$0.03&   0.02&   0.01&   0.00&   0.21&   0.21&    2\\
  BaII&$-$1.88&$-$0.83&$-$0.01&   0.01&   0.00&$-$0.01&   0.17&   0.17&    4&&$-$0.27&$-$0.13&$-$0.02&   0.01&   0.01&$-$0.02&   0.05&   0.06&    4\\
  LaII&\nodata&\nodata&\nodata&\nodata&\nodata&\nodata&\nodata&\nodata&    0&&\nodata&\nodata&\nodata&\nodata&\nodata&\nodata&\nodata&\nodata&    0\\
  EuII&\nodata&\nodata&\nodata&\nodata&\nodata&\nodata&\nodata&\nodata&    0&&$-$1.36&   0.44&$-$0.02&   0.03&   0.01&   0.00&   0.07&   0.08&    2\\
\enddata                                                                                               
\tablecomments{The table of elemental abundances and abundance errors for all program stars is published in its entirety in the machine-readable format. 
Six example objects are shown here for guidance regarding the level of errors and the content of the table.}
\end{deluxetable*}
\end{longrotatetable}

Since our program stars cover a rather wide range in the evolutionary status as well as metallicities, and include two sets of {\logg} estimations, 
we consider the following cases when selecting example stars to present the typical abundance uncertainties for the whole sample: 
VMP/EMP stars, giant/turnoff stars which shall present different line numbers and strengths, parallax-derived {\logg} and spectroscopic-corrected {\logg}. 
However, one should note that, as indicated in Figure ~\ref{fig:logg_correction}, the spectroscopic-corrected {\logg} case is in general 
only adopted on giants; therefore, six example targets have been selected to present the typical abundance uncertainties, as given in Table~\ref{tab:abun_error}. 
It should be noted that, for species whose abundances have been derived from strong lines (e.g., resonance lines for Sr and Ba), 
the systematic error caused by uncertainties in microturbulent velocities is rather significant. 

\subsection{Giants and turnoff stars}\label{subsec:RGBvsTO}

\begin{figure*}
\hspace{-1.5cm}\epsscale{1.25}
\plotone{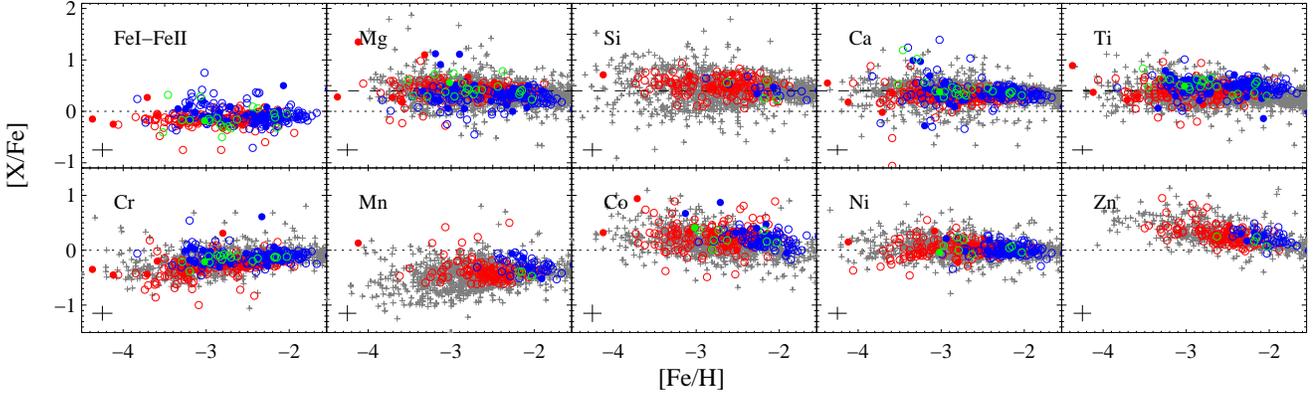}
\caption{Distribution of $\alpha$-element (Mg, Si, Ti, and Ca) and iron-group (Cr, Mn, Ni, Co, and Zn) abundances 
for the program stars (open and filled circles) along metallicities. Red, blue, and green symbols respectively refer to giants, turnoff, 
and horizontal-branch stars (defined as in Table~\ref{tab:class_def}), while CEMP and non-CEMP stars are shown with filled and open circles, respectively.
Literature results obtained from the SAGA database are presented with pluses for comparison. 
\label{fig:abun_RGBvsTO}}
\end{figure*}

To make sure that there is no significant offset between abundances derived for turn-off stars and giants 
due to technical problems in the abundance analysis, 
we here compare the elemental abundances of several key elements with relatively small internal scatters, 
including $\alpha$-elements (Mg, Si, Ca, and Ti) and iron-group elements (Fe, Cr, Mn, Ni, Co, and Zn). 
For these elements, the surface abundances are expected to be unchanged through the evolution from turnoff stars to giants.  
The comparison of the derived abundances for giants (red circles) and turnoff stars (blue circles) 
are shown in Figure~\ref{fig:abun_RGBvsTO}. 
In general, there is no systematic difference between the abundance distribution of giants and turnoff stars, 
while there are two exceptions, i.e., Mg and Cr, 
where offsets can be found in the abundance trends between giants and turnoff stars, 
and could mostly be explained by the NLTE effects. 
For Mg, the average \AB{Mg}{Fe} for giants is about 0.1\,dex higher than that of turnoff stars 
(see \S~\ref{subsubsec:alpha-elements}). 
Such difference has also been discovered in the FS sample (as shown in Fig. 4 in \citealt{Bonifacio2009AA}), 
though they have found much larger discrepancies where EMP turnoff stars have $0.2\sim0.3$\,dex lower \AB{Mg}{Fe} than EMP giants. 
The abundance difference observed in our sample is on average consistent with the NLTE correction difference 
between giants and turnoff stars at \FeHlt{-2.0} (e.g., \citealt{Osorio&Barklem2016AA}).
According to \citet{Sobeck2007ApJ}, a systematic difference between giants and turnoff stars for Cr 
is caused by metallicity dependence of \ion{Cr}{1} lines probably due to the NLTE effect (see \S~\ref{subsubsec:iron-peak-elements}). 
The dependence on effective temperatures is also reported and discussed in \citet{Suda2011MNRAS}.
Note that for Si, Mn, and Zn, the number of turnoff stars at low metallicity is insufficient for this comparison. 

\subsection{Comparison with literature}

\begin{figure*}
\plotone{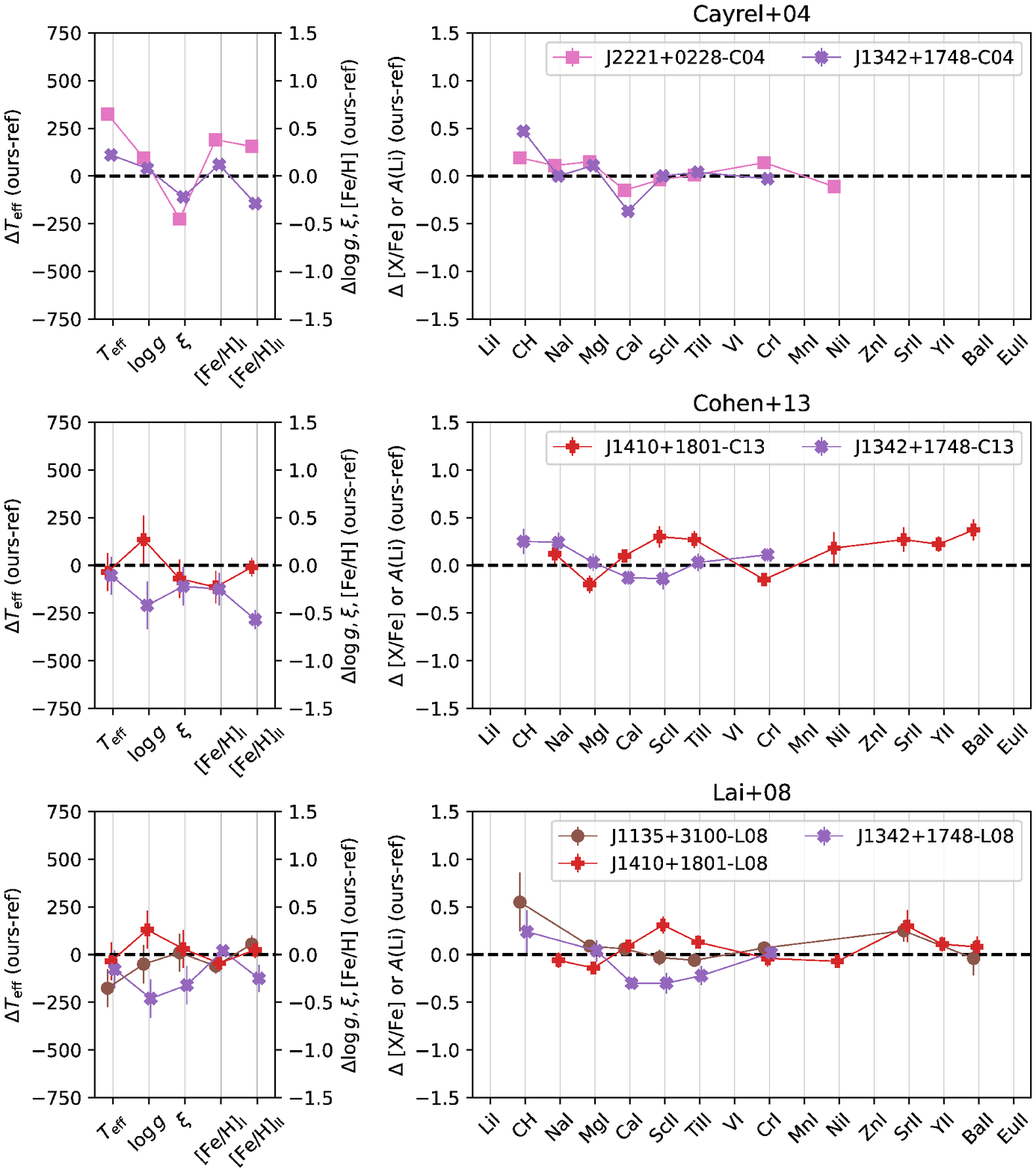}
\caption{Stellar parameters and chemical abundance comparison with \citet{Cayrel2004AA}, \citet{Cohen2013ApJ}, and \citet{Lai2008ApJ}. 
The same objects in different studies are shown with the same symbol in this figure. 
See Table~\ref{tab:literature_summary} for the analysis methods adopted in the literature.\label{fig:abcompare1}}
\end{figure*}

\begin{figure*}
\plotone{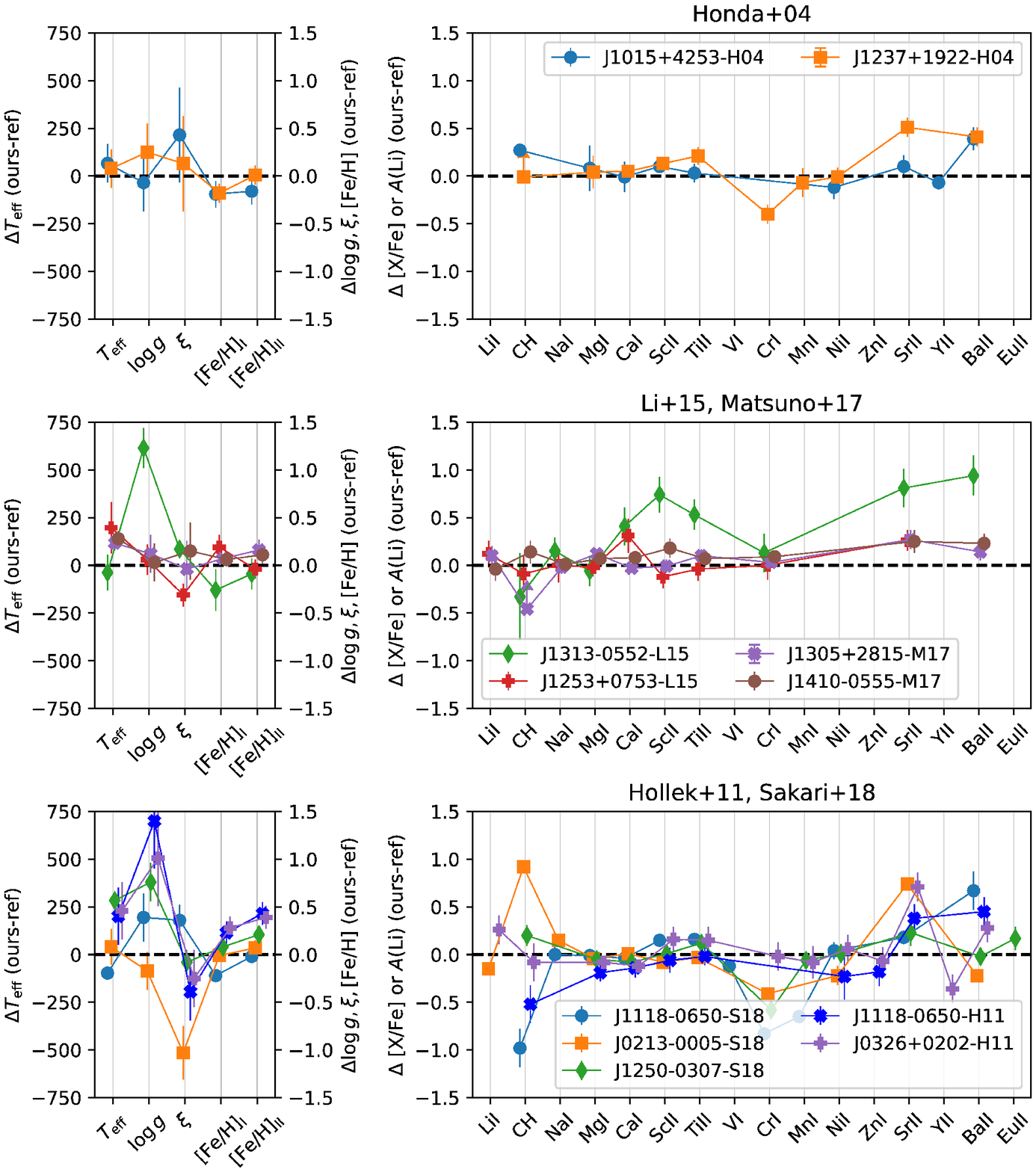}
\caption{Same as Figure~\ref{fig:abcompare1} but for comparisons with \citet{Honda2004ApJ,Honda2004ApJS}, \citet{Li2015PASJ}, \citet{Matsuno2017PASJ}, \citet{Hollek2011ApJ}, and \citet{Sakari2018ApJ}. Note that the same spectra from \citet{Li2015PASJ} and \citet{Matsuno2017PASJ} are used for the analysis in this paper. Unlike Figure~\ref{fig:abcompare1}, the same symbol does not correspond to the same object in this figure.\label{fig:abcompare2}}
\end{figure*}

\begin{figure*}
\plotone{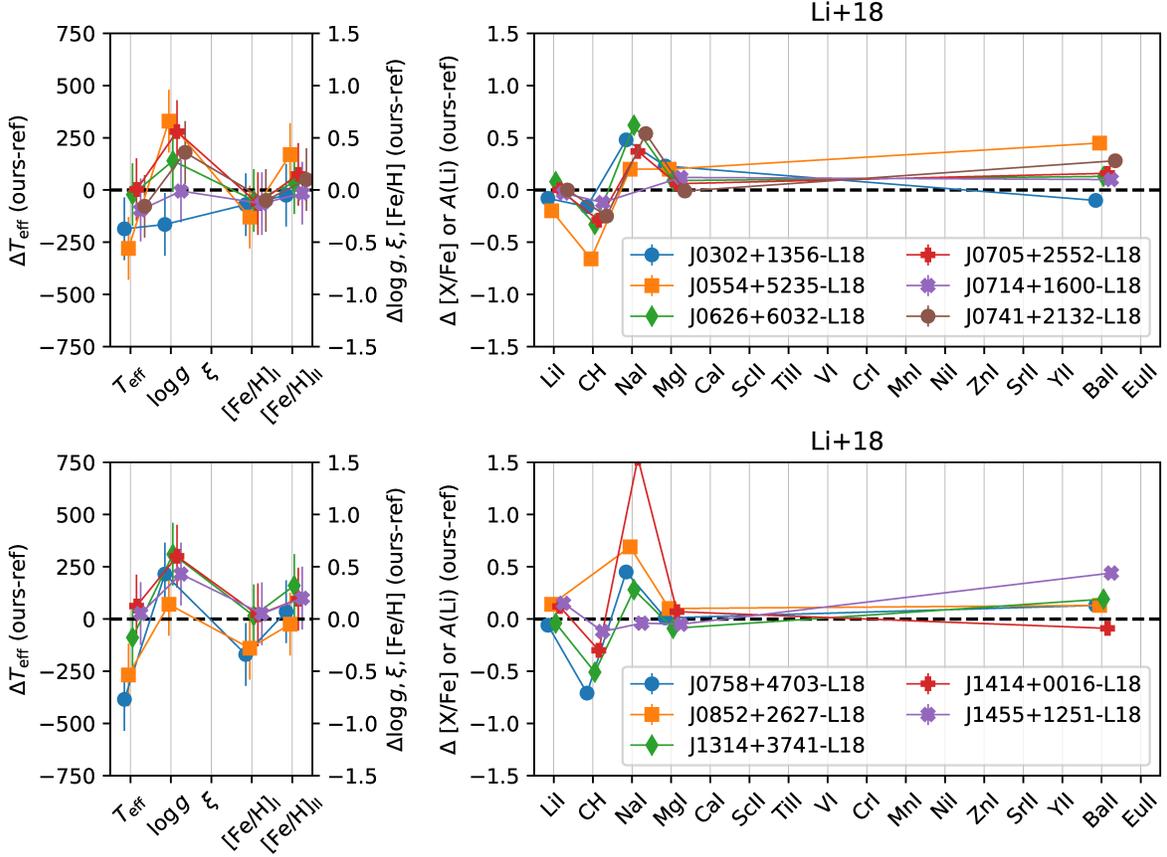}
\caption{Same as Figure~\ref{fig:abcompare1} but for comparisons with \citet{Li2018ApJL}. Note that the same spectra are used for the analysis in this paper and in \citet{Li2018ApJL}. Unlike Figure~\ref{fig:abcompare1}, the same symbol does not correspond to the same object in this figure.\label{fig:abcompare3}}
\end{figure*}

For 22 stars in our sample, we have found high-resolution 
spectroscopic analyses compiled by the SAGA database \citep{Suda2008PASJ,Suda2011MNRAS,Yamada2013MNRAS,Suda2017PASJ}. 
Figures~\ref{fig:abcompare1}--\ref{fig:abcompare3} present comparisons of stellar parameters and 
abundances for individual objects that are in common between the present study and literatures. 
The literatures considered here have measured abundances for multiple elements and have more than one common stars with our sample.
Readers may refer to Table~\ref{tab:literature_summary} for the methods adopted in the literature to derive stellar parameters.

There are generally good agreements in effective temperatures between our study and literatures that adopt photometric temperature
\citep{Cayrel2004AA,Honda2004ApJ,Cohen2013ApJ,Lai2008ApJ}. 
Poor agreements ($|\Delta T_{\rm eff}|>200\,\mathrm{K}$) are found for J2221+0228 (CS 29502--042) from \citet{Cayrel2004AA}.
We adopt $E(B-V) = 0.09$ for this object  
which is in the top 10 percentile from the largest among our sample.
Because of this large extinction, the photometric temperature estimate could be more likely affected by the uncertainty in $E(B-V)$, the extinction law, and the choice of color from which temperatures are estimated. 
The corresponding $E(B-V)$ value in the literature is set to 0.00.
The different $E(B-V)$ assumed is at least partially the reason for the discrepancy.

Literature that estimates effective temperatures from excitation balance of Fe~I lines reports lower values than this study \citep{Hollek2011ApJ}.
This systematic offset between spectroscopic and photometric temperatures has been known and is considered to be due to NLTE effects \citep[e.g.,][]{Frebel2013ApJ,Amarsi2016MNRAS}.
\citet{Sakari2018ApJ} adopts the same method but with line-by-line/mpstar $\langle 3D \rangle$ NLTE corrections by \citet{Amarsi2016MNRAS}, which, in fact, mitigates the offset. 
\citet{Li2015PASJ} and \citet{Li2018ApJL} also estimated temperature using Fe~I lines, but their $T_{\rm eff}$'s are in good agreement with the present study thanks to the correction by \citet{Frebel2013ApJ} that has been applied in these studies. 

A good agreement is found in comparisons with the results of \citet{Matsuno2017PASJ}. 
They derived $T_{\rm eff}$ based on the high-resolution spectra making use of the profiles of the Balmer line wings that depend on $T_{\rm eff}$ and $\log g $ of the stars. 
Such $T_{\rm eff}$ scale is not dependent on \ion{Fe}{1} lines.

Comparisons of surface gravity reveal a large scatter in $\Delta \log g$. 
Our $\log g$ values for these objects are determined using {\it Gaia} DR2 parallax, 
which could be more direct and reliable method than others (see \S~\ref{subsec:stellar_parameter}).
Since the literature work is done before {\it Gaia} DR2, 
they had to rely on ionization balance of Fe or theoretical isochrones that provide a relation between $T_{\rm eff}$ and $\log g$. 
Since both methods introduce a strong correlation between the estimated $T_{\rm eff}$ and $\log g$,  
the large scatter reflects not only the difference in the method to determine $\log g$ but also the difference in adopted $T_{\rm eff}$.
In addition, ionization balance of Fe can be affected by the NLTE effect, which can introduce additional systematic errors.

Figures \ref{fig:abcompare1}--\ref{fig:abcompare3} shows that the offset in metallicity, especially {\FeH} from \ion{Fe}{1} lines ([\ion{Fe}{1}/H]), 
correlates with that in $T_{\rm eff}$; underestimated $T_{\rm eff}$ leads to underestimated {\FeH}.
An available number of Fe~I lines in a spectrum is usually large enough that the choice of lines or the choice of oscillator strengths hardly affect the derived abundances. 
Instead, systematic uncertainties introduced by uncertain stellar parameters become the dominant source of uncertainties 
since the choice of stellar parameters affects the ionization and excitation states of Fe lines (see also Table~\ref{tab:abun_error}).

The typical difference between the abundances from this work and those from literature ($|\Delta([\mathrm{X/Fe}])|$) 
is $<0.13$ and $<0.22$ ~dex for 50\% and 68\%, respectively, 
of the sample presented in Figures~\ref{fig:abcompare1}--\ref{fig:abcompare3}. 
This level of difference is explained by the scatter seen in the comparisons of equivalent widths (Figures~\ref{fig:ewlite}) 
and the possible offset of stellar parameters as discussed above. 
A reason for large discrepancies found in some species is the difference of the spectral lines adopted for the analysis.
For example, for the Li-rich VMP stars reported by \citet{Li2018ApJL}, 
quite different Na abundances have been derived for a few objects as shown in the bottom panel of Figure~\ref{fig:abcompare3}. 
This is because, for these objects, the subordinate Na lines at 5682\,{\AA} and 5688\,{\AA} are detectable 
and thus the resonance NaD lines are not included in abundance estimation, 
while, for the purpose of a uniform analysis in this work, NaD lines have been used for all objects whenever they are measurable.   
The differences are mainly caused by the NLTE effect on the NaD lines. 
Indeed, the NLTE Na abundances derived from this work result in better agreement with those derived by \citet{Li2018ApJL}. 
We note that the relatively large differences are also seen in the abundance of Sr, 
which could be mainly due to its sensitivity to changes in microturbulent velocity (as discussed in \S~\ref{subsec:abun_error}), 
and also affected by the choice of different atomic data. 

\section{Interpretation of the results}\label{sec:discussion}

\begin{deluxetable}{r|rrcc}
\tablecaption{Definition of different classes of program stars\label{tab:class_def}}
\tablehead{
\colhead{Class} & \colhead{Definition} & \colhead{Number} & \colhead{Symbol} & \colhead{Figure}
}
\startdata
Red Giant Branch (RGB) & {\tefft} $< 5500$\,K & 194 & red & Figure~\ref{fig:abun_RGBvsTO}, \ref{fig:abun_trend_lithium}, \ref{fig:abun_trend_light_obs} through ~\ref{fig:abun_corr_heavy}\\
Horizontal Branch (HB) & {\tefft} $\ge 5500$\,K \& {\logg} $< 3.0$ & 19 & green & Figure~\ref{fig:abun_RGBvsTO}, \ref{fig:abun_trend_lithium}, \ref{fig:abun_trend_light_obs} through ~\ref{fig:abun_corr_heavy}\\
Turnoff (TO) & {\tefft} $\ge 5500$\,K \& {\logg} $\ge 3.0$ & 172 & blue & Figure~\ref{fig:abun_RGBvsTO}, \ref{fig:abun_trend_lithium}, \ref{fig:abun_trend_light_obs} through ~\ref{fig:abun_corr_heavy}\\
C-Enhanced (CEMP) & \citet{Aoki2007ApJ} criteria & 49 & filled & Figure~\ref{fig:abun_RGBvsTO}, \ref{fig:abun_trend_lithium}, \ref{fig:abun_trend_light_obs} through ~\ref{fig:abun_corr_heavy}\\
non-CEMP$^{a}$ & not a confirmed CEMP & 336 & open & Figure~\ref{fig:abun_RGBvsTO}, \ref{fig:abun_trend_lithium}, \ref{fig:abun_trend_light_obs} through ~\ref{fig:abun_corr_heavy}\\
\enddata
\tablenotetext{a}{This number does not exclude program stars whose C abundances have not been measured or with upper limits.}
\end{deluxetable}

\subsection{Abundance trends of large sample of VMP stars} \label{subsec:abundance_trend}

Elemental abundances of stars are characterized by the nucleosynthesis
history before the formation of the stars and, hence, a large sample
of Galactic VMP stars covering a wide metallicity range can provide important
clues in understanding the synthesis of elements by early generations
of stars and chemical evolution in stellar systems that have been formed in
the Milky Way.

The observed abundance trends of \AB{X}{Fe} along metallicities for individual species 
are presented in Figures~\ref{fig:abun_trend_lithium}, \ref{fig:abun_trend_light_obs}, \ref{fig:abun_trend_iron_obs}, and \ref{fig:abun_trend_heavy_obs} 
separately for Li and three groups of elements (light, iron-peak, and heavy). 
Figures \ref{fig:li_logg} and \ref{fig:CH_Teff} will be discussed in the subsections on Li and C abundances, respectively.
Note that for the observed Li abundance trend (Figure~\ref{fig:abun_trend_lithium}), 
fittings are separately derived for the two metallicity ranges divided at \FeHeq{-2.5}, 
and objects that show enhancement in Li are excluded for deriving the trend. 
For abundance trends (slope, average, dispersion, etc.) presented in Figures~\ref{fig:abun_trend_light_obs}, 
\ref{fig:abun_trend_iron_obs}, and \ref{fig:abun_trend_heavy_obs}, 
only program stars with SNR higher than 50 which is about the average SNR of our program stars (open circle + crosses) are adopted for the fitting, 
while those with relatively low SNR are also shown (open circles) in the figures for completeness. 
During the following discussions, the sample has been divided into six categories, 
i.e., non-CEMP and CEMP giants, non-CEMP and CEMP horizontal branch stars, 
and non-CEMP and CEMP turnoff stars, as defined in Table~\ref{tab:class_def}.  
Compared to the ``general trend'' of elemental abundances of the so-called ``normal population'' among VMP and EMP stars,
CEMP stars frequently present anomalous abundance patterns. Although the origins of carbon-excess would not be unique, 
the anomaly of the abundance patterns in CEMP stars would arise as the descendants of peculiar progenitors or as the contamination of their surface abundances with carbon.
In this paper, the general abundance trend for each species is evaluated based on non-CEMP stars (whose spectral SNR is better than 50), 
while the CEMP objects are presented separately. 
Note that we here adopt the criterion of CEMP stars of  \citet{Aoki2007ApJ} including the
effect of the evolution on the red giant branch, 
represented by the luminosity: \ABge{C}{Fe}{+0.7} for stars with $\log(L/L_{\odot})\le2.3$; 
and \ABge{C}{Fe}{3.0}$-\log(L/L_{\odot})$ for more luminous objects. 

\subsubsection{Lithium} \label{subsubsec:lithium}
Among all the elements, 
lithium is of special importance as it is with primordial origin in metal-poor stars 
and is considered as a key diagnostic to constrain our understanding of the Big Bang nucleosynthesis, 
as well as low-mass star formation and evolution in the early Milky Way. 
It has long been a problem that the primordial lithium abundance predicted by the standard Big Bang nucleosynthesis 
\citep{Cyburt2016a,Fields2019a} is nearly 0.5\,dex higher than 
what has been observed in the oldest stars of the Galaxy as the Spite plateau \citep{Spite1982AA,Charbonnel&Primas2005AA}. 

The trend of the Li abundances at lower metallicity is still unclear compared to the above metallicity range. 
It has been reported that significant fraction of stars show Li abundance below the Spite plateau value at \FeHlt{-2.5} \citep[e.g.][]{Thorburn1994ApJ,Norris1994ApJ,Ryan1999ApJ,Asplund2006ApJ,Aoki2009ApJ}, 
which is called meltdown of the Li Spite plateau for EMP stars \citep{Sbordone2010AA}. 
It is important to precisely understand how the Li abundance is distributed to understand astrophysical origin of this trend. The observational evidence is, however, sometimes confusing and contradicting each other. 
For example, while \citet{Melendez2010AA} suggest that stars form another plateau at \FeHlt{-2.5} with A(Li)$\sim 2.18$, which is lower than the Spite plateau value ($\sim2.27$ in their studies), \citet{Sbordone2010AA} detect a strong positive correlation between the lithium abundance and metallicity with a slope of 0.30 with $2-3\sigma$ of significance. 
Although \citet{Matsuno2017AJ} suggest that stars have almost constant Li abundance at \FeHlt{-3.5} with A(Li)$\sim 1.9$, \citet{Bonifacio2018AA} and \citet{Aguado2019ApJ} reported higher Li abundance for stars with \FeHsim{-4.0} and \FeHlt{-6.1}.

\begin{figure}
\epsscale{0.9}
\plotone{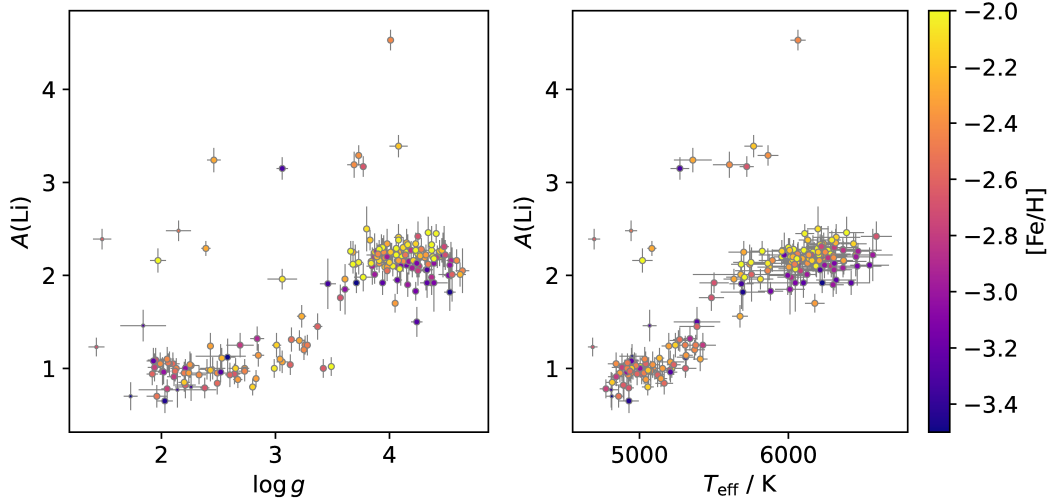}
\caption{Li abundance of stars as a function of surface gravity and temperature. The color represent its iron abundance. 
Stars whose surface gravity relies on \ion{Fe}{1}/\ion{Fe}{2} ionization balance are shown with smaller symbols. 
Horizontal branch stars (\tefft$>5500$~K and \logg$<3.0$) do not appear in this figure because Li is not detected in these stars. 
\label{fig:li_logg}}
\end{figure}

\begin{figure*}
\hspace{-1.5cm}\epsscale{1.25}
\plotone{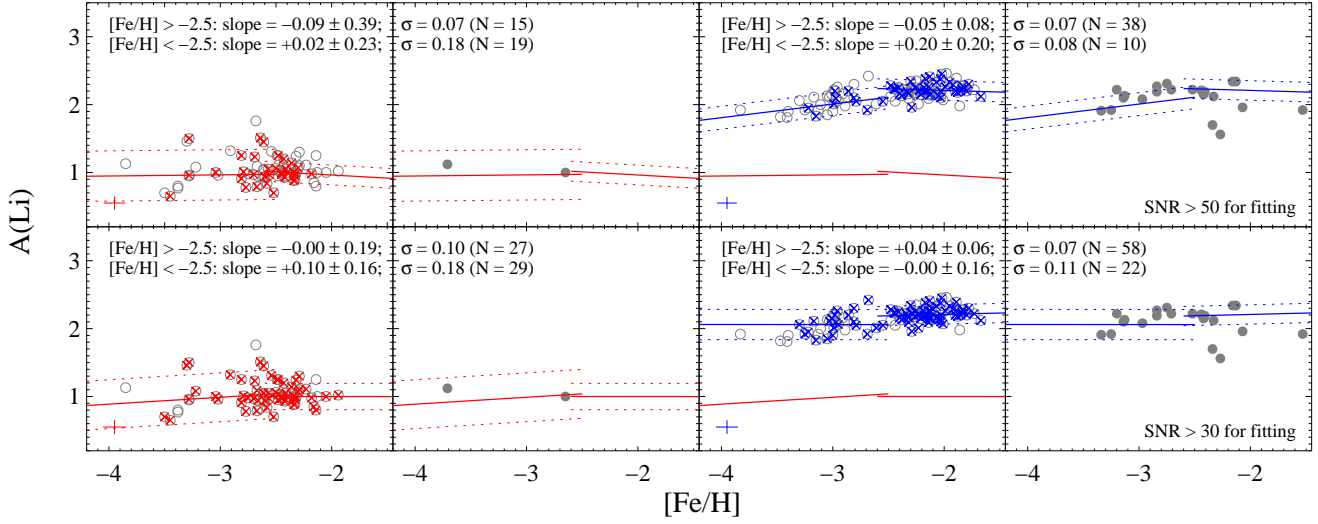}
\caption{
The abundance trend along metallicities is shown for lithium. 
The four columns, from the left to right, respectively refer to non-CEMP giants, CEMP giants, non-CEMP turnoff stars, and CEMP turnoff stars, as defined in Table~\ref{tab:class_def}. 
Open circles represent the non-CEMP objects, while filled circles for the CEMP stars. 
The typical error is shown in the lower left corner. 
Li-rich objects are not included in this plot or fitting. All non-CEMP objects which are not enhanced in lithium 
and have better spectral quality with SNR $> 50$ (top panels) and SNR $> 30$ (bottom panels) are marked with crosses 
and used to fit the abundance trend (red and blue solid lines), 
while the red and blue dotted lines show the $2\sigma$ scatter. 
In both cases, a segmented fitting has been derived respectively for the two metallicity regions 
divided at \FeHsim{-2.5}, which allows sufficient number of objects in both regions. 
The fitted trend for giants is also shown in panels of turnoff stars for comparison.
There are no horizontal-branch stars in this plot, 
since Li is not detected for these objects. 
See the main text in \ref{subsubsec:lithium} for more details. 
\label{fig:abun_trend_lithium}}
\end{figure*}

It is clearly seen in Figure \ref{fig:abun_trend_lithium} that the turnoff stars in our sample distribute around the plateau at \FeHsim{-2}. On the other hand, 
it shows a decreasing trend with decreasing metallicity from $A$(Li)=2.1\,dex at \FeHsim{-2.5} 
to $A$(Li)=1.8\,dex at \FeHsim{-4.0}. 
These apparent features are consistent with previous findings, 
i.e., the Li meltdown at \FeHlt{-2.5}.

\begin{deluxetable*}{c|rrcccrrcc}
\tablecolumns{10}
\tabletypesize{\scriptsize}
\tablecaption{The observed abundance trend of lithium \label{tab:abun_trend_fit_lithium}}
\tablewidth{0pt}
\tablehead{
\colhead{} & \multicolumn9c{SNR$ > 50$} \\
\cline{2-10} 
\colhead{} & \multicolumn4c{giant} & \colhead{} & \multicolumn4c{turnoff}\\
\cline{2-5}\cline{7-10}
\colhead{species}&\colhead{A}&\colhead{B}&\colhead{scatter}&\colhead{mean}&\colhead{}&
                 \colhead{A}&\colhead{B}&\colhead{scatter}&\colhead{mean}\\
\colhead{}&\colhead{}&\colhead{}&\colhead{dex}&\colhead{dex}&\colhead{}&
                 \colhead{}&\colhead{}&\colhead{dex}&\colhead{dex}
}
\startdata
     $A$(Li)&1.11$\pm$0.79&    0.05$\pm$0.30&0.26&1.03& &2.56$\pm$0.08&   0.17$\pm$0.03&0.08&2.18\\
\FeHgt{-2.5}&0.78$\pm$0.92& $-$0.09$\pm$0.39&0.07&1.02& &2.12$\pm$0.16&$-$0.05$\pm$0.08&0.07&2.23\\
\FeHlt{-2.5}&1.01$\pm$0.64&    0.02$\pm$0.23&0.18&1.03& &2.61$\pm$0.58&   0.20$\pm$0.20&0.08&2.04\\
\cline{1-10} 
\colhead{} & \multicolumn9c{SNR$ > 30$} \\
\cline{1-10} 
\colhead{} & \multicolumn4c{giant} & \colhead{} & \multicolumn4c{turnoff}\\
\cline{1-5}\cline{7-10}
     $A$(Li)&1.00$\pm$0.20&   0.00$\pm$0.08&0.14&1.04& &2.54$\pm$0.06&   0.16$\pm$0.03&0.08&2.17\\
\FeHgt{-2.5}&1.00$\pm$0.45&   0.00$\pm$0.19&0.10&1.03& &2.29$\pm$0.13&   0.04$\pm$0.06&0.07&2.22\\
\FeHlt{-2.5}&1.29$\pm$0.45&   0.10$\pm$0.16&0.18&1.05& &2.06$\pm$0.48&   0.00$\pm$0.16&0.11&2.07\\
\enddata
\tablecomments{The columns A and B correspond to the intercept and slope of the linear fitting of A(Li) = A + B$\times${\FeH} as shown in Figure~\ref{fig:abun_trend_lithium}}
\end{deluxetable*}

In the top panels of Figure~\ref{fig:abun_trend_lithium} and Table~\ref{tab:abun_trend_fit_lithium}, 
all non-CEMP (and not Li-enhanced) stars with spectral quality of SNR $> 50$ have been used to derive the abundance trend.
For the turnoff stars, the average of $A$(Li) is 2.23 for the plateau at \FeHgt{-2.5}, 
presenting no significant slope ($\Delta A$(Li)/$\Delta$[Fe/H]$=-0.05$), 
which well agrees with the result from \citet{Melendez2010AA}. 
However, unlike their conclusion of almost no slope for both plateaus, 
a rather steep slope of 0.20 is found for our sample in the metallicity range between \FeHsim{-3.5} and $-2.5$, 
which agrees well with what has been found by \citet{Sbordone2010AA}. 
It should be noted, however, that the significance of such a slope is around $1\sigma$ for our sample (Table~\ref{tab:abun_trend_fit_lithium}). 
Moreover, if we loosen the requirement of spectral quality to SNR$ > 30$ (bottom panels in Figure~\ref{fig:abun_trend_lithium}), 
the slope in [Fe/H]$<-2.5$ obtained by segmented fitting is unclear for the lower metallicity region in our study, 
presumably because most metal-poor stars are not well covered in our sample than previous studies that have focused on the measurements of Li abundances. 
There have been a number of other efforts on exploring the observed lithium plateau using extremely metal-poor stars \citep[][e.g.] {Matsuno2017AJ,Bonifacio2018AA,Aguado2019ApJ}; however, since their metallicity range (\FeHlt{-3.5}) is not sufficiently covered in our sample, comparisons with these studies are not discussed here.

It should be noticed that the derived lithium abundances are rather sensitive to the adopted stellar parameters, especially the effective temperature; therefore it is not surprising to find different slope or plateau among studies using different methods to derive stellar parameters, which makes a uniform sample with consistent data and methods very important when discussing the observed lithium trend.
In summary, based on our sample, we confirm that there exists a flat Li plateau at \FeHgt{-2.5},
and the average Li abundance is clearly lower when it comes to \FeHlt{-2.5}. 
However, the trend of $A$(Li) as a function of [Fe/H] in the extremely low metallicity region is still unclear. 
Future follow-up studies with higher-quality spectra for EMP and UMP turnoff stars found by this study would be helpful to clarify the trend. 
Just to clarify, the following discussion is based on the trend derived from the case of SNR $> 50$.

The scatter of $A$(Li) around the slope in \FeHlt{-2.5} is 0.08~dex, and that around the average value in \FeHgt{-2.5} is 0.07~dex. 
These are similar to the level of scatter found by \citet{Melendez2010AA} which derives 0.05 and 0.04 for the two metallicity ranges.
Such small scatter around the plateau or slope provides a strong constraint on the scenarios 
to explain the discrepancy between predictions of the primordial Li abundance and the observed values, 
in particular scenarios that assume depletion of Li inside stars we are observing or during the star formation processes. 

Several mechanisms have been proposed to explain the discrepancy between the Spite plateau and the expectation 
from the standard Big Bang nucleosynthesis by invoking atomic diffusion and turbulent mixing \citep{Richard2005ApJ,Korn2006a}, 
Li-destruction during the pre-main-sequence phase \citep{Fu2015a}, 
possible effect of chromospheric activity on the Li abundance measurement \citep{Takeda2019AA}.
There is also an observational indication that all the metal-poor stars observed in the current universe 
have experienced some Li depletion as their temperatures are not high enough \citep{Gao2020a}.
Although these studies have provided possible explanations for the discrepancy 
between the Spite plateau and the standard Big Bang nucleosynthesis, 
the breakdown of the Spite plateau at low metallicity needs further tuning of the models. 
For instance, although atomic diffusion is predicted to change the surface Li abundance differently for stars with different metallicity \citep{Richard2012MSAIS}, 
it is not yet clear how the metallicity plays a role when it is combined with turbulent mixing, 
which is considered to be needed to reproduce the constant Li abundance found for turnoff stars at $-2.5<$[{Fe}/{H}]$<-1.6$. 

The decreasing trend of $A$(Li) with decreasing metallicity in \FeHlt{-2.5} is found for both non-CEMP and CEMP turnoff stars, 
suggesting that the depletion of lithium shall not be relevant to the origin of carbon enhancement, 
which is in agreement with what has been suggested by \citet{Matsuno2017PASJ}. 
We note that stars for which the Li resonance line is not detected are not plotted in the diagram. 
Most of them are CEMP-s stars that would be affected by mass accretion from the companion AGB stars \citep{Masseron2012ApJ}.  
Hence, the trend found for CEMP stars here is primarily based on CEMP-no stars. 

Li abundance of giants also contains information about the Li-depletion during stellar evolution. 
Figure~\ref{fig:li_logg} shows $A$(Li) as a function of $\log g$ and {\tefft}. 
Here we adopt $\log g$ as the primary indicator of the evolutionary status from main-sequence to red giants, 
as the {\logg} values are well determined for most objects thanks to the parallaxes provided by {\it Gaia}. 
As stars leave the main-sequence, their surface Li abundances decrease due to the dilution 
caused by the first dredge-up in the evolutionary stage of $3\lesssim \logg \lesssim 4$. 
We note that the Li line is not detected in the horizontal branch stars with {\logg}$\sim 3$ and {\tefft}$>5500$~K which would have already experienced Li-depletion in the red giant phase (see below).
Comparing the Li abundance between the turnoff stars and giants observed in our sample, 
the degree of depletion can be as large as 1.0$\sim$1.2\,dex, 
suggesting that the surface material is diluted by the first dredge-up by about 15 times. 
This is consistent with the theoretical models of low-mass stars that deplete surface lithium abundances by $\sim 1.5$ dex \citep{Iben1967ApJ}.
Li abundance of giants remains more or less constant after the first dredge-up up to the RGB bump 
\citep[e.g., ][]{Lind2009AA}, which is also seen in Figure~\ref{fig:li_logg}.
This indicates that Li is not further destroyed at this stage.
Instead the Li abundance is mainly determined by the total Li content of the stellar surface 
and the maximum depth of the surface convective layer during the first dredge-up \citep[e.g.,][]{Mucciarelli2012MNRAS}.
It is also clear from Figures~\ref{fig:abun_trend_lithium} and \ref{fig:li_logg} 
that Li abundances do not show a distinctive trend in giants with metallicity down to [{Fe}/{H}]$\sim-3.5$: 
The average Li abundances in the two metallicity ranges of \FeHgt{-2.5} and \FeHlt{-2.5} 
are almost identical (1.02 and 1.04, respectively). 
This result is not inconsistent with the breakdown of the Spite plateau seen among turnoff stars if their surface lithium has been depleted by diffusion or gravitational settling \citep[e.g., ][]{Richard2005ApJ}. 
The Li abundance of these lower red giant branch stars are similar to the values seen in) stars in globular clusters with similar evolutionary status \citep[e.g.,][]{Lind2009AA,Mucciarelli2018AA}.

Further decrease of Li is found in more evolved stars along the red giant branch above the RGB bump ($\log g \lesssim 2$), which suggests extra mixing. 
The Li line is not detected in most of these stars in our sample though the detection limit is lower in stars with low temperatures, 
indicating that their $A$(Li) is clearly lower than the values of red giants below the RGB bump. 

As found in Figure~\ref{fig:li_logg}, the Li abundances show a rapid drop at $\log  g = 3.5 - $  3.7, 
and settle down to $A$(Li)=1.0 – 1.2 at $\log  g = \sim 3.2$. This range is well covered by studies for globular cluster stars. 
The $\log g$ values corresponding to this transition well agrees with that found for the globular cluster NGC~6397 \citep{Lind2009AA}. 
There are three objects in the transition stage in our sample (J0006+0123, J0600+2301 and J0834+2307) that have $T_{\rm eff}$ of 5400 -- 5650~K and  $\log  g = 3.2$ – 3.4. 

Figures~\ref{fig:li_logg} contain stars with unusually high Li abundance. 
We have excluded these Li-rich objects in previous discussions on the plateaus and the general trend. 
These low-metallicity Li-rich stars are already reported in \citet{Li2018ApJL}.
With updated $\log g$ estimates utilizing the {\it Gaia} DR2 parallax measurements, 
we confirm our previous findings that Li-rich stars are found at all evolutionary stage including a main-sequence turnoff star (Figure~\ref{fig:li_logg}). 

Recent studies of Li-enhanced stars in metal-rich evolved stars have reported that most Li-enhanced objects are found in clump stars rather than in red giant branch stars \citep[e.g.,][]{Kumar2020NatAs, Yan2021NatAs}. 
By contrast, at low metallicity, Li-rich objects distribute in a wide range of evolutionary status. 
Moreover, no Li-enhanced object is found in horizontal branch stars, which are metal-poor counterpart of clump stars, 
in our sample, although the sample size of such stars is still small (21 objects). 
This indicates that the mechanism of Li enhancement in very metal-poor stars could be different  from that in Li-rich stars with high metallicity.  

\begin{figure*}
\epsscale{1.2}
\plotone{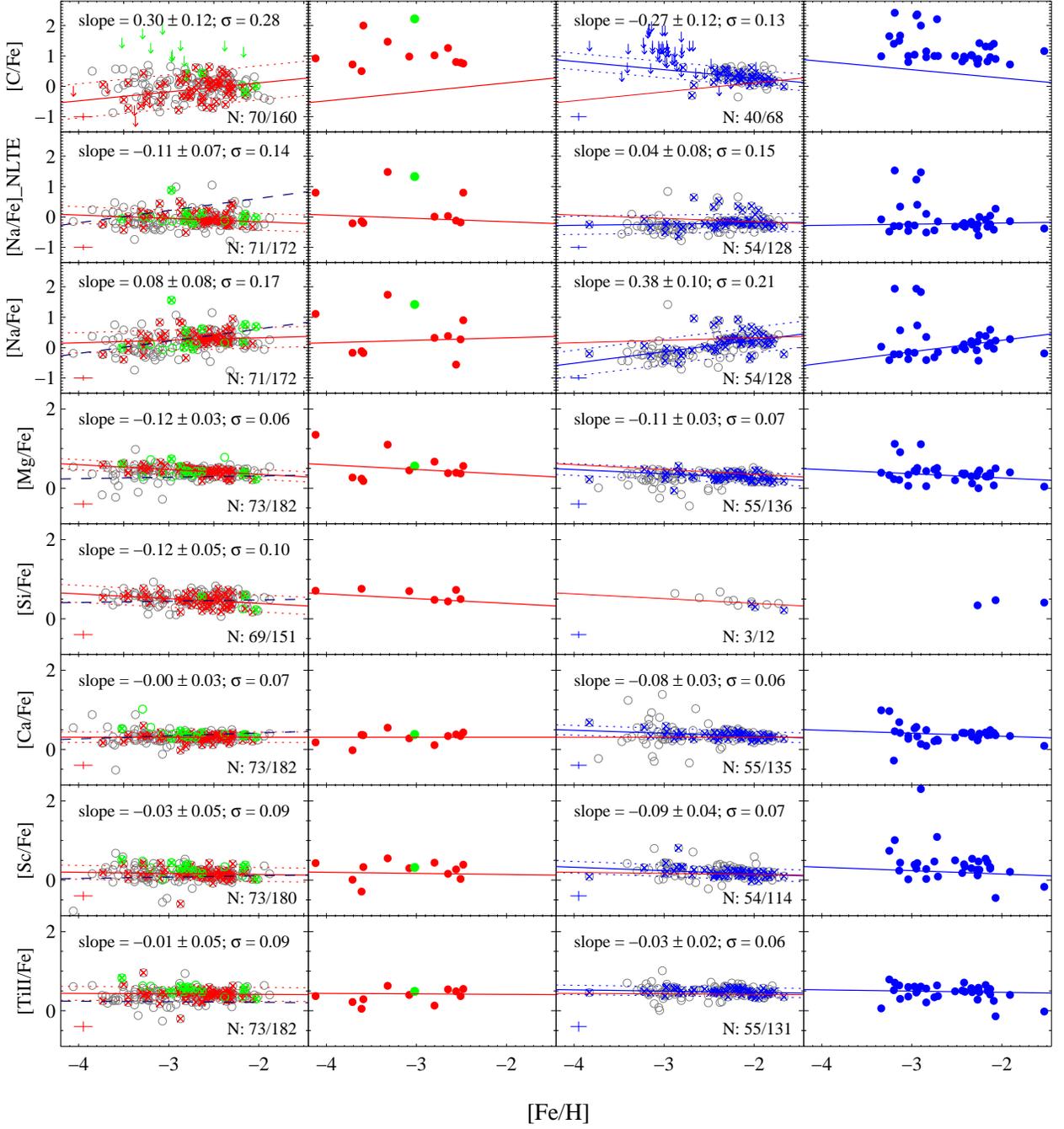}
\caption{Abundance trend along metallicities for light elements. 
The four columns from the left to right respectively refer to non-CEMP giants (open circle), CEMP giants (red filled circle), 
non-CEMP turnoff stars (open circle), and CEMP turnoff stars (blue filled circle). 
Only program stars with SNR $> 50$ (open circle with cross) in the panels of non-CEMP giants and turnoff stars have been used to fit the trends (red and blue solid lines), 
while the red and blue dotted lines show the $2\sigma$ scatter. 
Horizontal-branch stars are also shown with green symbols in the panels of giants for reference. 
Slope of the fitting and the standard deviation ($\sigma$) around the fitting is shown in each panel. 
The two numbers in the lower right corner correspond to the number of objects with SNR $> 50$ (numerator) and all objects with measurement (denominator). 
The fitted trend for giants is also shown in panels of turnoff stars for comparison.
Observational trends from the FS sample (dark blue dashed lines) are also presented for comparison. 
The steep slope found for \AB{C}{Fe} of non-CEMP turnoff stars would be due to incompleteness of the sample (see text).
\label{fig:abun_trend_light_obs}}
\end{figure*}

\subsubsection{Carbon} \label{subsubsec:carbon} 

Here we investigate stars excluding carbon-enhanced stars, which are discussed separately later in this section. 
As can be seen from top panels in Fig.~\ref{fig:abun_trend_light_obs},
the distribution of the abundance ratios of \AB{C}{Fe} exhibits two remarkable features: 
the average values for non-CEMP giants and turnoff stars are around the solar abundance, 
and the observed dispersion is larger than that of other light elements.

Decreases of carbon abundances in red giants have been reported for field and globular cluster stars, in particular objects above the red giant bump \citep[e.g., ][]{Gratton2000AA, Shetrone2019ApJ}. Hence, we need to discuss carbon abundances in red giants and main-sequence turn-off stars separately.
Figure~\ref{fig:abun_trend_light_obs} shows the abundances for turnoff stars and giants in different panels.
The carbon abundance ratios in turn-off stars would provide better constraint on chemical evolution models. 
However, the CH molecular bands are weak in
general in turnoff stars and are only measurable in stars with
relatively high C abundance. This limits the metallicity range for
which the abundance trend is investigated, and also results in
significant bias of the abundance trend because objects with higher [C/Fe] are sampled at lower metallicity. 

\begin{figure*}
\epsscale{0.8}
\plotone{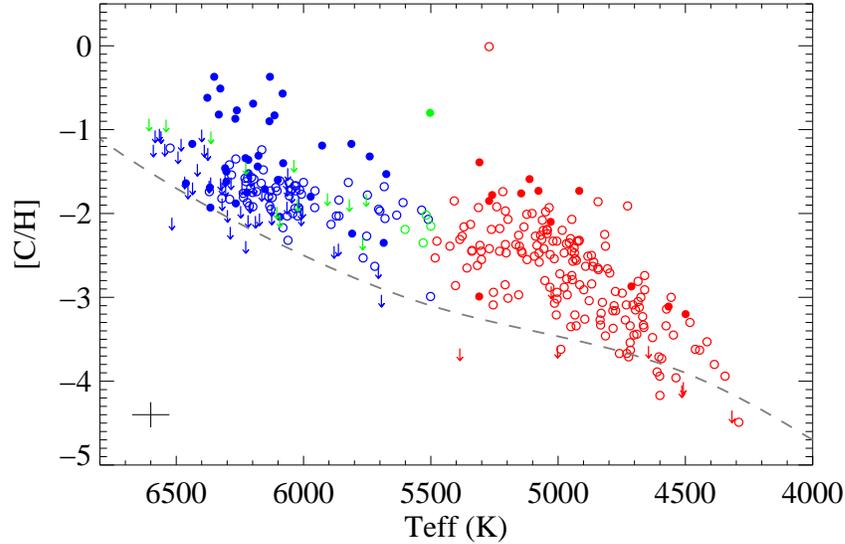}
\caption{\AB{C}{H} as a function of the effective temperature for the program stars. 
Red, blue, and green symbols refer to giant, turnoff, and horizontal-branch stars as defined in Table~\ref{tab:class_def}. 
The filled and open symbols respectively represent CEMP and non-CEMP stars, 
and the downward arrows refer to the upper limits.
The dashed line indicates the value of [C/H] for which the depth of the CH G-band 
reaches to about 3\% below the local continuum, as a function of temperature, estimated from synthetic spectra.
\label{fig:CH_Teff}}
\end{figure*}

Figure~\ref{fig:CH_Teff} shows the [C/H] values obtained for our
sample as a function of \tefft. The line indicates the C abundances
that produces CH molecular absorption with 3\% depth in a high
resolution spectra at the G-band estimated by the spectrum synthesis, which is the typical detection limit
in our spectral data. If stars have [C/Fe]$\gtrsim
0$, their carbon abundance could be measured for giants
(\tefft$<5500$~K) with [Fe/H]$>-3.5$, whereas many warm stars could be
missed, in particular for lower metallicity ([Fe/H]$<-3$).

Hence, although turnoff stars in our sample show decreasing trend in [Fe/H] with increasing metallicity with a slope of $-0.34$ (Figure~\ref{fig:abun_trend_light_obs}), 
this could be due to the incompleteness of the sampling of stars with lower metallicity. On the other hand, 
since the C abundances are measured in 95\% of giant stars, the increasing trend of [C/Fe] with increasing metallicity 
is not due to the temperature dependent detection limit of the CH band (see also \S\ref{subsec:cemp}). 

The average carbon abundance for giants are about 0.4\,dex lower than that for turnoff stars at
a metallicity of \FeHsim{-2.5}. Such difference has also been noticed in the FS sample \citep{Bonifacio2009AA}. 
They suspect that it may be caused by the granulation (3D) effects. 
A quite thorough discussions on the 3D correction on C abundances in metal-poor stars 
have been presented by \citet{Norris&Yong2019ApJ}, which indicate that for VMP/EMP giants, 
the correction will slightly increase the C abundance (no more than 0.1\,dex), 
while for turnoff and dwarf stars, it will decrease the abundance by 0.3-0.4\,dex; 
therefore, adopting 3D correction could partially explain the observed discrepancy.
This is, however, not sufficient for fully explaining the observed
difference in our sample. The difference between giants and turnoff stars
could be affected by the difference of the detection limit at least in
our study, as inspected above. This indicates that there could be turnoff stars with lower [C/Fe] than the current results. 
Another factor that affects the average [C/Fe] values is the evolutionary effect in highly evolved red giants 
in which carbon could be significantly depleted by extra mixing \citep{Spite2006AA,Aoki2007ApJ}. 

The giant stars in our sample present larger scatter (almost twice) in
[C/Fe] values than that of turnoff stars, as well as that in abundance ratios
of other elements (e.g., Mg).  This could partially be caused by the
mixing during stellar evolution of low-mass stars along the giant
branch, where carbon abundances could have been altered by the CNO cycle (i.e., C transformed into N). 
Indeed, C/H ratios are lower on average in cooler giants (Figure~\ref{fig:CH_Teff}), in which larger effects of internal mixing are expected (e.g., \citealt{Spite2005AA}). We note that \citet{Placco2014ApJb} apply corrections of carbon abundances taking into account the depletion due to CN processing on the upper red giant branch. We do not apply the corrections in this work, but include the effect in the criterion of CEMP stars.

\subsubsection{$\alpha$ elements}\label{subsubsec:alpha-elements}

All the four measured $\alpha$ elements Mg, Si, Ca, and Ti are enhanced relevant to Fe at similar level, 
where the averages of [Mg/Fe], [Si/Fe], [Ca/Fe], and [Ti/Fe] (for non-CEMP stars with spectral SNR $> 50$) 
are 0.42, 0.47, 0.30, and 0.40, respectively for giants, and 0.29, 0.30, 0.36, and 0.48 for turnoff stars. 
The $\sim 0.1$~dex offset of the \AB{Mg}{Fe} between giants and turnoff stars has been discussed in \S~\ref{subsec:RGBvsTO}, 
and could be attributed to the different NLTE effect on Mg abundances between giants and turnoff stars. 
For \AB{Si}{Fe}, turnoff stars with measured Si abundances are very limited in number 
and only covers \FeHsim{-2.0}, which makes it impossible to compare with that for giants. 
We inspect stars having very high or low abundances of $\alpha$ elements separately in \S~\ref{subsec:alpha_peculiar}. 
These $\alpha$-peculiar objects are shown in Figure~\ref{fig:abun_trend_light_obs}, 
but most of the $\alpha$-deficient objects are excluded 
in the above statistics due to limited SNR of corresponding spectra, 
and while most $\alpha$-enhanced objects are also C-enhanced and thus are neither included in the statistics. 

Since there is only one line of Si around 4103\,{\AA} detectable in our spectra, 
which is blended with the wing of H$\delta$ for warmer stars, 
we could derive Si abundances mostly for giants, and only for 15 turnoff and four horizontal-branch stars. 
The standard deviation around the average abundance ratio is not larger than 0.1\,dex for all the four species, 
which is as small as the measurement uncertainties. 
The abundance ratios for giants and turnoff stars are in good agreement with previous LTE measurements 
(e.g., FS sample, Y13 sample, \citealt{Reggiani2017AA}), though they are derived from different samples and analyses.

\begin{figure*}
\hspace{-1cm}\epsscale{1.1}
\plotone{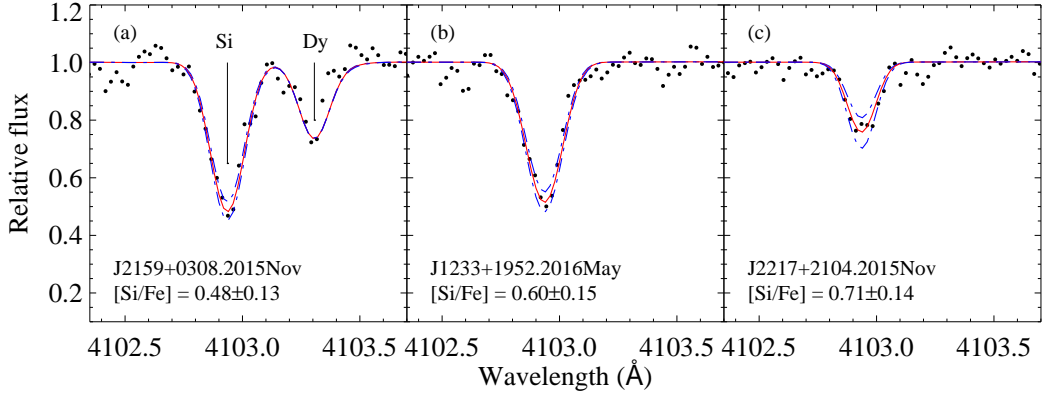}
\caption{Samples of Subaru/HDS spectra near the 4103\,{\AA} Si line, respectively for a VMP (a), an EMP (b), and a UMP (c) program stars. 
Symbols are the same to that for Figure~\ref{fig:Li_fitting}.
\label{fig:Si_fitting}}
\end{figure*}

The observed abundance ratios of \AB{Ca}{Fe} and \AB{Ti}{Fe} in general agree well between giants and turnoff stars.  
Such agreement has been found for Ca in previous studies, e.g, FS sample (as shown with the purple line), Y13 sample, etc. 
On the other hand, \citet{Bonifacio2009AA} have found that 
the \AB{Ti}{Fe} values in turnoff stars are 0.2\,dex higher than those of EMP giants, 
which cannot be explained by uncertainties in atomic data or stellar parameters. 
The trends obtained by fitting to our sample support the direction of these differences, 
however, as shown in Tables~\ref{tab:abun_trend_fit_RGB} and \ref{tab:abun_trend_fit_TO},
the observed discrepancies are negligible for our sample stars in the region of \FeHgt{-2.5} and about 0.1\,dex at \FeHle{-2.5}. 
Both works have used \ion{Ti}{2} and \ion{Fe}{1} lines to derive the abundance ratio of \AB{Ti}{Fe}. 
A difference is that the {\logg} has been determined based on parallaxes for our sample, 
while the FS sample is based on the ionization balance of Fe and Ti lines for the giant, 
and on the ionization balance of Fe for the turnoff/dwarf. 
This would lead to different \ion{Ti}{2} and \ion{Fe}{1} abundances, 
and thus partly explain the discrepancy in the \AB{Ti}{Fe} between the two works. 
Note that the NLTE correction is negligible for \ion{Ti}{2} lines for metal-poor stars \citep[e.g.,][]{Sitnova2016MNRAS}, 
while the level of NLTE correction for \ion{Fe}{1} lines are comparable for giant and turnoff stars in our sample \citep[e.g.,][]{Lind2012MNRAS}. 

The over-abundances similarly found for the four $\alpha$-elements suggest that the yields of Fe and these species hold similar ratios 
at low-metallicity environment, even though the detailed production site of these elements would be different (i.e., synthesis during stellar evolution, 
incomplete and complete burning during supernova explosions). 
Meanwhile, being a homogeneously observed and analyzed sample, 
the small dispersion of the observed abundance ratios of the $\alpha$ elements 
indicates that the stellar yields of their progenitor SN are rather homogeneous, 
and/or that the ejecta of the SN are sufficiently well-mixed with the surrounding ISM \citep{Francois2007AA}. 
Such small scatter could also be used to constrain the stellar mass range of their progenitors. 

The slope of the trend along {\FeH} ($|\Delta$[X/Fe]$/\Delta$[Fe/H]$|$) is about 0.1 or shallower. 
It should be noted, however, that the slopes are negative in any cases. 
A more detailed inspection indicates that the slopes of [Mg/Fe] both for turnoff stars and giants, 
which are relatively large among these three elements, 
are mostly caused by the decreases of the ratios at relatively high metallicity. 
Indeed, the average values of \AB{Mg}{Fe} for giants and turnoff stars are 0.38 and 0.28 in \FeHgt{-2.5},
and 0.44 and 0.34 in the region of \FeHle{-2.5} (as seen in Table~\ref{tab:abun_trend_fit_RGB} 
and ~\ref{tab:abun_trend_fit_TO}).        
Our sample extends to \FeHsim{-2.0}, which is not well covered by the FS sample. 
The difference of the metallicity coverage could be a reason 
for the difference of the average values and slopes between the FS sample and ours. 
Moreover, as pointed out by \citet{Bonifacio2009AA}, the Mg abundance derived for the FS giants 
has been underestimated by about 0.15\,dex since \citet{Cayrel2004AA} has neglected the wings of strong lines.   
It is also very likely that, at the metallicity region close to \FeHsim{-2.0}, 
some of the VMP stars are originated from progenitors that have experienced chemical evolution with a longer timescale, 
which results in low-$\alpha$ element abundances.
This is, for example, seen among stars with large retrograde motion \citep{Matsuno2019ApJL}.
\citet{Reggiani2017AA} made a similar discussion, when the 
traditional Galactic chemical evolution models usually only consider the evolution of in-situ halo but not including the accreted populations
cannot fit the low-$\alpha$ population discovered by \citet{Nissen&Schuster2010AA}. 
We inspect the stars with low-$\alpha$ element abundances in more details in \S~\ref{subsec:alpha_peculiar}, 
and will discuss the connection with their kinematics in a forthcoming paper.

\subsubsection{Light odd-Z elements} 

We have derived abundances for two light odd-$Z$ elements, Na and Sc. 
The abundance ratios of \AB{Na}{Fe} exhibit a rather large scatter of about 0.2\,dex for both giants and turnoff stars, 
which is very much consistent with what has been found in the FS sample.
On the other hand, conformed to previous LTE abundance analysis, 
the \AB{Na}{Fe} ratio follows quite different trends between metal-poor giants and turnoff stars.  
It should be noticed, however, that, for VMP and EMP stars, the Na abundance is usually derived from 
the Na\,D resonance lines at 5890\,{\AA} and 5896\,{\AA}, and thus usually suffer rather strong NLTE effect. 
For example, the NLTE corrections for Na abundance estimated from the Na\,D lines can be as large as $-0.5$\,dex 
for EMP stars \citep{Lind2011AA}. 
As NLTE effects are usually related to the metallicity, taking NLTE correction into consideration 
should then derive a different but more reliable abundance trend. 
Therefore, we have derived the NLTE Na abundances for all program stars whose Na lines are measurable, 
using the correction provided by \citet{Lind2011AA} 
through the INSPECT database\footnote{\url{http://inspect-stars.com/}}.

Depending on stellar parameters and spectral quality, 
the Sc abundances have been measured from 1 to 14 \ion{Sc}{2} lines covered by our Subaru/HDS spectra.
And the derived Sc abundance ratios of our sample are close to or only slightly overabundant compared to the solar value, and show rather small dispersion. 
Unlike what has been found in the FS sample, 
there is no clear discrepancy in Sc abundances between giants and turnoff stars in our sample, 
though the slope of the trend is slightly different. 

\subsubsection{Iron-peak elements} \label{subsubsec:iron-peak-elements}
Iron-peak elements are synthesized in thermonuclear explosions of supernovae (Type Ia supernovae), 
as well as in incomplete or complete Si-burning during explosive burning of core-collapse supernovae \citep{Woosley&Weaver1995ApJS,Chieffi1998ApJ,Kobayashi2006ApJ}
For our sample, abundances of six iron-group elements other than Fe have been measured, 
i.e., V, Cr, Mn, Co, Ni, and Zn. 
The V abundance has been derived from the \ion{V}{1} 4379\,{\AA} line, 
and Zn has been determined from the two \ion{Zn}{1} lines at 4722 and 4810\,{\AA}.
There are quite a number of usable \ion{Cr}{1} and \ion{Ni}{1} lines 
in our spectra, ranging from one to dozens of, to determine the Cr and Ni abundances.
One to three lines have been used to derive Mn (4055, 4783, and 4823\,{\AA}) 
and Co (4092, 4118, and 4121\,{\AA}).

\begin{figure*}
\epsscale{1.2}
\plotone{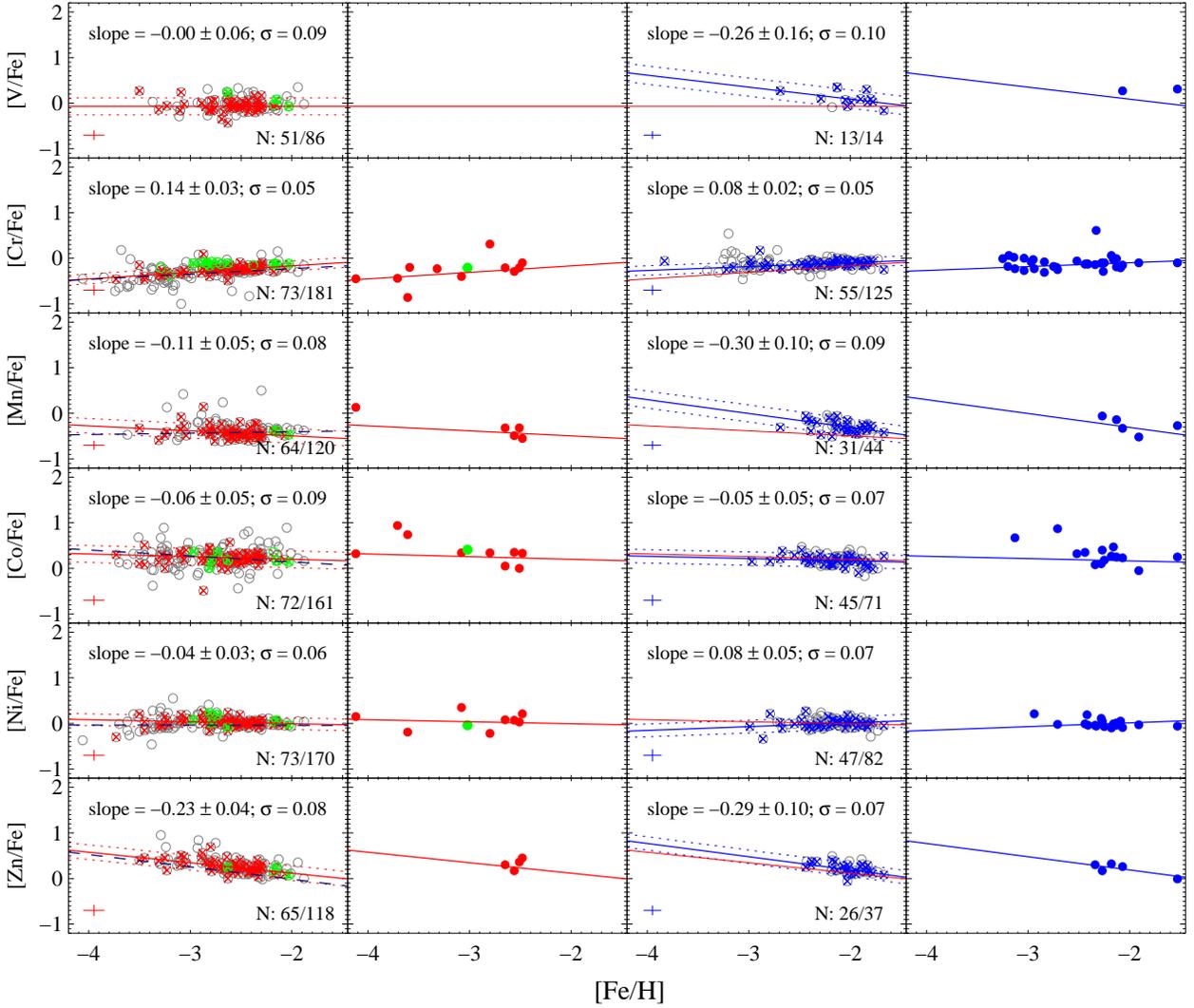}
\caption{Abundance trend along metallicities for iron-group elements. 
The meaning of the lines and symbols are the same as in Figure~\ref{fig:abun_trend_light_obs}.
\label{fig:abun_trend_iron_obs}}
\end{figure*}

As shown in Figure~\ref{fig:abun_trend_iron_obs}, for all these six elements, the dispersion of the abundance ratio is smaller than 0.10\,dex. Such small dispersion can also be seen among CEMP stars. 
Abundance trends are more or less in agreement with the FS sample. 
Especially for the case of Cr, the observed trend agrees well with previous studies, 
and moreover, extremely small scatter over different metallicity regions can be found, 
which also supports previous observations (e.g., FS sample, \citealt{Reggiani2017AA}),
demonstrating that the origin of Cr and Fe may be very closely linked. 
It should be noticed, however, that decreasing Cr abundance along decreasing metallicities found for giants 
could be due to the NLTE effect of the Cr\,I lines. 
For example, as \citet{Bergeman2010AA} have pointed out, for metal-poor stars, 
the NLTE abundance corrections to Cr\,I lines can be as large as $+0.3\sim+0.5$\,dex, 
which becomes larger for lower {\logg} and {\tefft}. 
Taking the NLTE correction into consideration, the abundance ratio of \AB{Cr}{Fe} derived from Cr I is approximately solar, 
and consistent with the measurement based on Cr\,II lines. 
This has also been verified by previous studies on Cr abundance for VMP stars 
(e.g., \citealt{Lai2008ApJ}, \citealt{Reggiani2017AA}).

Both giants and turnoff stars have similar abundance trends for the iron-group elements, 
except for the low Cr abundances in giants obtained by our LTE analysis as mentioned above. 
A very good agreement is found for Ni, which is consistent with both FS and Y13 sample. 
While both literature samples indicate that the abundances of Cr and Co in giants 
are $0.2-0.3$\,dex lower than those in turnoff stars. As observed in our sample, 
there is no clear discrepancy in the Co abundance, 
while for the case of Cr, the discrepancy can be as large as 0.2\,dex at \FeHsim{-3.0}, 
which can be explained by NLTE effect as discussed above.
For V, Mn, and Zn, due to the weakness of the absorption features, 
reliable abundances cannot be derived for EMP turnoff stars, 
which makes the slope of the trend uncertain. 
If we limit the comparisons of the average abundances to the VMP region, 
giants and turnoff stars are consistent within the measurement uncertainties for V and Zn, 
while for Mn, a systematic abundance difference of about 0.2\,dex is found. 
Such offset has been found in a number of previous studies \citep{Lai2008ApJ,Bonifacio2009AA}, 
and is not fully explained by current studies on NLTE effects.

Among all the measured iron-peak elements, Zn is of special importance to understand 
the physics of core-collapse supernovae as its main isotope is enhanced in the deepest region of HNe  \citep{Umeda&Nomoto2002ApJ}. 
Zn is believed to be mainly produced by complete Si burning at the low-metallicity region \citep{Woosley&Weaver1995ApJS,Kobayashi2006ApJ,Hirai2018ApJ},
and the \AB{Zn}{Fe} ratio exhibits an increasing trend when the metallicity decreases. 
The average abundance ratios of \AB{Zn}{Fe} for our sample in \FeHgt{-2.5} are 0.20 and 0.21 
for giants and turnoff stars, respectively, in excellent agreement with the average value of 0.19 from \citet{RoedererBarklem2018ApJ} 
derived from UV Zn\,II lines that are less sensitive to NLTE effects to derive the Zn abundance. 
The observed slope of Zn is quite steep, with $-0.23$ for giants and $-0.29$ for turnoff stars, 
consistent with the measurement from \citet{Reggiani2017AA}, with an average ratio of 0.29.
A quite high Zn abundance (\ABeq{Zn}{Fe}{0.8}) is determined for the hyper metal-poor star HE~1327--2326 
(\FeHeq{-5.2} including NLTE correction) by the observation of the UV range \citep{Ezzeddine2019ApJ}. 
Interestingly, this data point is on the line of the increasing trend of \AB{Zn}{Fe} 
with decreasing {\FeH} estimated from our sample (Figure~\ref{fig:abun_trend_iron_obs}). 


\begin{figure*}
\epsscale{1.2}
\plotone{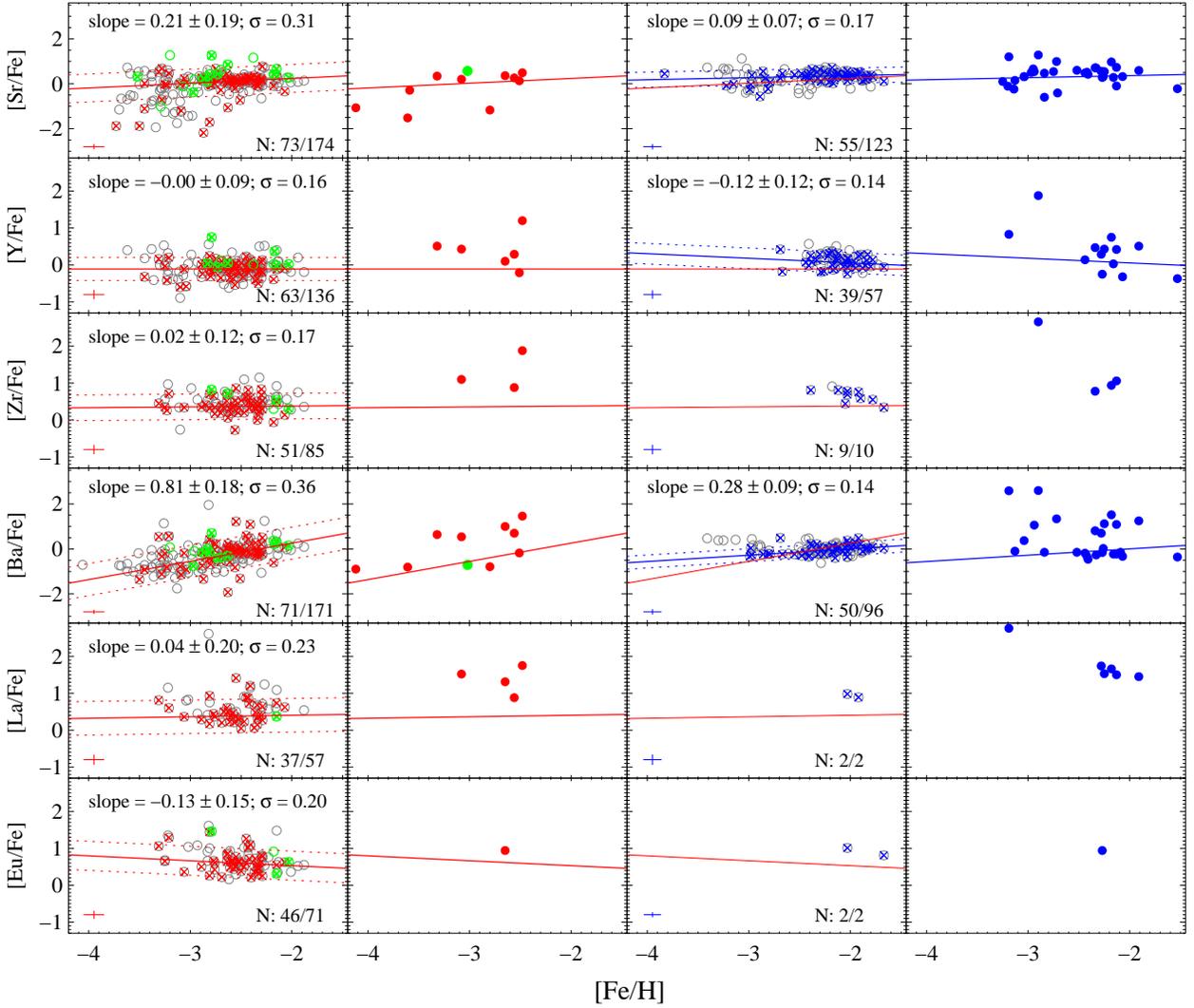}
\caption{Abundance trend along metallicities for heavy elements. 
The meaning of the lines and symbols are the same as in Figure~\ref{fig:abun_trend_light_obs}. 
\label{fig:abun_trend_heavy_obs}}
\end{figure*}

\subsubsection{Heavy elements} \label{subsubsec:heavy-elements}
Elements heavier than iron-peak elements are mostly synthesized by neutron-capture processes. The s-process (main s-process) in AGB stars is a major contributor to the neutron-capture elements in solar-system material. However, the long timescale of the evolution of low- and intermediate-mass stars results in minor contributions of the main s-process to chemical evolution at very low metallicity \citep[e.g., ][]{Kobayashi2011MNRAS}.
It should be noted that some very metal-poor stars show large excess of s-process elements as well as carbon, which are recognized as a result of mass transfer from companion AGB stars that should have already evolved to white dwarfs (see \S~\ref{subsec:cemp}).  The rapid neutron-capture process (r-process) produces heavy elements including actinides. A small fraction of very metal-poor stars shows a large excess of heavy elements with abundance pattern similar to that of the solar-system r-process component \citep{Sneden2003ApJ}, indicating the contribution of the r-process to the early chemical evolution. Mergers of binary neutron stars are considered as the major sites of the r-process, whereas other phenomena including magneto-rotational supernovae are also suggested \citep{Cowan2021RvMP}. To explain the light neutron-capture elements (e.g., Sr, Y, Zr) in very metal-poor stars, other processes have been proposed. They have been called LEPP \citep{Travaglio2004ApJ} or weak r-process \citep{Wanajo2006NuPhA}, but their mechanism and nucleosynthesis sites are not well determined. Mass loss from rotating massive stars, which could include light neutron-capture elements produced by the s-process during stellar evolution, is also proposed as a source at low metallicity \citep{Choplin2018AA}.

Among a large number of neutron-capture elements, only limited species 
can be detected in a sizable sub-sample of our program stars. 
Others are only found in a small number of stars in which neutron-capture elements 
are significantly enhanced. 
In this paper, we focus on six neutron-capture elements, Sr, Y, Zr, Ba, La, and Eu. 
Abundances for these elements have been derived from one to a few lines of \ion{Sr}{2} (4077 and 4215\,{\AA}), 
\ion{Y}{2} (4374, 4854, 4883, and 4900\,{\AA}), \ion{Zr}{2} (4161 and 4209\,{\AA}), \ion{Ba}{2} (4554, 4934, 6141, and 6497\,{\AA}), 
\ion{La}{2}(4077, 4086, 4196, and 4921\,{\AA}), and \ion{Eu}{2} (4129 and 4205\,{\AA}).
There are about 50 objects for which abundances are derived for all the six elements, 
and the number becomes 80 if limited to detection of Sr, Ba, and Eu. 
The sample discussed here adds quite a number of EMP stars with measurement of 
these heavy elements, especially of Sr, Y, and Ba. 
Other neutron-capture elements that are measurable for a limited number of stars will be studied separately in future work. 

Abundances of Sr, Ba and Eu are discussed in detail in \S~\ref{subsec:ncap}. 
We here provide a brief overview of the abundance trend of the six heavy elements. 
The objects for which Eu abundances are determined show abundance ratios of [Ba/Eu] similar to that of the r-process component in solar-system material (see \S~\ref{subsec:ncap}). 
We need to take account of the bias in the subsample that abundances of Eu, as well as Zr and La, are only measured in a
very limited number of turnoff stars, because corresponding absorption lines 
are generally too weak. No strong constraints on the origins of neutron-capture elements in stars with no detection of Eu have been obtained. 

\begin{figure*}
\hspace{-1cm}\epsscale{1.05}
\plotone{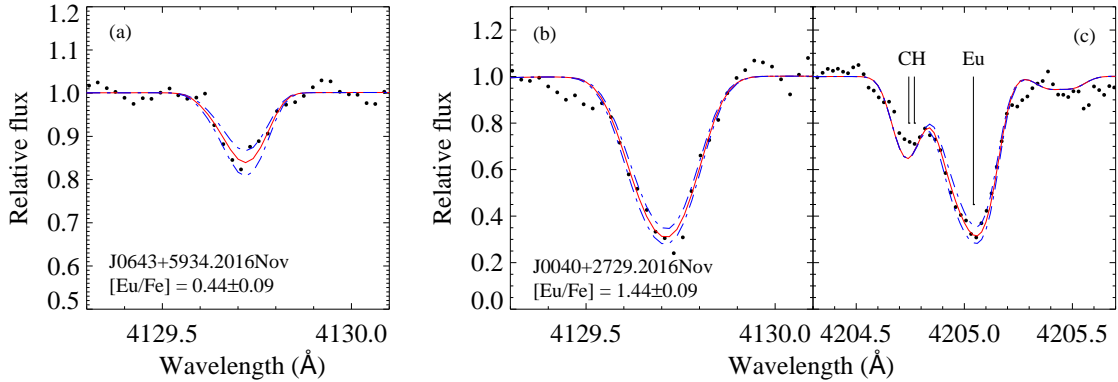}
\caption{Examples of spectral fitting of the Eu 4129\,{\AA} line, 
for program stars with normal Eu abundance (a) and extreme enhancement (b). In a few case where the Eu 4205\,{\AA} line is measurable, 
Eu abundance has also been derived from fitting that line (c). 
Definition of symbols are the same as Figure~\ref{fig:Li_fitting}.
\label{fig:Eu_fitting}}
\end{figure*}

Sr is an important element for exploring the early enrichment of neutron-capture elements, 
since the two resonance lines are strong enough to be detected even in EMP/VMP stars. For our sample,  
we could derive the Sr abundances for about 96\% program stars, including 185 giants, 157 turnoff stars, and 19 horizontal-branch stars. Exceptions are a dozen EMP turnoff stars with too weak lines and a few objects with blends of other species to the Sr lines.
 

The trend of Sr observed in our sample shown in Figure~\ref{fig:abun_trend_heavy_obs} is similar to those found by previous studies, 
e.g., a rather flat distribution around the solar values in [Fe/H]$\gtrsim -3.0$, 
and a drop in the abundance ratios below that metallicity range,
indicating that the progressive enrichment of this element reaches the solar ratio 
when it comes to the metallicity of $-3.0\sim-2.8$ on average. 
The other prominent feature of the Sr abundance is the large dispersion, especially in [Fe/H]$\lesssim -2.5$, 
which is clearly larger than the typical measurement uncertainties and the scatter found for $\alpha$ and iron-peak elements, 
indicating  the existence of rare events that have provided the early universe with a significant amount of Sr.

The other two light neutron-capture elements Y and Zr show similar trend to Sr, 
in the sense of presenting a relatively constant distribution at \FeHgt{-3.0}, which is consistent with previous studies \citep[e.g.][]{Zhao&Magain1991AA}, but with a much larger sample. 
However, at lower metallicity region, 
these species neither exhibit such large dispersion as Sr nor drop in the abundance ratio as claimed by the FS sample. 
This would be basically due to the limited number of EMP stars (especially EMP turnoff stars) with Y and Zr measurements in our sample.  

The three heavier elements Ba, La and Eu exist near the second abundance peak of neutron-capture elements ($Z\sim 52$ and $Z\sim 56$ for thr r- and s-processes, respectively). 
We note that elements at the r-process peak, e.g., Te, have been little measured in stellar atmospheres \citep{Roederer2012ApJ}. 
Eu is a useful element with relatively high abundance that represent the contributions of the r-process. 
The Ba abundance has also been derived for the majority of our sample, including 181 giants, 122 turnoff stars, and 15 horizontal-branch stars.
The dispersion of Ba abundances is as large as that of Sr. 
The objects with highest [Ba/Fe] ratios are carbon-enhanced stars, which could be explained by mass transfer from companion AGB stars (\S~\ref{subsec:cemp}). 
Some Ba-enhanced stars are explained as the large contribution of the r-process, taking account of their [Ba/Eu] ratios ($\sim -0.7$, as presented in \S~\ref{subsec:ncap}).
The large excess of r-process elements in small fraction of extremely metal-poor stars has been interpreted as a result of insufficient mixing in interstellar matter polluted by r-process events 
from which low-mass metal-poor stars were formed. Recent discoveries of ultra-faint dwarf galaxies with large excess of r-process elements \citep[e.g., ][]{Roederer2016AJ,Ji2016Nature}, 
however, suggest that the dispersion could be attributed to the difference of the r-process abundances 
between small stellar systems in which metal-poor stars were formed and later accreted to the Milky Way \citep[e.g., ][]{Hirai2015ApJ}. 

It should be noted that there are several CEMP stars exhibiting significant enhancement in heavy elements, 
in particular Ba (Figure~\ref{fig:abun_trend_heavy_obs}), which are usually explained by mass-transfer from its companion AGB star in a binary system. 
The excess of heavy elements in these stars is attributed to the s-process in AGB stars rather than the r-process (\S~\ref{subsec:cemp}).

\begin{deluxetable*}{l|rr|rrrrrrrr}
\tablecolumns{8}
\tabletypesize{\scriptsize}
\tablecaption{Observed abundance trends from C through Eu for giants \label{tab:abun_trend_fit_RGB}}
\tablewidth{0pt}
\tablehead{
\colhead{} & \multicolumn2c{Regression line} & \multicolumn2c{All} & \multicolumn2c{\FeHgt{-2.5}} & \multicolumn2c{\FeHle{-2.5}}\\
\cline{2-11}
\colhead{species}&\colhead{A}&\colhead{B}&\colhead{mean}&\colhead{$\sigma$}&\colhead{mean$_{1}$}&\colhead{$\sigma_{1}$}
&\colhead{mean$_{2}$}&\colhead{$\sigma_{2}$}&\colhead{$N_{1}$}&\colhead{$N_{2}$}
}
\startdata
          \AB{C}{Fe}&     0.70$\pm$0.32&    0.30$\pm$0.12&$-$0.05& 0.28&   0.01& 0.23&$-$0.09& 0.31& 25& 44\\
\AB{Na}{Fe}$_{\mathrm{NLTE}}$&  $-$0.38$\pm$0.19& $-$0.11$\pm$0.07&$-$0.05& 0.14&$-$0.07& 0.15&$-$0.04& 0.13& 24& 46\\
         \AB{Na}{Fe}&     0.49$\pm$0.22&    0.08$\pm$0.08&   0.28& 0.17&   0.31& 0.16&   0.28& 0.17& 24& 46\\
         \AB{Mg}{Fe}&     0.11$\pm$0.08& $-$0.12$\pm$0.03&   0.42& 0.06&   0.38& 0.08&   0.44& 0.06& 25& 47\\
         \AB{Si}{Fe}&     0.15$\pm$0.13& $-$0.12$\pm$0.05&   0.47& 0.11&   0.46& 0.15&   0.48& 0.09& 24& 44\\
         \AB{Ca}{Fe}&     0.31$\pm$0.08&    0.00$\pm$0.03&   0.30& 0.07&   0.31& 0.06&   0.29& 0.07& 25& 47\\
         \AB{Sc}{Fe}&     0.09$\pm$0.13& $-$0.03$\pm$0.05&   0.16& 0.09&   0.14& 0.07&   0.16& 0.10& 25& 47\\
         \AB{Ti}{Fe}&     0.39$\pm$0.12& $-$0.01$\pm$0.05&   0.40& 0.09&   0.40& 0.06&   0.41& 0.10& 25& 47\\
         \AB{ V}{Fe}&  $-$0.07$\pm$0.17&    0.00$\pm$0.06&$-$0.06& 0.09&$-$0.04& 0.08&$-$0.07& 0.10& 21& 30\\
         \AB{Cr}{Fe}&     0.12$\pm$0.07&    0.14$\pm$0.03&$-$0.25& 0.05&$-$0.21& 0.04&$-$0.27& 0.06& 25& 47\\
         \AB{Mn}{Fe}&  $-$0.71$\pm$0.14& $-$0.11$\pm$0.05&$-$0.42& 0.08&$-$0.45& 0.06&$-$0.40& 0.09& 25& 39\\
         \AB{Co}{Fe}&     0.08$\pm$0.13& $-$0.06$\pm$0.05&   0.22& 0.09&   0.24& 0.06&   0.21& 0.11& 25& 46\\
         \AB{Ni}{Fe}&  $-$0.09$\pm$0.08& $-$0.04$\pm$0.03&   0.02& 0.06&$-$0.00& 0.03&   0.04& 0.07& 25& 47\\
         \AB{Zn}{Fe}&  $-$0.34$\pm$0.12& $-$0.23$\pm$0.04&   0.27& 0.08&   0.20& 0.06&   0.31& 0.09& 25& 40\\
         \AB{Sr}{Fe}&     0.66$\pm$0.51&    0.21$\pm$0.19&$-$0.03& 0.31&   0.16& 0.15&$-$0.09& 0.37& 25& 47\\
         \AB{ Y}{Fe}&  $-$0.11$\pm$0.23&    0.00$\pm$0.09&$-$0.12& 0.16&$-$0.12& 0.14&$-$0.13& 0.16& 25& 38\\
         \AB{Zr}{Fe}&     0.42$\pm$0.30&    0.02$\pm$0.12&   0.39& 0.17&   0.40& 0.19&   0.38& 0.16& 23& 28\\
         \AB{Ba}{Fe}&     1.88$\pm$0.49&    0.81$\pm$0.18&$-$0.22& 0.36&   0.02& 0.27&$-$0.35& 0.40& 25& 46\\
         \AB{La}{Fe}&     0.48$\pm$0.53&    0.04$\pm$0.20&   0.49& 0.23&   0.54& 0.26&   0.45& 0.20& 16& 21\\
         \AB{Eu}{Fe}&     0.26$\pm$0.39& $-$0.13$\pm$0.15&   0.63& 0.20&   0.61& 0.20&   0.65& 0.19& 21& 25
\enddata
\tablecomments{The columns A and B correspond to the intercept and slope of the linear fitting of \AB{X}{Fe} = A + B$\times${\FeH} as shown in the first column of Figure~\ref{fig:abun_trend_light_obs}, \ref{fig:abun_trend_iron_obs}, and \ref{fig:abun_trend_heavy_obs}.}
\end{deluxetable*}

\begin{deluxetable*}{l|rr|rrrrrrrr}
\tablecolumns{8}
\tabletypesize{\scriptsize}
\tablecaption{Observed abundance trends from C through Eu for turnoff stars \label{tab:abun_trend_fit_TO}}
\tablewidth{0pt}
\tablehead{
\colhead{} & \multicolumn2c{Regression line} & \multicolumn2c{All} & \multicolumn2c{\FeHgt{-2.5}} & \multicolumn2c{\FeHle{-2.5}}\\
\cline{2-11}
\colhead{species}&\colhead{A}&\colhead{B}&\colhead{mean}&\colhead{$\sigma$}&\colhead{mean$_{1}$}&\colhead{$\sigma_{1}$}
&\colhead{mean$_{2}$}&\colhead{$\sigma_{2}$}&\colhead{$N_{1}$}&\colhead{$N_{2}$}
}
\startdata
                  \AB{C}{Fe}&  $-$0.27$\pm$ 0.26& $-$0.27$\pm$ 0.12&   0.28& 0.13&   0.29& 0.11&   0.12&   0.44& 37&  3\\
\AB{Na}{Fe}$_{\mathrm{NLTE}}$&  $-$0.12$\pm$ 0.18&    0.04$\pm$ 0.08&$-$0.15& 0.15&$-$0.13& 0.14&$-$0.24&   0.19& 41& 12\\
                 \AB{Na}{Fe}&     1.00$\pm$ 0.23&    0.38$\pm$ 0.10&   0.16& 0.21&   0.24& 0.21&$-$0.09&   0.22& 41& 12\\
                 \AB{Mg}{Fe}&     0.05$\pm$ 0.08& $-$0.11$\pm$ 0.03&   0.29& 0.07&   0.28& 0.06&   0.34&   0.09& 41& 13\\
           \AB{Si}{Fe}$^{a}$&            \nodata&           \nodata&   0.30& 0.08&   0.30& 0.08&\nodata&\nodata&  3&  0\\
                 \AB{Ca}{Fe}&     0.18$\pm$ 0.06& $-$0.08$\pm$ 0.03&   0.36& 0.06&   0.34& 0.05&   0.39&   0.08& 41& 13\\
                 \AB{Sc}{Fe}&  $-$0.02$\pm$ 0.09& $-$0.09$\pm$ 0.04&   0.19& 0.07&   0.15& 0.06&   0.32&   0.12& 41& 12\\
                 \AB{Ti}{Fe}&     0.40$\pm$ 0.05& $-$0.03$\pm$ 0.02&   0.48& 0.06&   0.47& 0.06&   0.51&   0.06& 41& 13\\
                 \AB{ V}{Fe}&  $-$0.44$\pm$ 0.32& $-$0.26$\pm$ 0.16&   0.11& 0.10&   0.10& 0.11&   0.27&   0.00& 12&  1\\
                 \AB{Cr}{Fe}&     0.07$\pm$ 0.06&    0.08$\pm$ 0.02&$-$0.12& 0.05&$-$0.11& 0.04&$-$0.14&   0.07& 41& 13\\
                 \AB{Mn}{Fe}&  $-$0.92$\pm$ 0.20& $-$0.30$\pm$ 0.10&$-$0.30& 0.09&$-$0.30& 0.09&$-$0.31&   0.21& 30&  1\\
                 \AB{Co}{Fe}&     0.06$\pm$ 0.11& $-$0.05$\pm$ 0.05&   0.17& 0.07&   0.16& 0.07&   0.24&   0.08& 40&  5\\
                 \AB{Ni}{Fe}&     0.18$\pm$ 0.11&    0.08$\pm$ 0.05&   0.00& 0.07&   0.01& 0.06&$-$0.05&   0.12& 41&  6\\
                 \AB{Zn}{Fe}&  $-$0.39$\pm$ 0.20& $-$0.29$\pm$ 0.10&   0.21& 0.07&   0.21& 0.07&\nodata&\nodata& 26&  0\\
                 \AB{Sr}{Fe}&     0.55$\pm$ 0.17&    0.09$\pm$ 0.07&   0.30& 0.17&   0.36& 0.13&   0.09&   0.29& 41& 13\\
                 \AB{ Y}{Fe}&  $-$0.19$\pm$ 0.26& $-$0.12$\pm$ 0.12&   0.08& 0.14&   0.08& 0.13&   0.12&   0.30& 37&  2\\
           \AB{Zr}{Fe}$^{a}$&            \nodata&           \nodata&   0.64& 0.17&   0.64& 0.17&\nodata&\nodata&  9&  0\\
                 \AB{Ba}{Fe}&     0.56$\pm$ 0.20&    0.28$\pm$ 0.09&$-$0.03& 0.14&$-$0.00& 0.15&$-$0.14&   0.14& 41&  9\\
           \AB{La}{Fe}$^{a}$&            \nodata&           \nodata&   0.94& 0.06&   0.94& 0.06&\nodata&\nodata&  2&  0\\
           \AB{Eu}{Fe}$^{a}$&            \nodata&           \nodata&   0.91& 0.14&   0.91& 0.14&\nodata&\nodata&  2&  0
\enddata
\tablenotetext{a}{For species with too limited number of objects to derive the fitting trend, 
the $\sigma$ refers to the standard deviation of all measurements.}
\tablecomments{The columns A and B correspond to the intercept and slope of the linear fitting of \AB{X}{Fe} = A + B$\times${\FeH} as shown in the third column of Figure~\ref{fig:abun_trend_light_obs}, \ref{fig:abun_trend_iron_obs}, and \ref{fig:abun_trend_heavy_obs}.}
\end{deluxetable*}

\subsection{Comparison with chemical evolution models} \label{subsec:comparison_GCE_model}

A large sample of VMP stars is very useful to examine and constrain the Galactic chemical evolution models. 
While previous studies are usually based on comparisons with results from different sets of observational data, 
which could inevitably involve offset between different data sets and analysis methods. 
Our sample is thus especially valuable in the sense that it is uniform and the observed abundance dispersion 
is able to better reflect chemical inhomogeneity in the interstellar medium at very early epochs of the Galactic evolution. 

\begin{figure*}
\hspace{-1cm}\epsscale{1.2}
\plotone{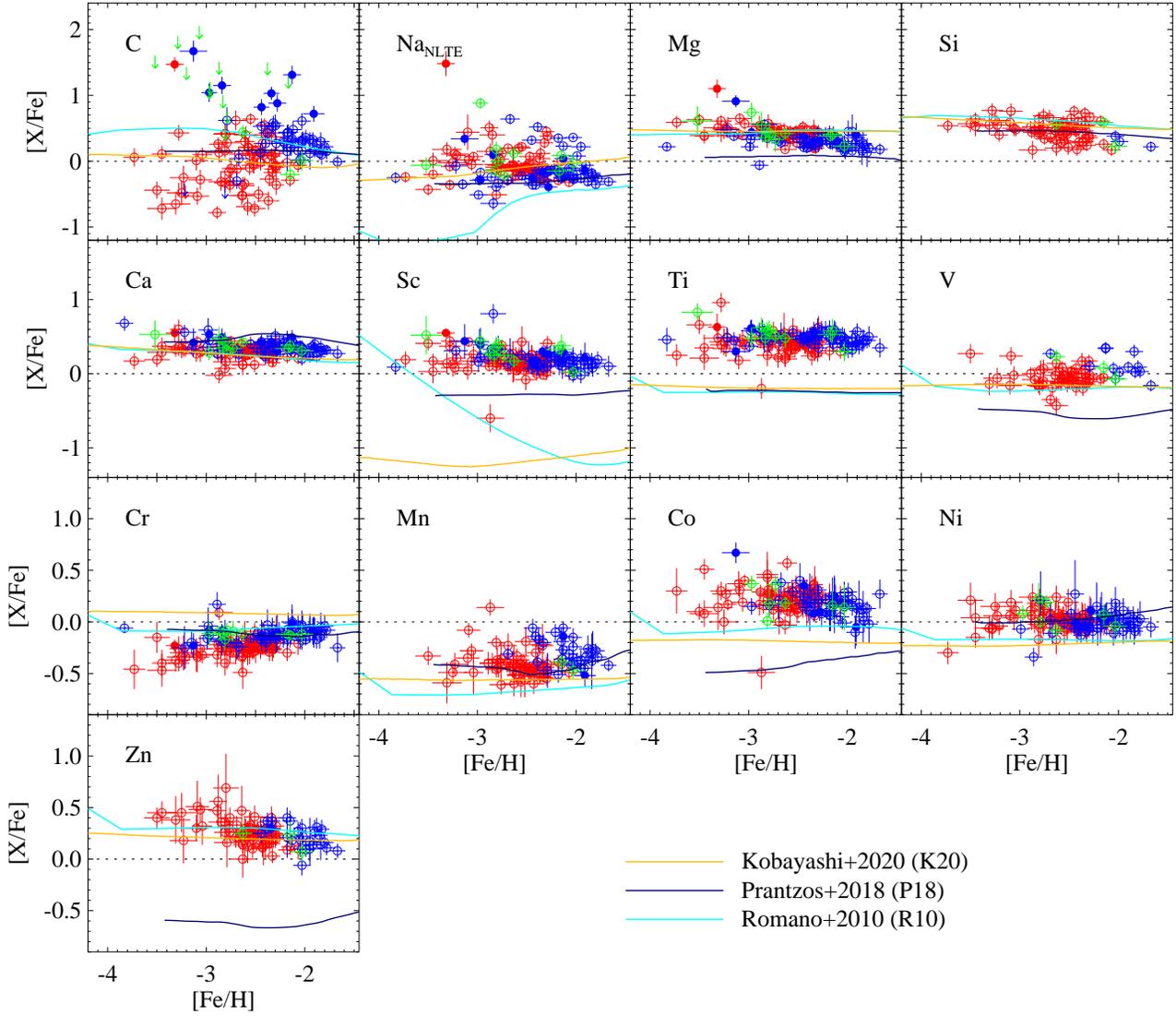}
\caption{Abundance trend along metallicities for light and iron-peak elements. 
The meaning of symbols are the same as in Figure~\ref{fig:abun_trend_light_obs}, 
and note that only objects with SNR $> 50$ are included for comparisons with theoretical models.  
Theoretical predictions from P18, K20, and R10 are shown in solid lines with 
navy, gold, and cyan, respectively. See the main text in \ref{subsec:comparison_GCE_model} for more details. 
\label{fig:abun_trend_model}}
\end{figure*}

Model predictions for chemical enrichment in the Galaxy are compared with our observational trends in Figure~\ref{fig:abun_trend_model}. 
The selected models shown are R10 \citep{Romano2010A&A}, 
P18 \citep{Prantzos2018MNRAS}, and K20 \citep{Kobayashi2020ApJ}. 
R10 has computed 15 models with different sets of nucleosynthesis prescriptions, 
and the best yield choice Model 15 has been selected for our comparison, 
which uses the yields by \citet{Karakas2010MNRAS} for low-to-intermediate mass (LIM) stars, 
the yields for massive stars limited to the pre-supernova stage, but computed with both mass loss and rotation \citep{Hirschi2007AA,Ekstrom2008AA}, 
and the results of explosive nucleosynthesis of \citet{Kobayashi2006ApJ} for SN~II below 20 $M_{\sun}$ and hypernovae above that limit. 
P18 is based on \citet{Goswami2000A&A}, and update by \citet{Kubryk2015A&A}, 
which adopts the stellar IMF of \citet{Kroupa2002Sci} in the mass range $0.1-120 M_{\sun}$, 
the metallicity-dependent yields of \citet{Cristallo2015ApJS} for LIM stars 
and of \citet{Limongi2018ApJS} for massive stars, and \citet{Iwamoto1999ApJS} for SNIa by interpolating models for $Z=0$ and $Z_{\odot}$. 
Their baseline model with rotating massive star yields has been selected for our comparison. 
K20 adopts the \citet{Kroupa2008ASPC} initial mass function at $0.01-50 M_{\sun}$, 
and the SN Ia model based on the single-degenerate scenario \citep{Kobayashi&Nomoto2009ApJ} with the metallicity effect \citep{Kobayashi1998ApJ}. 
The nucleosynthesis yields from 1D models are taken from \citet{Kobayashi2006ApJ} and \citet{Kobayashi2011MNRAS} for SN and hypernovae
(with the metallicity-dependent fraction of hypernovae tested), 
and from \citet{Lugaro2012ApJ}, \citet{Fishlock2014ApJ}, and \citet{Karakas2018MNRAS} for low-metallicity AGB stars.
Readers may refer to their papers for more details about the above models.

The three models predict solar abundance ratio or over-abundance of carbon, i.e., [C/Fe]$\gtrsim$0. R10 best fits the observed trend for TO stars, 
while the other two models predict relatively flatter trend.
Here we primarily focus on comparison with turnoff stars, since carbon abundances in giants 
shall partly reflect results of internal mixing (\S\ref{subsubsec:carbon}). 
R10 exhibit the best agreement with the decreasing trend of \AB{C}{Fe} ratios found for turnoff stars (Figure~\ref{fig:abun_trend_light_obs}),
although our sample is not sufficient to constrain the behavior of carbon at the EMP region. 

The odd-Z element, Na is synthesized mostly during hydrostatic carbon burning, 
partly during hydrogen burning through the NeNa cycle, as well as the $s-$process. 
As discussed in \S~\ref{subsec:abundance_trend}, the LTE approximation can lead to substantial overestimation of Na abundances, 
and thus the NLTE corrected Na abundances are used for comparison with model predictions.  
K20 can fit the observed abundance trend in general.
P18 also reproduce the trend, although the [C/Fe] values predicted are slightly lower than the observational trend.
while the sharp decline of Na abundances with decreasing metallicity predicted by R10 is not supported by the result of this work. 
As discussed in R10, only a set of model without HNe predicts a \AB{Na}{Fe} ratio 
in agreement with the halo data, while models (including Model 15 that is adopted here for comparison) 
including nucleosynthesis from HNe tend to underestimate the \AB{Na}{Fe} in the low-metallicity region. 
The discrepancy between R10 and the observation could be as large as 1.4\,dex at the extremely low metallicity region, 
which are mainly driven by the adopted mass cut and energy of the explosion in models of massive stars, 
and shall reduce to $\sim 0.2$\,dex when it goes up to \FeHsim{-2.0}.

Among the $\alpha$-elements, R10 and K20 predicted rather similar abundance trend, 
as they have adopted the similar sets of yields of massive stars and LIM stars. 
Both models reproduce the observational results well for Ca, and marginally for Mg and Si.  
P18 fits the trend best for \AB{Si}{Fe} and partially for Ca, while tends to underestimate the Mg abundance in general. 
This would be due to the yields adopted by P18, 
which also underproduce the Mg abundances obtained from Mg isotopes in solar-system material. 
However, the situation for Ti is very different. We discuss this problem below.
All the three models predict a plateau behavior 
of Ti as generally found for $\alpha-$elements, but the predicted values are much lower than observations, 
even though efforts have been made to account for higher Ti abundances, 
e.g., inclusion of hypernova jets by \citep{Kobayashi2015ApJ}. 
In particular, the very similar level of deficiency resulted from various sets of yields is quite intriguing. 
The deficiency is unable to be explained by uncertainties of the measurements, 
including the choice of the species (i.e., \ion{Ti}{1} or \ion{Ti}{2}) to determine Ti abundance. 
Although the abundances derived from \ion{Ti}{2} lines 
are in general higher than those from \ion{Ti}{1} in the present work, the difference is at most $0.1-0.2$\,dex. 
Moreover, as discussed in \S~\ref{subsubsec:alpha-elements}, 
the NLTE effect on abundances derived from \ion{Ti}{2} lines is negligible, 
while the NLTE effect of \ion{Fe}{1} which would also affect the adopted \AB{Ti}{Fe} is also about $0.1-0.2$\,dex. 
Hence, the overabundance of Ti derived by the present work is robust.
Previous studies including FS and Y13 also indicate similar discrepancy between observations and models. 
P18 suggest the importance of further investigation of physical condition where the major Ti isotope $^{48}$Ti is produced though $^{48}$Cr, 
in particular the net production out of the equilibrium (some kind of $\alpha$-rich freeze out). 
Such condition could be realized by multi-dimensional models of supernovae, including jet-induced explosions \citep{Tominaga2009ApJ}.

Meanwhile, the slope found for \AB{$\alpha$}{Fe}, reported in \S~\ref{subsec:abundance_trend}, is not well reproduced by chemical evolution models. 
The discrepancy is distinct in \AB{Mg}{Fe} (also seen in \AB{Si}{Fe}) at \FeHgt{-2.5}, 
where the abundance ratios could be affected by contributions of small stellar systems like dwarf galaxies 
that usually contain stars with lower \AB{$\alpha$}{Fe} formed with longer time-scale.


The predicted abundance trends of Sc and V are in overall underabundant for all three selected models. 
Especially pointed out by R10, the predicted \AB{Sc}{Fe} ratios severely disagree with observations 
if the \citet{Kobayashi2006ApJ} yields for massive stars are used, independently of the assumed hypernova (HN) fraction.
Tests carried out by P18, which include the contribution of rotating massive stars, 
show that consideration of this effect could reduce the observation-model discrepancy to a certain degree, 
but cannot fully solve the overall underproduction. 
Both R10 and K20 have mentioned that the $\nu-$process can increase the production of these elements. 
Besides, it is suggested that nucleosynthesis calculations with 2D supernova explosion models are able to solve the problem (Maeda \& Nomoto 2003), 
which might also solve the underproduction problem for Ti. 

For iron-peak elements, model prediction can more or less reproduce the observed trends 
for the cases of Cr, Mn, and Ni, while shows larger discrepancies for Co and Zn. 
All three models predict an almost constant \AB{Cr}{Fe} distribution, which agrees well with the turnoff sample. 
As discussed in \S~\ref{subsubsec:iron-peak-elements}, the decreasing trend of [Cr/Fe] with decreasing [Fe/H] found for red giants could probably be removed 
if NLTE correction on the Cr abundances from \ion{Cr}{1} lines is included. 
It is noticed that P18 tends to 
reproduce the increase of \AB{Mn}{Fe} with metallicity at \FeHgt{-2.0}, which is due to the metallicity-dependent yields of Mn from massive stars for the P18 model. 

The Co abundance trends seem to be the most difficult case among the iron-peak group, 
where neither model succeeds to explain the observed trend well. 
In previous studies, one major puzzle of Co evolution that theoretical models 
fail to reproduce is the observed increasing trend at \FeHlt{-2.5} with decreasing [Fe/H]. 
However, as shown in Figure~\ref{fig:abun_trend_iron_obs}, 
the slopes of the observed trend in both giants and turnoff stars in our sample are respectively $-0.06$ and $-0.05$, 
not indicating very clear rising trends as observed in previous studies (e.g., FS and other observations compared in R10, K20, and P18).  
Therefore, the main discrepancy seen here is the Co underproduction in the models, 
which could become even more significant after considering NLTE corrections (e.g., Figure 15 in R10 paper). 
The difference may be weakened through enhanced explosion energies which may lead to higher Co production 
in low-metallicity and high-mass stars \citep[see][etc., for the discussion]{Umeda&Nomoto2005ApJ,Nomoto2013ARAA}

P18 fails to reproduce the observed trend for Zn, presenting significant underproduction, 
which seems to be a common feature of the yields from \citet{Woosley&Weaver1995ApJS} and \citet{Nomoto2013ARAA}. 
In general, all compared models tend to predict a rather flat trend of \AB{Zn}{Fe} 
towards the extremely low-metallicity region. 
This is inconsistent with the observed increasing trend with decreasing [Fe/H], 
which may imply that the fraction of HNe/high-energy SN has been larger in the early stages of the Galaxy evolution. 
For example, as tested by \citet{Kobayashi2020ApJ}, if large fraction of HNe is applied, 
the GCE model predicts a typical \AB{Zn}{Fe} ratio of about 0.2 in a wide range of metallicity, 
which is in good agreement with our observation (as described in \S~\ref{subsubsec:iron-peak-elements}). 
On the other hand, with a metallicity-dependent HNe fraction, 
the predicted \AB{Zn}{Fe} ratio may continuously decrease towards \FeHgt{-1.0}, 
presenting much lower values than observations (e.g., Saito et al. 2009). 
In that sense, more realistic modeling of synthesis of Zn in supernova explosions, 
as well as chemodynamical simulations (e.g., Kobayashi \& Nakasato 2011),     
may be required to better reproduce the observation. 
Also, synthesis of Zn in a variety of supernova explosions 
(e.g. hypernovae, electron-capture supernovae) have been explored to solve the problem, \citep[e.g., ][]{Nomoto2013ARAA, Hirai2018ApJ}.
For example, such trend can be explained with the dependence of the swept-up H mass on the explosion energy 
(e.g., \citealt{Umeda&Nomoto2005ApJ}, \citealt{Nozomu2007ApJ}) 
and also indicate that the environment is not well mixed, 
which cannot be reproduced with a one-zone model in all models adopted here.


To briefly summarize, the models are in agreement with the abundance trend for some key elements, 
including Na, Mg, Si, Ca, Cr, Mn, and Ni, though there remain difficulties to fully explain the observed slope. 
For some cases, such discrepancy may have been caused by technical issues of abundance measurements, 
such as NLTE effect (e.g., for the positive trend of \AB{\ion{Cr}{1}}{Fe} among giants). Stellar evolution on the upper red-giant branch could affect surface carbon abundances.
Some other cases like the decreasing trend of \AB{Mg}{Fe} and \AB{Si}{Fe} at \FeHgt{-2.5} would be explained by taking  
the assembly (dynamical) process of the Milky Way formation into consideration.
On the other hand, serious disagreements exist between observations and models for some elements, 
such as Sc, Ti, V, and Co, which would require further updates of nucleosynthesis yields. 
It should be kept in mind that the three selected models (the majority of theoretical studies)
have been made under the framework of one-zone model (for the local halo), 
which is based on the assumption of a unique relation between the metallicity (e.g., \FeH) and time, 
and is thus clearly an oversimplification of the real situation. 
For example, due to the well-known large scatter in the age-metallicity relation of field stars, 
it is difficult to define at what metallicity a long-lived source, e.g., AGB stars, started to enrich the Galactic interstellar medium. 
Comparisons of the observations with the results of recent chemo-dynamic simulations will be helpful to examine these effects. 

\subsection{Abundance correlations} \label{subsec:abun_correlation}

\begin{figure*} 
\hspace{-1.2cm}\epsscale{1.2}
\plotone{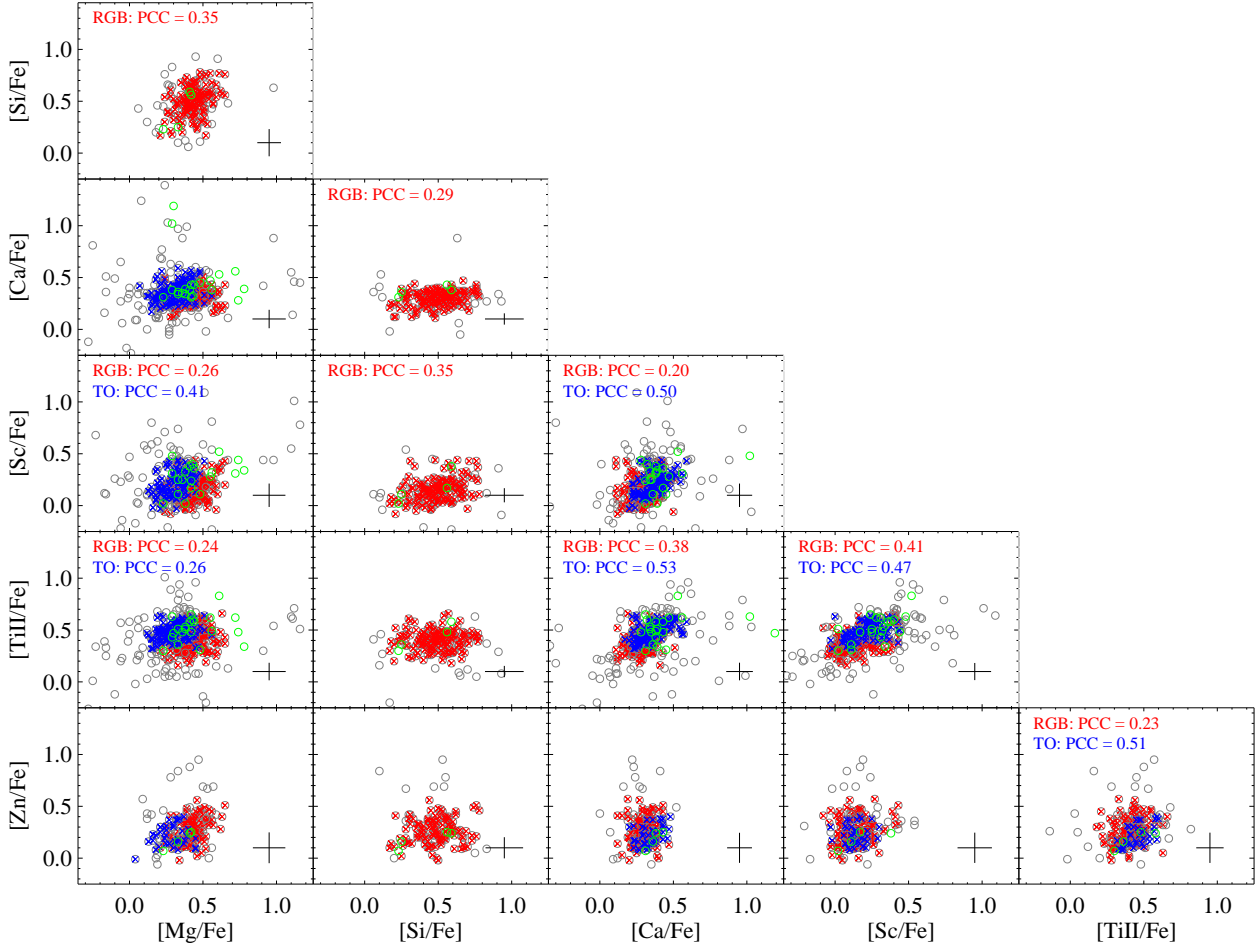}
\caption{Abundance correlation between $\alpha$-elements, Sc and Zn. 
The Pearson correlation coefficient (PCC) calculated for each set is presented in the plot in cases that exhibit potential correlation, 
e.g., PCC $> 0.20$, separately for giant (red) and turnoff (blue) stars. Open circles indicate objects whose abundance ratios are beyond the 3$\sigma$ of the general trend of \AB{X}{Fe} and are thus excluded for the calculation of the correlation (see text). Horizontal-branch stars (green) are also shown here for reference. 
\label{fig:abun_corr_alpha_Zn}}
\end{figure*}

\begin{figure*}
\hspace{-1.2cm}\epsscale{1.2}
\plotone{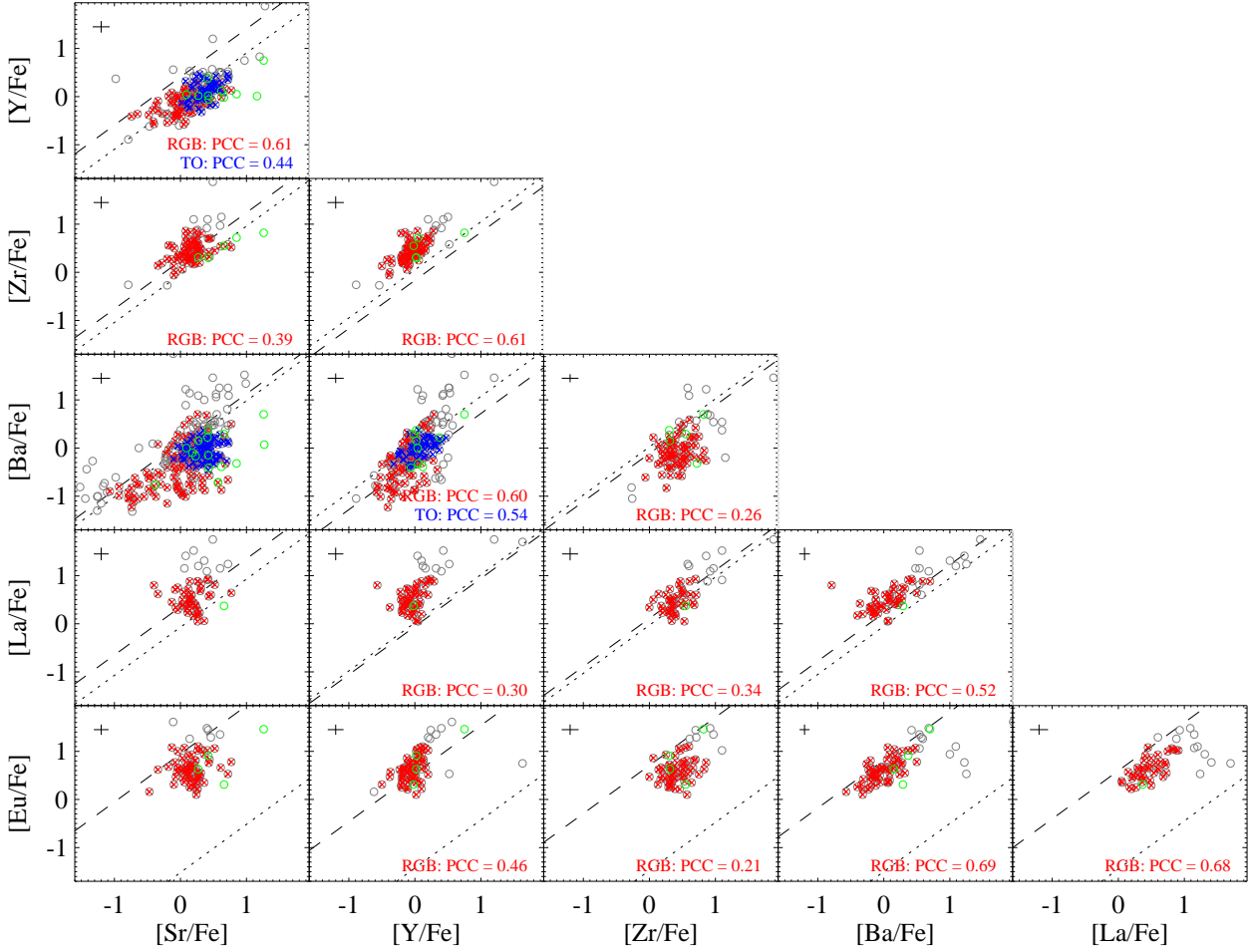}
\caption{Abundance correlation between heavy elements Sr, Y, Zr, Ba, La, and Eu. 
Definition of symbols are the same as that of Figure~\ref{fig:abun_corr_alpha_Zn}. 
The dashed and dotted lines represent the abundance ratios of the $r-$process and $s-$process components in solar-system material, respectively. 
\label{fig:abun_corr_heavy}}
\end{figure*}

Correlations among different species are able to reflect the relation between the site or mechanisms 
that produce these elements. We here present the abundance correlations among $\alpha$-elements, Sc, and Zn, as well as correlations among the six heavy elements that have been measured for this sample, aiming at exploring the relations and differences of the origins of these species (Figures \ref{fig:abun_corr_alpha_Zn} and \ref{fig:abun_corr_heavy}). 
Note that when discussing about the abundance correlations, 
we have excluded program stars that are located beyond 3$\sigma$ of the abundance trend as defined in \S~\ref{subsec:abundance_trend}, 
and thus the following discussion is based on the bulk of the sample which are regarded the 
so-called ``general'' population of VMP/EMP stars. In Figures \ref{fig:abun_corr_alpha_Zn} and \ref{fig:abun_corr_heavy}, however, objects excluded from the statistics are also shown by open circles for completeness. 

Figure~\ref{fig:abun_corr_alpha_Zn} shows correlations of abundances between $\alpha-$elements, Sc and Zn, which are key elements to constrain the early star formation history 
and the properties of core-collapse supernovae. 
Correlations between the four $\alpha$-elements (Mg, Si, Ca and Ti) are clearly seen, 
which is expected from the (relevant) production sites of  these species. 
More detailed inspection reveals that correlations between Mg and Si, and between Ca and Ti are stronger, whereas correlations between Mg (or Si) and Ca (or Ti) are weaker or insignificant. This suggests that the  origins between the two lighter elements (Mg and Si) and heavier ones (Ca and Ti) are similar with each other, but slightly different between the two groups. This is also expected from theories of massive star evolution and supernova explosions: Mg mainly comes from hydrostatic carbon burning and explosive Ne burning; 
Si is mostly produced during hydrostatic burning and explosive O burning; Ca is directly contributed from explosive incomplete Si and O burning; the main isotope of Ti ($^{48}$Ti) is mainly synthesized during explosive complete and incomplete Si burning \citep{Woosley&Weaver1995ApJS,Chieffi1998ApJ,Kobayashi2006ApJ}. 

Interestingly, Sc abundances show relatively strong correlations with all the four $\alpha$-elements. 
This correlation would support the prediction by nucleosynthesis models that Sc is produced both during Ne burning and in explosive O and Si burning, which shares similar origins with these $\alpha$-elements. 
The slight overabundance in Sc is also in agreement with the general overabundance in $\alpha$-elements 
as seen in Figure~\ref{fig:abun_trend_light_obs}.
A correlation is also found between Zn and Ti, whereas the correlation is not clear between Zn and other $\alpha$-elements. 
Note that Zn is mainly produced during explosive complete Si burning, which is similar to the production site of Ti. 
Moreover, with the higher explosion energy, the explosive O and incomplete Si burning layers just shift, 
while the complete Si burning layer increases. This may explain the unclear correlation between Zn and the other three elements (Mg, Si, and Ca).

Abundance correlation among Sr, Y, Zr, Ba, La, and Eu are presented in Figure~\ref{fig:abun_corr_heavy}. 
Correlations are found between most of these six heavy elements, 
and a stronger correlation is usually detected between elements having similar atomic numbers, 
i.e., relatively stronger correlation between the lighter neutron-capture elements Sr and Y, and that between the heavier ones Ba and Eu. 
This strong correlation between elements with similar atomic (mass) numbers reflect the nuclear physics. 
The weak, or non-detectable, correlations between lighter and heavier neutron-capture elements 
suggest variations of conditions in the nucleosynthesis sites, or contributions of more than one processes. 
We note again that the correlation between Ba and Eu abundances can be studied only for objects with Eu measurements, 
i.e. those having relatively high abundances of neutron-capture elements. Whereas the Ba and Eu abundances of most of these stars are well explained by the r-process, 
no useful information is obtained for other stars from our sample.

As reported by previous studies, abundance ratios between Ba and lantanides show clear correlations following the $r-$process ratios, 
where \AB{Ba}{Eu} and \AB{La}{Eu} are excellent indicators to distinguish contributions of the $r-$ and $s-$processes.  
Among them, the \AB{La}{Eu} ratios show a small departure from the solar $r-$process ratio (Fig.~\ref{fig:abun_corr_heavy}, 
which suggests partial contribution from the $s-$process. 
No signature of s-process contribution is, however, found in the \AB{Ba}{Eu} ratios. 
Further studies are required to understand the departure of \AB{La}{Eu} ratios in these stars. 

Among the abundance ratios between light neutron-capture elements, 
\AB{Sr}{Y} ratios better agree with the solar $s-$process ratio than that of the $r-$process 
\citep[e.g., see][for the solar $s-$ and $r-$ process ratios]{Simmerer2004ApJ}.
On the other hand, the \AB{Zr}{Y} ratios are even higher than those of the $s-$process as well as of the $r$-process. 
Hence, we are not able to conclude which process can better explain the abundance ratios of these light neutron-capture elements. 
A clear observational result is that Y is underabundant compared to Zr. 
This could be a constraint on nucleosynthesis models that is responsible for light neutron-capture elements in the early stage of chemical evolution, 
which would be different from the usual $r-$process 
\citep[e.g., ][]{Travaglio2004ApJ, Wanajo2006NuPhA}. 

\subsection{CEMP stars}\label{subsec:cemp}

\begin{figure*}
\gridline{\fig{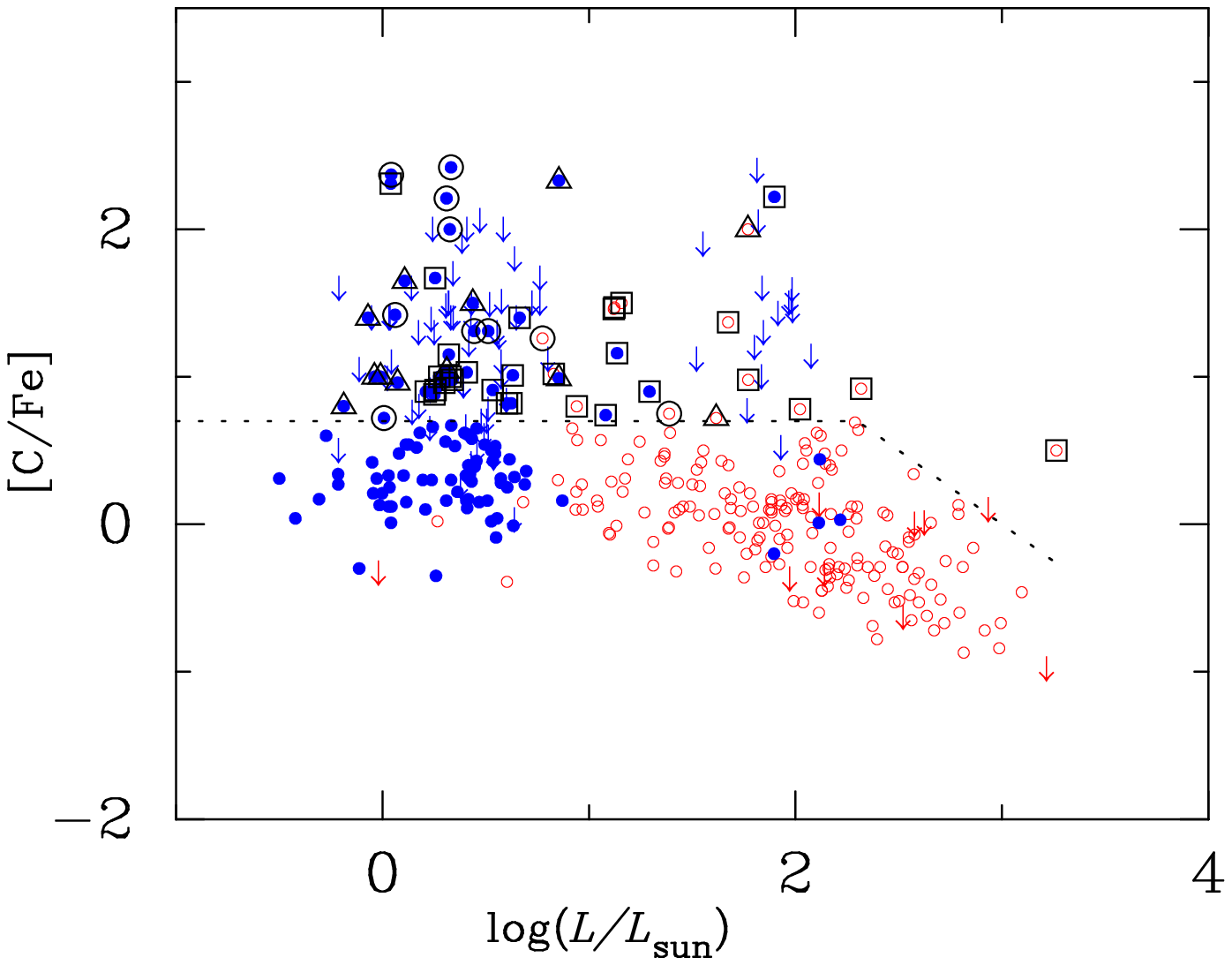}{0.4\textwidth}{}}
\gridline{\fig{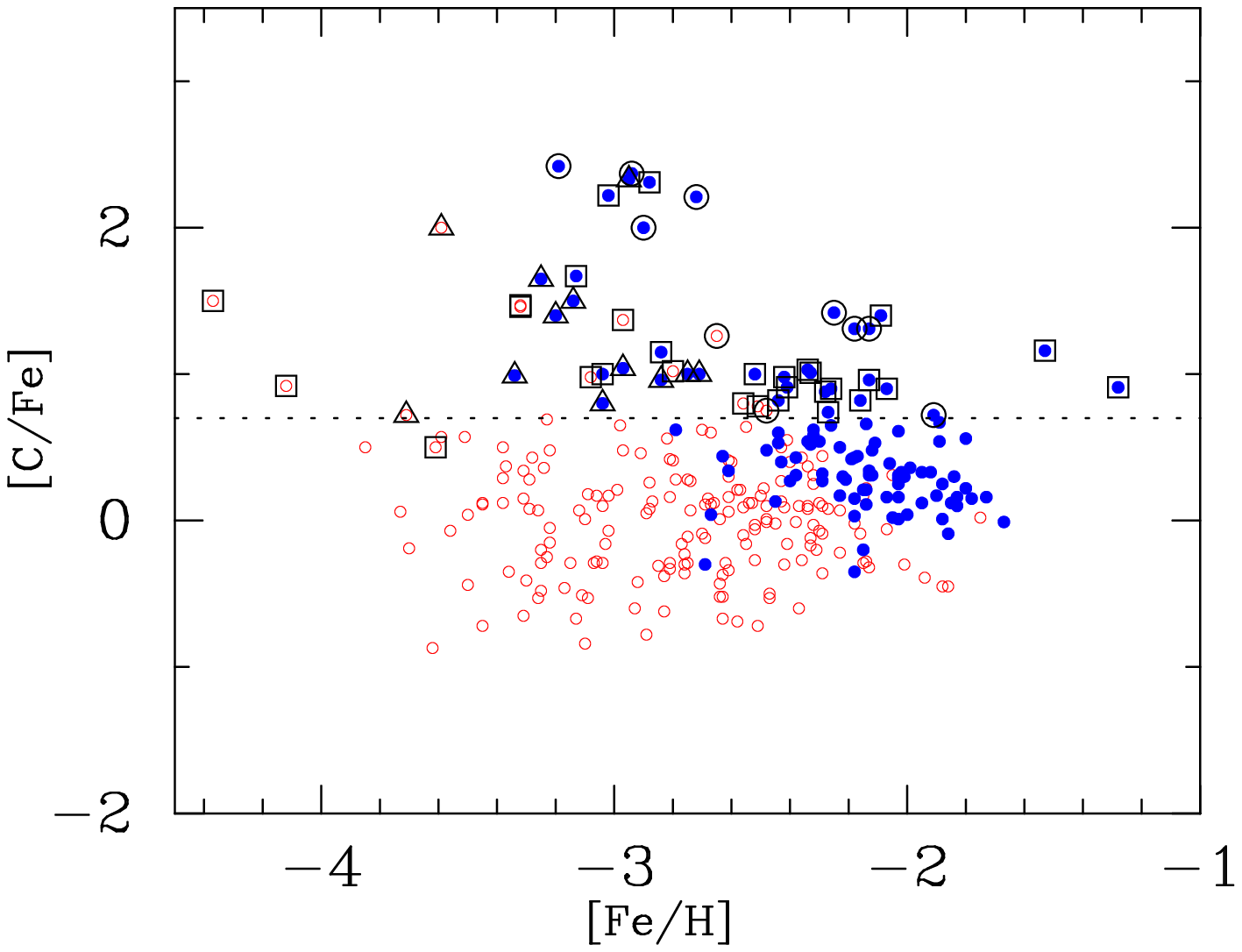}{0.4\textwidth}{}}
\gridline{\fig{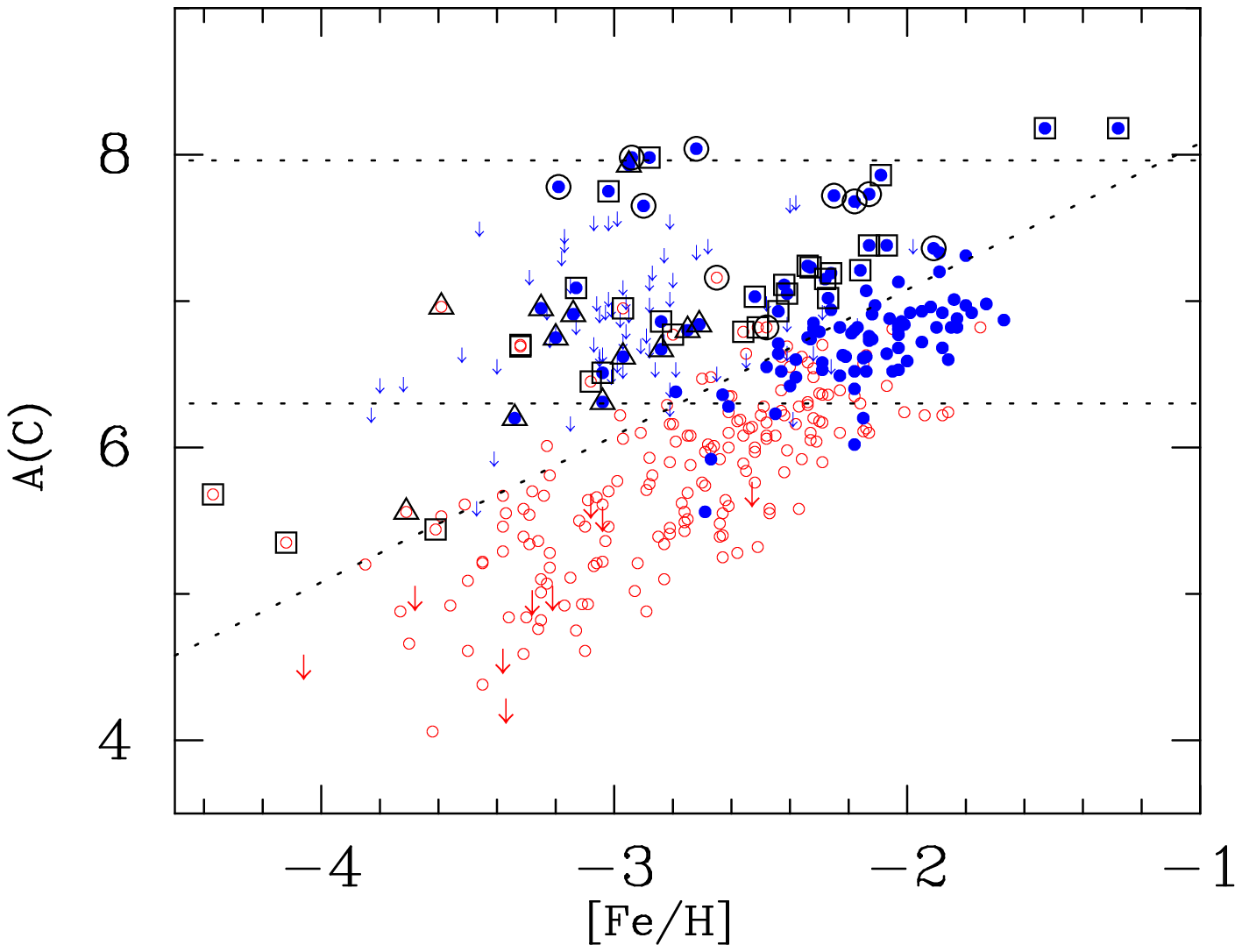}{0.4\textwidth}{}}
\caption{(top) [C/Fe] as a function of luminosity for our sample. Filled (blue) and open (red) circles indicate warm turnoff stars (\tefft$>5500$~K) and cool giants, respectively. CEMP-s and CEMP-no stars are indicated by large circles and squares, respectively. 
Triangles mean CEMP stars for which classification is not available. 
Arrows mean upper limits. The dotted line indicates that criterion of CEMP stars. 
(middle and bottom) [C/Fe] and A(C) as a function of [Fe/H] for our sample with the same symbols for the top panel. 
\label{fig:cfe}}
\end{figure*}


A remarkable feature found in very metal-poor stars is the high frequency of
objects with high carbon abundances, called carbon-enhanced metal-poor
(CEMP) stars. Previous studies have reported that the frequency is as high
as 20\% in the metallicity range of [Fe/H]$<-2$, and it increases with
decreasing metallicity \citep[e.g., ][]{Placco2014ApJa}. 
The dependence of the CEMP frequency on the distribution in the Milky Way halo has been also studied based on large spectroscopic survey data, 
reporting higher frequency in the outer halo region \citep{Lee2017ApJ}.
CEMP stars are classified into s-process-enhanced ones (CEMP-s) characterized by high Ba abundances, those with low Ba abundance (CEMP-no), and other minor sub-classes \citep[e.g., ][]{Beers&Christlieb2005ARAA}. 

Figure \ref{fig:cfe} (top and middle) shows carbon abundance ratio ([C/Fe]) as a function of luminosity (Table~\ref{tab:stellar_param}) and 
metallicity ([Fe/H]), separating the objects into turnoff stars (filled circles) and red giants (open circles). 
The CEMP criterion given in \S~\ref{subsubsec:carbon} is shown by the dotted line in the top panel. 

Carbon abundances are determined from the CH bands at around 4315 and 4322~{\AA} for 281 objects in the present
work. The numbers of turnoff stars and giants for which carbon
abundances are determined are 109 and 172, respectively (Table~\ref{tab:cemp}).
The CH molecular band is not detected in spectra of the
remaining stars, for which upper-limit is estimated by comparison of
synthetic spectra with the observed ones. The molecular absorption is
weaker in general in turnoff stars, and, hence, the upper limit of
carbon abundances estimated for objects with no detection of the CH bands are relatively high, in particular for extremely
metal-poor stars. The limit of carbon abundances ([C/H]) for which the
CH band is detectable is shown in Figure~\ref{fig:CH_Teff}.
Among the objects for which upper limits of carbon abundances are determined, 
most of the giants and some turnoff stars have the upper limits lower than the
criterion for CEMP stars (see Table~\ref{tab:cemp}).

In Figure~\ref{fig:cfe} (middle and bottom), CEMP stars with large excess of Ba ([Ba/Fe]$>1$; CEMP-s stars) 
are indicated by large open circles whereas objects with lower Ba abundances, 
including those having upper-limit of [Ba/Fe] lower than this criterion, are shown by large squares. 
The numbers of CEMP-s and CEMP-no stars are given in Table~\ref{tab:cemp}. 
CEMP stars with no useful constraint on Ba abundances are indicated in the figure by large open triangles overplotted. 
Most of these objects are main-sequence turn-off stars with extremely low metallicity ([Fe/H]$\lesssim -3$), 
for which high SNR spectra are required to determine Ba abundances if Ba is not enhanced ([Ba/Fe]$<1$).

   

The numbers of CEMP-s turnoff stars and giants are 8 and 2, respectively. 
Four of the turnoff stars have very high [C/Fe] values ($>2$). 
The [C/Fe] values of others are not extreme, but some of these stars have  high $A$(C) values 
given the relatively high metallicity (the lower panel of Figure~\ref{fig:cfe}). 
Their high carbon abundance corresponds to the CEMP Group I stars in the classification by \citet{Yoon2016ApJ}.

By contrast, most of the CEMP-no stars (20 turnoff stars and 9 giants) have [C/Fe] (and [C/H]) not as high
as those of CEMP-s stars. This is known as the feature of Group II and III
in \citet{Yoon2016ApJ}. The distribution of these stars extends to extremely
low metallicity ([Fe/H]$<-3.5$), whereas that of CEMP-s stars show
lower bound at around [Fe/H]$=-3$. Many CEMP-no stars are also found
in [Fe/H]$>-2.5$ in our sample. This is not the case of the sample of
\citet{Aoki2007ApJ}, who discussed that CEMP-no stars appear at lower
metallicity. This would be due to a bias in their sample selection. 
The present work includes many turnoff stars with [Fe/H]$>-2.5$, 
some of which have slightly higher [C/Fe] than the CEMP criterion. 
Such stars are not well covered by \citet{Aoki2007ApJ} who focused
on stars with strong CH bands. Indeed, ten of the 14 CEMP-no turnoff stars with [Fe/H]$>-2.5$ have $0.7<$[C/Fe]$<1.0$ in our sample. 
These stars have similar abundance ratios (e.g., $\alpha$/Fe) to those of non-CEMP stars. 


Most of the CEMP stars for which Ba abundances are not determined are
turnoff stars having [Fe/H]$\sim -3$ and [C/Fe]$<2$. The distribution
of Fe and C abundances overlaps with that of CEMP-no stars, suggesting
that most of them are CEMP-no objects.

The fraction of CEMP stars among giants is 14/179=7.8\%. We note that six objects among
the seven for which CH band is not detected have upper
limits of carbon abundance lower than the CEMP criterion. Hence, the estimate is not affected by the detection limit of CH bands for giants. 

The fraction of CEMP turnoff stars is nominally 37/120=31\%. Nine objects with the upper limit of [C/Fe]$<0.7$ are included in C-normal stars in this calculation. If all other stars for which CH bands are not detected are assumed not to be carbon-enhanced, the fraction is 37/172=22\%. This could be the lower limit of the frequency of CEMP stars in turnoff stars.
It should also be noted that there are many CEMP stars with [C/Fe] close to the criterion. 
This is found in the previous sample as the Group II stars of \citet{Yoon2016ApJ}.
If we adopt [C/Fe]$=+1$ as the criterion, the CEMP fraction in turnoff stars decreases to 11\%.  

Even if these factors are taken into consideration, the frequency of
CEMP stars is higher in turnoff stars than in giants. 
In particular, the frequency of CEMP-s stars is clearly high in turnoff stars, 
and the excess of carbon in most of them is remarkable (i.e. Group I stars). 
The origin of carbon excess of CEMP-s stars is presumed to be mass transfer from companion AGB stars. 
The accreted carbon-rich material could be significantly diluted in the stellar envelope in giants. 
If the mixing during the turn-off phase is not effective, carbon abundances are expected to be higher in turn-off stars than in giants. 
An extreme example of CEMP main-sequence turn-off star is J0119-0121 which has [C/Fe]=+2.4 and [Fe/H]$=-3.2$. 
This star was studied by \citet{Zhang2019PASJ} as an extreme CEMP-s star having the largest excess of
neutron-capture elements ([Ba/Fe]$=+2.6$) with the abundance pattern well explained by the s-process. 
There are three other CEMP-s turnoff stars that have similarly large excess of carbon (J0824+3025,
J0949+2707 and J1020+4046), though the excess of Ba shows wide distribution. 
No such extreme object is found among CEMP-s giants in our sample: 
the giant having the highest [C/Fe] is J2257+3859 ([C/Fe]$=1.3$).
We note that mixing during turn-off phase is proposed for carbon-enhanced stars by mass accretion 
(e.g., thermohaline mixing, \cite{Stancliffe2007AA}). Our result does not deny existence of such mixing, 
but suggests that such mixing is not effective in many cases of carbon-enhanced turn-off stars.

The frequency of CEMP-no turnoff stars is also higher than CEMP-no giants, 
although the contrast is not as significant as found for CEMP-s stars. 
CEMP-no stars have been interpreted as stars formed from C-enhanced cloud to which some progenitors 
(e.g., faint supernovae, rotating massive stars) have contributed \citep[e.g., ][]{Meynet2006AA,Tominaga2014ApJ}.  
Even in such cases, the surface C abundances might be reduced in highly evolved red giants 
as suggested from C-normal red giants (\S~\ref{subsec:abundance_trend}). 
The C abundances of about half of CEMP-no turnoff stars is $0.7<$[C/Fe]$<1.0$. 
This relatively small excess could be erased by mixing  with internal material affected by the CNO-cycle, 
resulting in lower frequency of CEMP-no stars in evolved giants. 

An alternative possibility is that some of the CEMP-no stars have been formed 
by mass-accretion of C-enhanced material from companion AGB stars in which s-process is not active \citep{Suda2004ApJ}. 
The surface C-enhanced material could be diluted through the first dredge-up as in the case of CEMP-s stars. 
To confirm this possibility, radial velocity monitoring to examine the binarity for these stars is useful. 
Paper I reports radial velocity variations from the measurements with LAMOST and Subaru. 
Although the measurements are not sufficient to definitively identify binary systems for most of the stars in our sample, 
the fraction of stars with radial velocity variations would provide some hint on the binary frequency of each class of CEMP stars. 
The numbner of stars that show radial velocity variations larger than 20~km~s$^{-1}$ is given in Table~\ref{tab:cemp}. 
More than half of CEMP-s stars show radial velocity variations, 
supporting the scenario that carbon-excesses are caused by mass transfer in binary systems. 
Interestingly, the fraction of stars showing radial velocity variations is about half also in CEMP-no stars. 
This is higher than the faction for the whole sample reported in Paper I (35\%). 
This result might support the scenario that some of CEMP-no stars are also formed through binary mass transfer. 
It should be noted that the fraction is also quite high in CEMP stars that are yet classified (8/12). 
Binarities of CEMP-s and CEMP-no stars have also been investigated in previous studies, 
e.g., \citet{Hansen2016AAa} and \citet{Hansen2016AAb} which have discovered a binary fraction of 82\% and 17\% 
respectively for extremely metal-poor CEMP-s and CEMP-no stars. Our results are consistent in the fact that 
such fraction is higher in CEMP-s than CEMP-no stars, though showing different fractions for each class, 
which may be caused by different metallicity coverage of the two works, and inclusion of more moderately C-enhanced stars in our sample.

There are two CEMP-no turnoff stars with [C/Fe]$>2$ showing no excess of Ba (J0150+2149 and J1313$-$0552).
The carbon-excess of these stars is as large as the ``physically motivated'' criterion of C-Enhanced EMP stars ([C/Fe]$=+2.3$) proposed by \citet{Chiaki2017MNRAS}, 
above which carbon grains are dominant coolant in the process of low-mass star formation. 
Further studies of detailed abundance patterns of these stars would be useful 
to constrain the condition of star formation in the early Galaxy.



\begin{table}
\caption{Numbers of CEMP and C-normal stars \label{tab:cemp}}
\begin{center}
\begin{tabular}{rrrr}
\hline\hline
   & Giants$^{a}$ & turnoff$^{a}$ & \\
\hline
CEMP-s & 2(1) & 8(5) & \\
CEMP-no & 8(5) & 19(8) & \\
CEMP (unclassified) & 2(2) & 10(6) & \\ 
C-normal$^{b}$ & 160 + 6 & 72 + 11 &  \\
C upper-limit$^{c}$ & 1 & 52 & \\
\hline
\end{tabular}
\end{center}
\tablenotetext{a}{Numbers of objects showing radial velocity variations in parenthesis}
\tablenotetext{b}{Including objects with [C/Fe] upper-limit lower than the CEMP criterion}
\tablenotetext{c}{Objects with [C/Fe] upper-limit higher than the CEMP criterion }
\end{table}

\subsection{$\alpha$-peculiar stars}\label{subsec:alpha_peculiar} 

\begin{figure*}
\hspace{2.0cm}
\epsscale{1.0}
\plotone{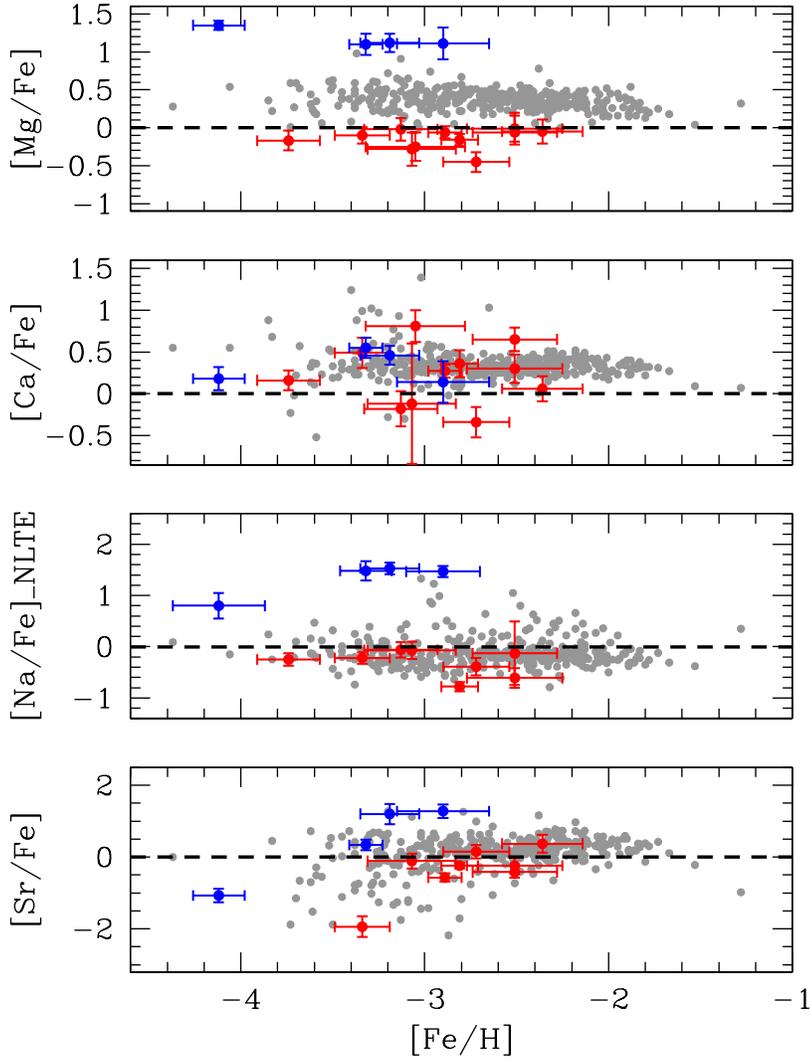}
\caption{Comparison of Mg, Ca, Na (NLTE corrected) and Sr of Mg-poor stars (filled red circles) 
and Mg-rich stars (filled blue circles) with those of other metal-poor stars (filled gray circles) in our sample.\label{fig:ap1}}
\end{figure*}

As shown in Figure~\ref{fig:ap1}, the bulk of metal-poor stars in our sample present similar overabundance of $\alpha$ elements (e.g., [Mg/Fe]$\sim$0.35) with small scatter. 
However, 
a small number of stars deviate from the majority by showing extreme enhancement or significant deficiency in $\alpha$-element abundances.
For example, there are 11 program stars showing sub-solar Mg abundances with [Mg/Fe]$<0$ (hereafter the ``Mg-poor'' stars), 
and four objects showing extreme enhancement in Mg with [Mg/Fe]$>1$ (hereafter the ``Mg-rich" stars). 
The Mg-poor stars cover the metallicity range from \FeHeq{-3.7} to $-2.3$, overlapping with 
the majority of the sample which shows typical enhancement of \ABsim{Mg}{Fe}{0.3\sim0.4} (as shown in \S~\ref{subsec:abundance_trend}]). 
Our discovery of three Mg-poor stars with [Fe/H]$<-3.1$ extends the distribution of such objects to lower metallicity region. 
Although no Mg-poor stars with [Fe/H]$<-3.8$ have yet been found in either our sample or previous studies, 
this could probably be due to the limited sample size of such extremely low-metallicity stars as well as the rarity of Mg-poor stars. 
Larger sample of EMP stars with \FeHlt{-3.8} will be necessary to determine or better constrain 
whether there exist a lower limit in metallicity for such Mg-poor stars.

\begin{figure*}
\epsscale{1.0}
\plotone{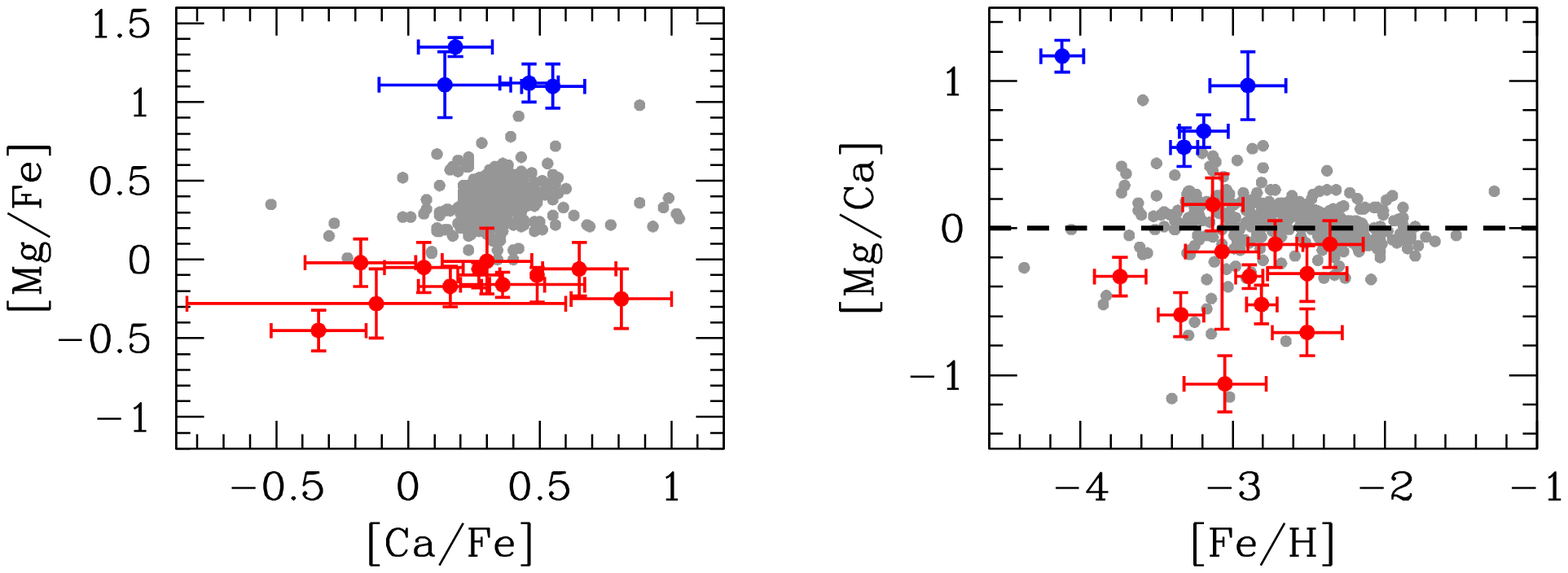}
\caption{Left: [Mg/Fe] as a function of [Ca/Fe]. Right: [Mg/Ca] versus [Fe/H]. 
The same symbols are used as in Figure~\ref{fig:ap1}. 
\label{fig:ap2}}
\end{figure*}

The Mg-rich stars with \ABgt{Mg}{Fe}{1.0} distribute in the metallicity range from $-4.1$ to $-2.9$ dex, 
consistent with previously known metal-poor stars with extremely enhanced Mg (e.g., CS22949--037: \citealt{Norris2002ApJ}; CS~29498--043:\citealt{Aoki2002ApJ}). 
It is interesting that most outliers for Mg have normal Ca abundances, resulting in a large scatter in the plot of \AB{Mg}{Ca} as a function of metallicity (see Figure~\ref{fig:ap2}). 
The \AB{Ca}{Fe} of Mg-rich stars are only slightly higher than that of the comparison stars (see below for more details). 
Only three Mg-poor stars have sub-solar \AB{Ca}{Fe}, one of which presents the lowest Mg abundance among our sample (J1220+1637: \ABeq{Mg}{Fe}{-0.45}), 
as well as a relatively low Ca abundance (\ABeq{Ca}{Fe}{-0.34}) in comparison with other stars around this low metallicity ([Fe/H]=$-2.72$). 
The other two Mg-poor stars (J1542+2115 and J1349+1551) also have sub-solar [Ca/Fe] ratios. 
For J1349+1551, however, the Ca abundance is more deficient than that of Mg, making it a unique Mg-poor star with a positive \AB{Mg}{Ca}.

On the other hand, there are a few Ca-peculiar stars showing relatively high or sub-solar \AB{Ca}{Fe} ratios but with normal Mg abundances. 
However, most of their Ca abundances are determined from the single \ion{Ca}{1} line at 4226\,\AA, 
which are not as reliable as abundances determined from more than one Ca lines. 
Therefore for this work, we only consider Mg-poor and Mg-rich stars as candidate $\alpha$-peculiar objects. 
The absorption lines of Mg have been visually checked for each object to ensure their distinctive \AB{Mg}{Fe} ratios.

The 11 Mg-poor program stars include five EMP and six VMP stars. 
Stars with similarly deficiency in Mg have been found in the field halo \citep[e.g., ][]{Ivans2003ApJ, Aoki2014Science} 
and nearby dwarf spheroidal galaxies \citep[e.g., ][]{Tolstoy2009ARAA, Shetrone2003AJ}. 
\citet{Ivans2003ApJ} suggests that these objects would have been formed in environment with larger contributions from SN~Ia yields 
in comparison with other halo stars with similar metallicities. 
Since Mg is mainly produced during massive star evolution while Ca is produced by both type Ia and type II SN, 
Mg is expected to be more deficient than Ca in this scenario. 
As shown in the right panel of Figure~\ref{fig:ap2}, most of the Mg-poor stars in our sample have sub-solar \AB{Mg}{Ca} ratios, 
which is consistent with increased contribution from SN~Ia. 
However, J0908+3119, the most metal-poor object in our Mg-poor stars has a metallicity of \FeHeq{-3.74}, 
and there are four other Mg-poor EMP stars in our sample. 
It is not clear whether SN~Ia could have significantly contributed to stars with such extremely low metallicity, 
because the time-scale of the contribution of SN~Ia to chemical evolution is much longer than that of SN~II. 
A possible scenario is that these Mg-poor stars are originated from classical or ultra-faint dwarf galaxies that had very low star formation rates. 
In such stellar systems, the onset of SN~Ia might occur at much lower metallicities than that in the MW. 
Indeed, EMP $\alpha$-poor stars have been 
found in some dwarf galaxies \citep[e.g., ][]{Aoki2009AA}. 
The \AB{Mn}{Fe} ratios are expected to be low for stars formed from cloud enriched only by SN~II \citep[e.g., ][]{Tominaga2014ApJ}. 
J1542+2115 ([Fe/H]$=-3.07$) is the only one with Mn measurement in our Mg-poor sample. 
Its relatively high \AB{Mn}{Fe} in comparison with other EMP stars implies that SN~Ia may have enriched the gas before the formation of some EMP stars. 
Similar low Mg abundances have been found in two EMP stars by \citet{Caffau2013AA}. 
They also suggest that contributions from SN~Ia are responsible for the Mg deficiencies. 
On the other hand, \citet{Kobayashi2014ApJ} argued that the yields of core-collapse supernovae, i.e. 13-25 M$_\odot$ SN, 
could explain the low \AB{Mg}{Fe} ratios of EMP stars from \citet{Caffau2013AA} without contribution from SN~Ia, 
although very low \AB{Mn}{Fe} ratios are still predicted by their models. 
In addition, Pair-instability supernovae (PISN) could also lead to sub-solar \AB{Mg}{Fe} in VMP stars 
by contributing larger amount of iron than core-collapse supernovae \citep{Aoki2014Science}. 
\citet{Takahashi2018ApJ} demonstrated that high \ABsim{Ca}{Mg}{0.5-1.3} abundance ratios could be prominent PISN signatures 
compared with abundances of normal metal-poor star, which have been found in four of our Mg-poor stars. 
However, extremely low [C/Fe] and [Na/Fe] abundances are expected be produced by PISN in the meanwhile, 
which have not been discovered in our Mg-poor sample stars.

The Mg-poor sample collected by \citet{Ivans2003ApJ} shows unusually low abundances of Na, Sr, and Ba with respect to Fe, 
e.g., the average [(Sr,Ba)/Fe] is more than 1.0 dex lower than the typical values, 
indicating possible association between the low-[Mg/Fe] ratios and the low [Sr/Fe] (and [Ba/Fe]). 
However, recent discoveries on Mg-poor stars \citep[e.g., ][]{Xing2019NatAs, Sakari2019ApJ, Ezzeddine2020ApJ} 
suggest that these objects do not necessarily to be deficient in Sr and Ba, 
and they could be with normal abundances of neutron-capture elements, or even $r-$process enhanced.
As shown in Figure~\ref{fig:ap1}, the Mg-poor stars in our sample consists of objects with both normal and deficient Na and Sr abundances. 
No $r-$process enhanced Mg-poor stars have been found in our sample, which may be due to the rarity of such objects.

The fraction of Mg-rich stars in our sample is about 1.0\% (four out of 385). 
All of them are enhanced in Na (with the NLTE corrected [Na/Fe]$>1$ for three of them,  
as seen in Figure~\ref{fig:ap1}), while their Ca abundances seem to be normal, leading to high [Mg/Ca] ratios. 
Such high [Mg/Ca] and [Na/Fe] have been found by previous studies for a couple of halo stars: 
e.g., CS~22949--037 ([Mg/Ca]=1.2:\citealt{Norris2002ApJ}, [Na/Fe]=1.57:\citealt{Andrievsky2007A&A}), HE~1327--2326 ([Mg/Ca]=0.99, [Na/Fe]=0.94; \citealt{Frebel2008ApJ}), 
and J~2217+2104 ([Mg/Ca]=1.3, [Na/Fe]=1.33; \citealt{Aoki2018PASJ}) that is also included in our sample, 
suggesting that there exist a population of EMP Mg-rich stars. 
Several stars with high [Mg/Ca] ratios have also been discovered in the ultra faint dwarf galaxy Hercules \citep{Koch2008ApJ} and Grus II \citep{Hansen2020ApJ}. 
Both of these two studies argue that the property of metal-poor stars with high [Mg/Ca] ratios may be caused by massive ($\geq$ $20$M$_\odot$) stars. 
Moreover, all of the four Mg-rich stars in our sample are also highly enhanced in C, with the [C/Fe] ratios ranging from 0.92 to 2.42 dex. 
Two of them are CEMP-s stars with enhanced Ba abundances ([Ba/Fe]$>1$). 
Only one of these four stars, J2217+2104 does not possess large [Sr/Fe] or [Ba/Fe], 
but rather shows significantly low [Sr/Fe] and [Ba/Fe] in comparison with other known Mg-rich CEMP stars, 
suggesting that the excess of C and Mg may not be caused by the event related to neutron-capture processes \citep{Aoki2018PASJ}.

\subsection{Abundance distributions of neutron capture elements}\label{subsec:ncap}

We here focus on the three neutron-capture elements Sr, Ba and Eu, 
which are measured in a relatively large number of stars in our sample. 
These are also key elements to constrain the nucleosynthesis processes that produce heavy elements in the universe. 
Detailed abundance patterns of neutron-capture elements that are determined for a small number of stars 
will be discussed separately in a future work in this series. 

\subsubsection{Eu-rich stars}

Eu abundances are determined for 79 stars in our sample. 
Since Eu is not as abundant as Sr or Ba, it is detectable only for cool giants or Eu-enhanced warm stars among our sample. 
Figure~\ref{fig:ncap}(a) shows the [Eu/Fe] values as a function of [Fe/H]. 
As seen in this figure, Eu is detected in only several stars in [Fe/H]$<-3$.

Twelve stars are identified to have [Eu/Fe]$>+1$ and [Ba/Eu]$<0$ (Figure~\ref{fig:ncap}b). 
There is no evidence of carbon-excess for these stars. 
These stars are classified into ``r-II stars'' \citep{Beers&Christlieb2005ARAA} whose heavy
elements are enhanced by the r-process (filled symbols in Figure~\ref{fig:ncap}a,b). 
Similarly, the objects with [Eu/Fe]$>+0.5$ and [Ba/Eu]$<0$ are classified into ``r-I stars'' (plus symbols in the figure). 
There are 34 such objects in our sample. 

The [Fe/H] values of the four most Eu-enhanced stars are lower than $-2.5$. 
J~1109+0754 is one of the most metal-poor example of r-II stars known to date ([Eu/Fe]$=+1.4$ and [Fe/H]$=-3.3$). 
The detailed abundances of this object will be reported separately (Honda et al., in preparation). 
The abundances of this object have also been reported by \citet{Li2015RAA} and \citet{Mohammad2020ApJ}.
Very recently, \citet{Yong2021Nature} reported the discovery of an r-II star with [Fe/H]$=-3.5$, 
for which they suggest a magnetorotational hypernova as the progenitor, 
taking the long delay time expected for binary neutron star merger events into consideration. 
The low-matallicity tail of the metallicity distribution of r-II stars would be a key to constraining the origin of the r-process in the early Universe.

Other four r-II stars have $-2.5<$[Fe/H]$<-2.0$, and the [Eu/Fe] values are between 1.0 and 1.2. 
Such less metal-poor r-II stars have been identified by recent studies \citep[e.g.,][]{Roederer2018ApJ, Xing2019NatAs}, 
indicating the existence of efficient processes 
to enrich r-process elements in ISM at this metallicity. 

The frequency of r-II stars estimated by previous studies is 3-5\% \citep[e.g., ][]{Barklem2005AA}. 
The frequency of r-II stars in our sample is 12/380$\sim 3$\%,
which is similar to, or slightly lower than the previous estimates. 
The value is, however, should be a lower limit 
because there could be many warm stars that have high Eu abundances but their Eu lines are yet detected due to the limitation of data quality. 
The frequency could be better estimated for giants, in which Eu is detectable in most cases if it is enhanced. 
The frequency of stars with [Eu/Fe]$>+1$ is 10/211 ($\sim 5$\%) for giants. 
We note that the two r-II main-sequence turn-off stars 
found in this study are J0948+0000 ([Eu/Fe]$=+1.5$ and [Fe/H]$=-2.8$) 
and J1044+3713 ([Eu/Fe]$=+1.0$ and [Fe/H]$=-2.0$).

\subsubsection{Sr and Ba abundances}

\begin{figure*}
\epsscale{0.5}
\plotone{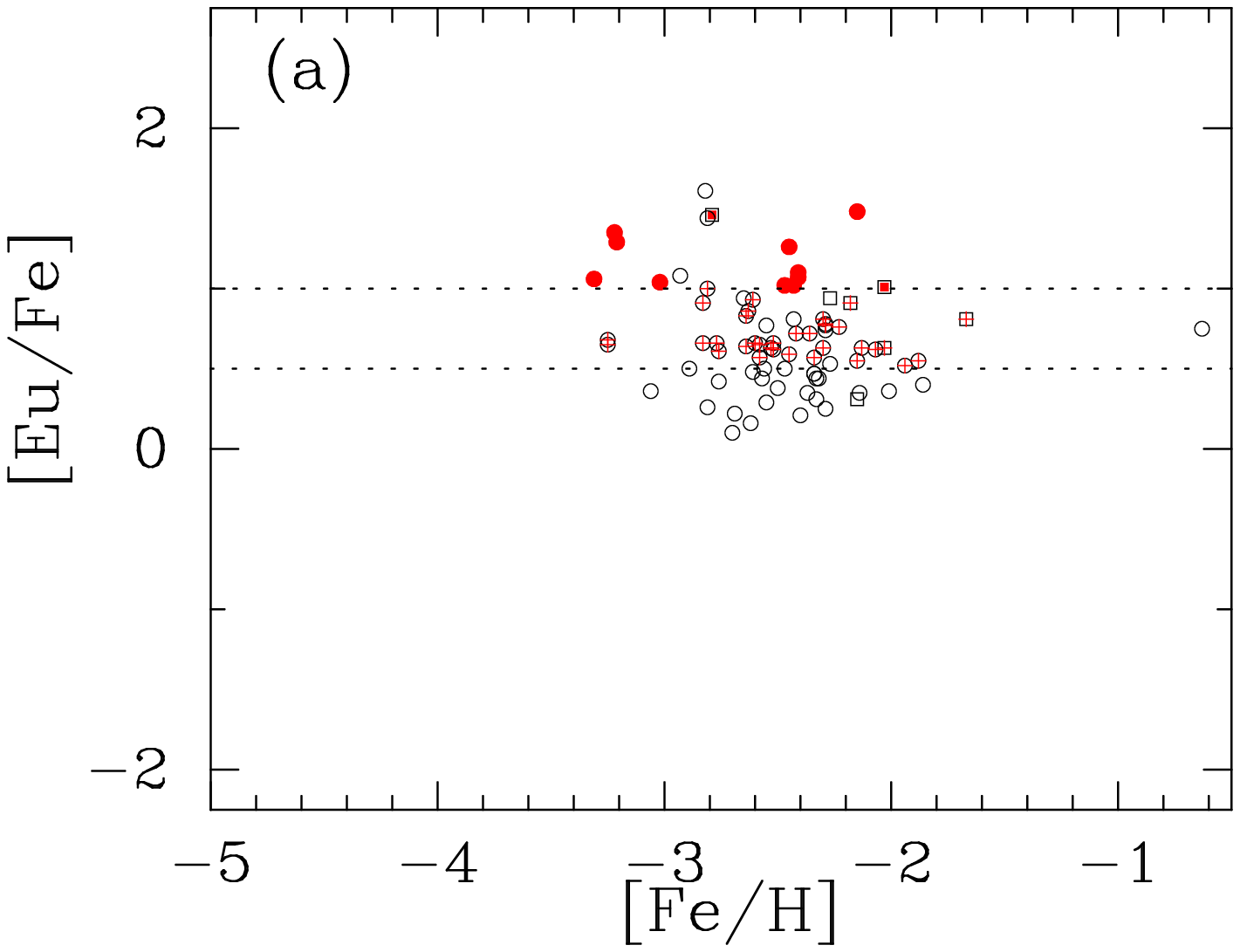}
\plotone{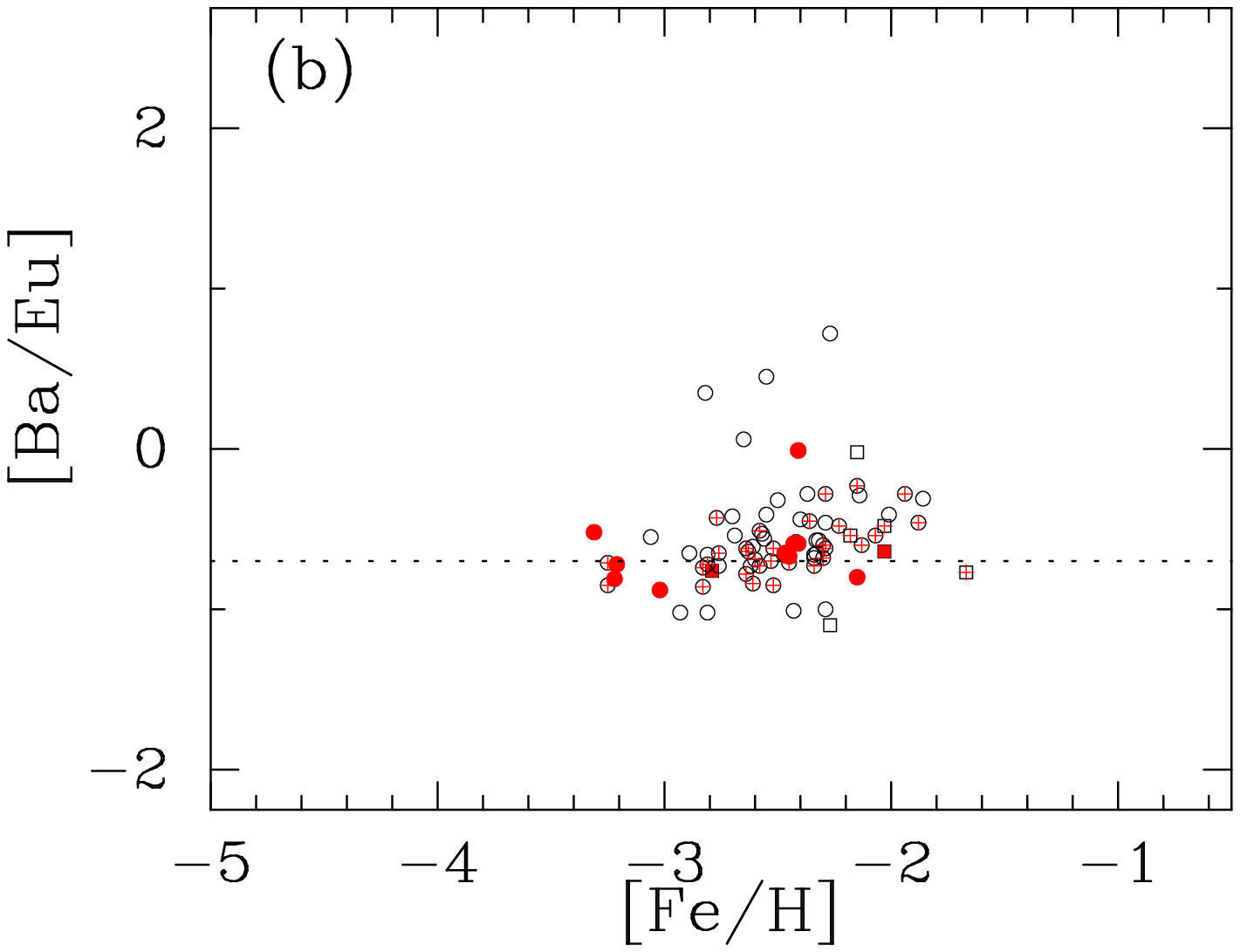}
\plotone{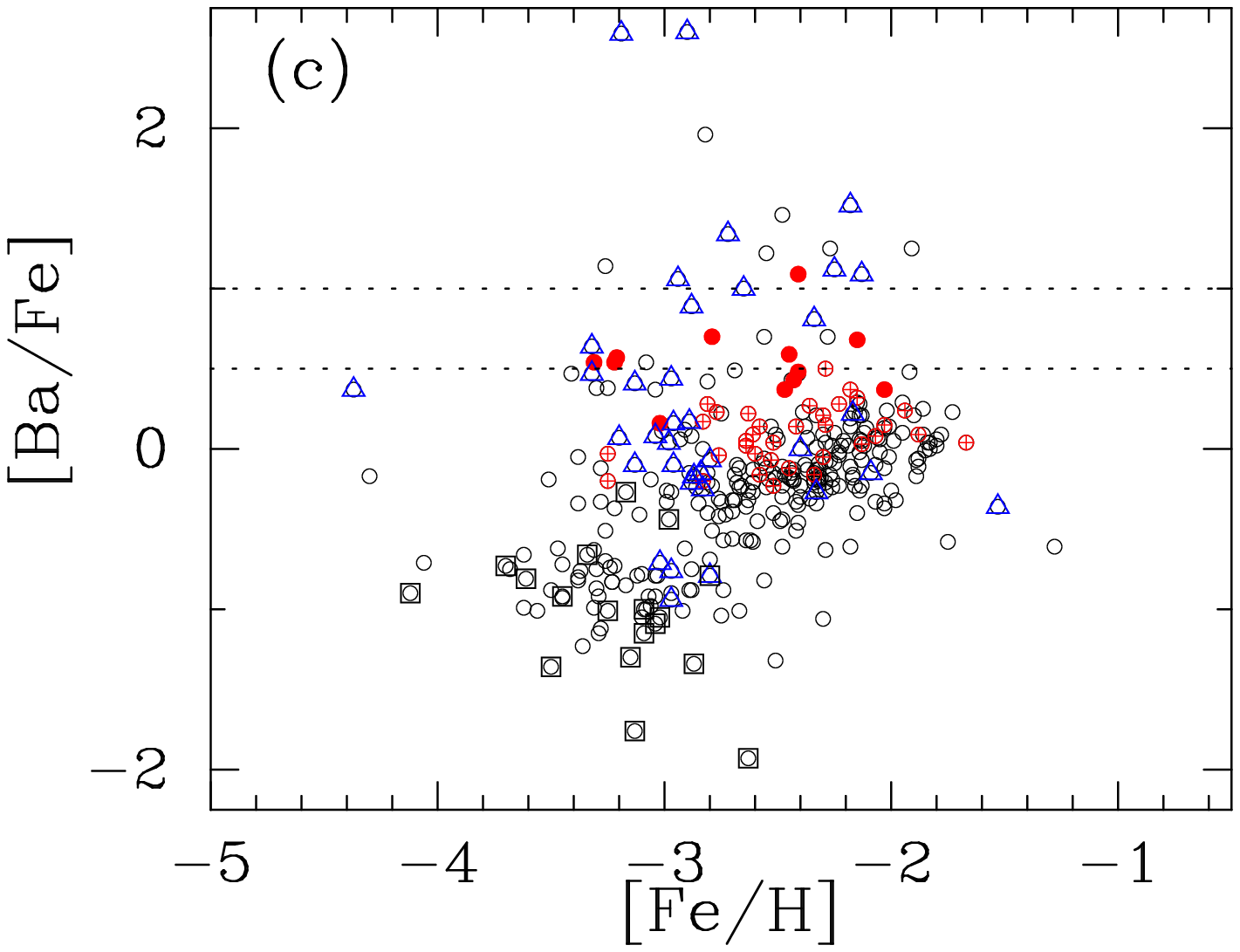}
\plotone{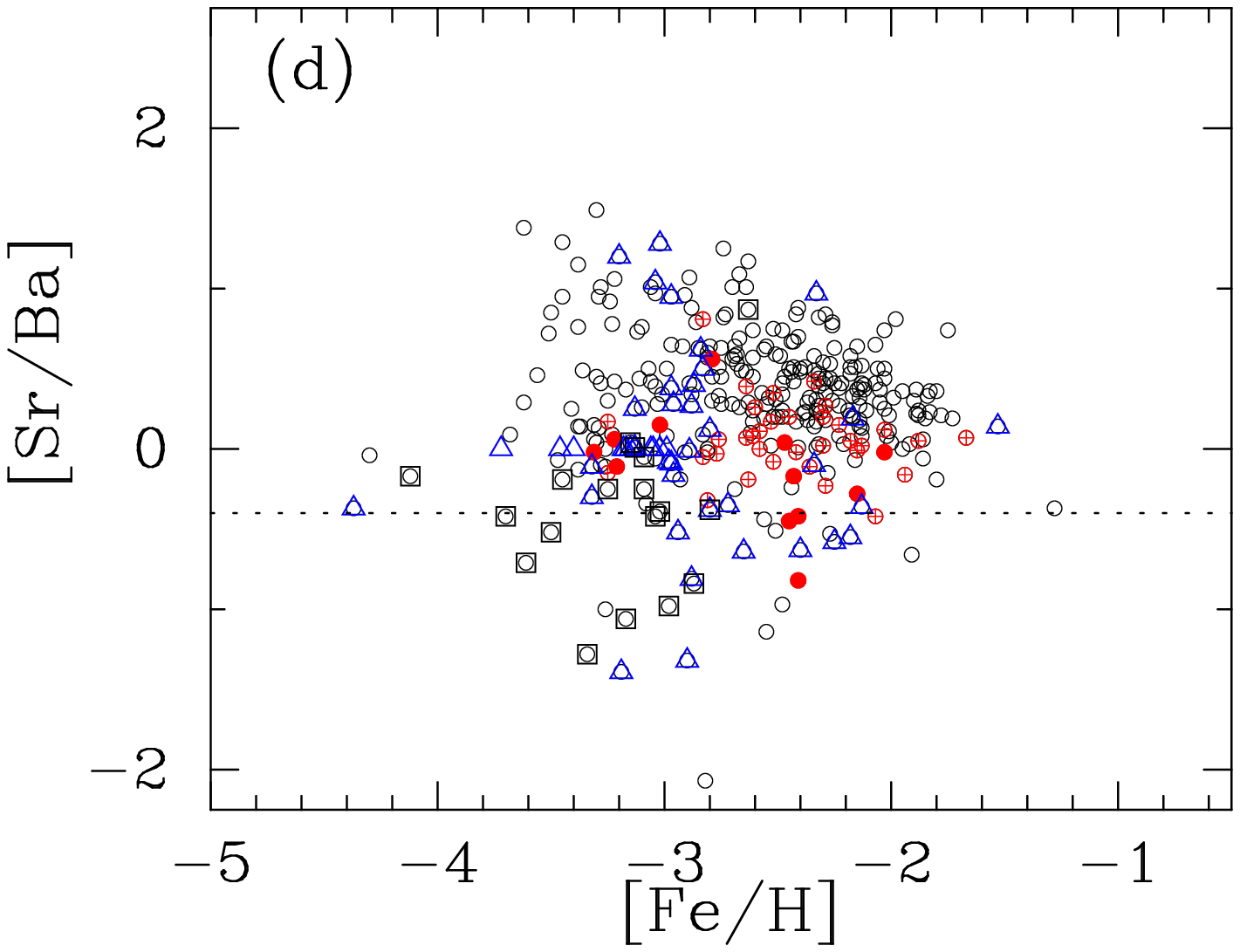}
\plotone{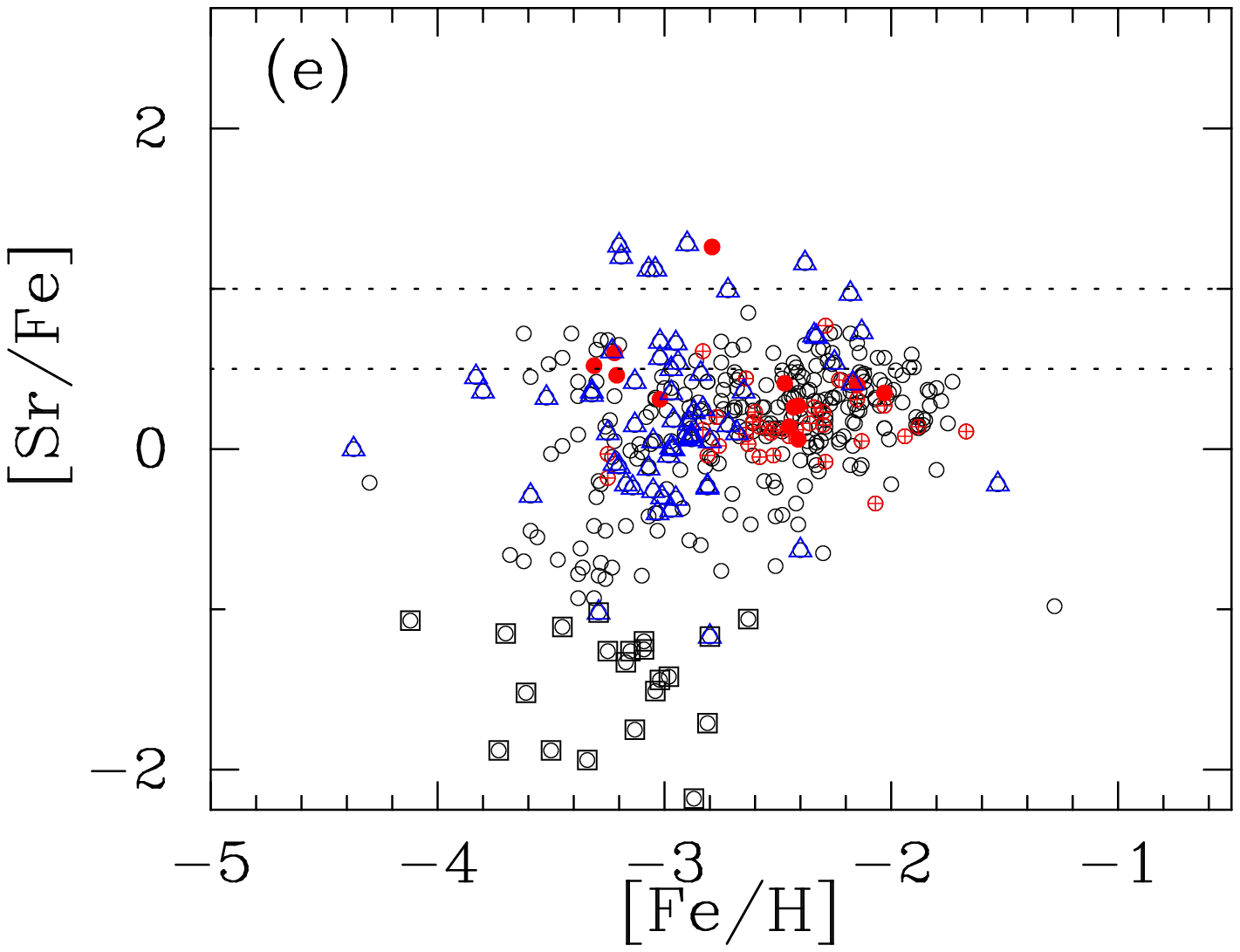}
\plotone{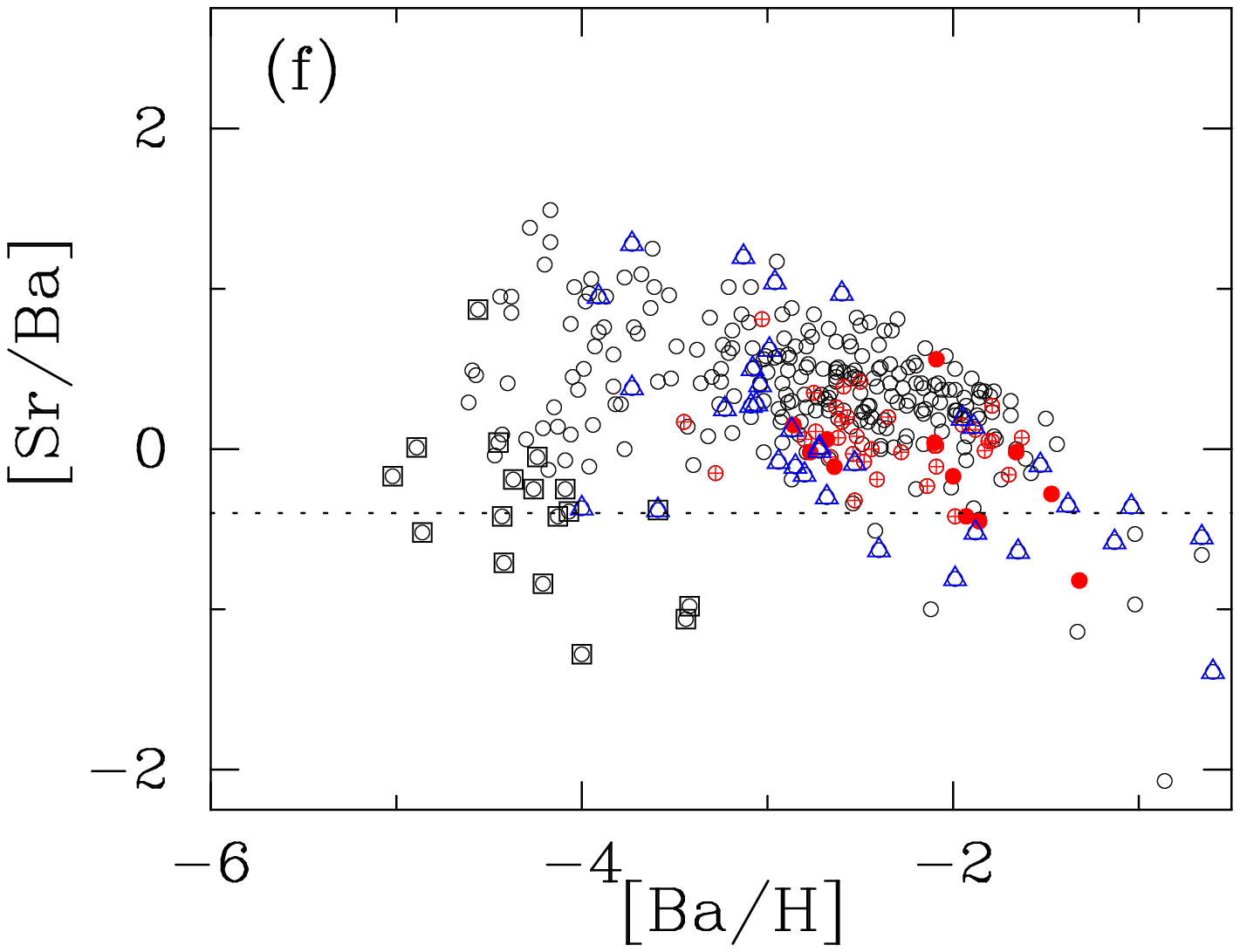}
\caption{Abundance ratios of neutron-capture elements.  
Left panels: Abundance ratio with respect to Fe as a function of [Fe/H] for Eu, Ba, and Sr. 
(a) Circles and squares mean turnoff and giant stars, respectively. 
r-II and r-I stars are shown by (red) filled symbols and (red) plus symbols, respectively. 
The dotted lines indicate the criteria of r-I and r-II stars. 
(c, e) r-I and r-II stars are shown by the same symbols as the top panel. 
CEMP stars and Ba-deficient stars ([Ba/Fe]$<-1$) are presented by overplotting large triangles and squares, respectively. 
Right panels: (b) [Ba/Eu] as a function of [Fe/H]. The symbols are the same as for the panel for [Eu/Fe]. 
The dotted line indicates the value expected for the r-process. (d and f) [Sr/Ba] as a function of [Fe/H] and [Ba/H], respectively.
The symbols are the same as for the panel for [Ba/Fe] and [Sr/Fe] in the left panels. 
The dotted line is the value of the r-process component in solar-system material. 
\label{fig:ncap}}
\end{figure*}

Ba and Sr are detected in most of the stars in our sample, i.e., 318 and 362 stars for Ba and Sr, respectively, of the 385 stars, 
thanks to the existence of strong resonance lines in the optical range.
Note that we include the remaining stars in the discussion on Sr and Ba deficient stars below. 
The Ba and Sr abundances as a function of [Fe/H] are shown in Figure~\ref{fig:ncap}(c). 
The objects with highest Ba abundances ([Ba/Fe]$>2$) are CEMP-s stars (triangles). 
The r-II stars discussed above (filled circles) have $0.5<$[Ba/Fe]$<1.0$ as expected from the high Eu abundance
([Eu/Fe]$>1$) and the r-process abundance ratio ([Ba/Eu]$\sim -0.7$). 
The r-I stars (plus symbols with open circles) have Ba abundances that
are almost indistinguishable from most of the other stars.

Excluding the CEMP-s and r-II stars, a majority of stars in our sample have [Ba/Fe]$\sim 0$ 
with some increasing trend with increasing metallicity as already pointed out in \S~\ref{subsec:abundance_trend}. 
There are, however, some fraction of stars that have very low Ba abundances ([Ba/Fe]$<-0.5$). 
These stars are discussed below. 
The Sr abundances also show a wide distribution like Ba (Figure~\ref{fig:ncap}(e)). 
Majority of stars have [Sr/Fe]=0.0--0.5 and show no clear trend for metallicity. 
The CEMP-s stars and r-II stars are not well separated from the bulk of the other stars in the [Sr/Fe] distribution.
This indicates that light neutron-capture elements are not enhanced as significantly as heavy neutron-capture elements by the r-process and the main s-process, 
and that another source of these light neutron-capture elements plays more important role at such very low metallicity.
There are also a small fraction of stars that have very low [Sr/Fe] as found for [Ba/Fe].

Figures~\ref{fig:ncap}(d and f) show [Sr/Ba] as functions of [Fe/H] and [Ba/H], respectively. 
The large scatter of [Sr/Ba] at lower metallicity is evident. 
The scatter is primarily caused by the dependence of [Sr/Ba] on [Ba/H] in [Ba/H]$>-4$ found in Figure~\ref{fig:ncap}(f), 
that reflects decreasing [Sr/Ba] with increasing Ba by the r-process. 
The large scatter of [Sr/Ba] found in objects with low Ba abundances ([Ba/H]$\lesssim -4$) is another factor. 

Most of the stars have [Sr/Ba] higher than the value of r-process component in solar-system material ([Sr/Ba]$=-0.4$). 
Objects with high abundances of heavy neutron-capture elements including Ba, 
which have [Ba/Eu] explained by the r-process (panel b), have relatively low [Sr/Ba] ratios close to the r-process value (panel f). 
This indicates that light neutron-capture elements including Sr in these stars are also originated from the r-process. 
On the other hand, many objects with lower Ba abundances ([Ba/H]$<-2$, see panel f) exhibit large scatter in [Sr/Ba] ratios. 
To explain objects with high [Sr/Ba], contributions of processes that efficiently produce light neutron-capture elements than r-process and main s-process (\S~\ref{subsubsec:heavy-elements}). 
We note that such processes could also have contributed to objects with stars with high [Ba/H] ratios, 
however, the larger contributions of the r-process would dominate in [Sr/Ba] ratios of these stars.
The remaining small number of stars with [Sr/Ba]$<-0.4$ need different explanations. A possible source is the main s-process at low metallicity, 
for which production of large amount of heavy neutron-capture elements like Ba is expected \citep{Aoki2013ApJ}. 
Indeed, about a half of objects with [Sr/Ba]$<-0.4$ are CEMP-s stars presented by triangles in the figure. 
Although no clear excess of C is found for the remaining stars, small contribution of the s-process is not excluded as a possible source of the low Sr/Ba ratios. 
Further studies for them based on better quality spectra will be useful to examine the possible contributions of the s-process. 

\begin{figure*}
\epsscale{0.5}
\plotone{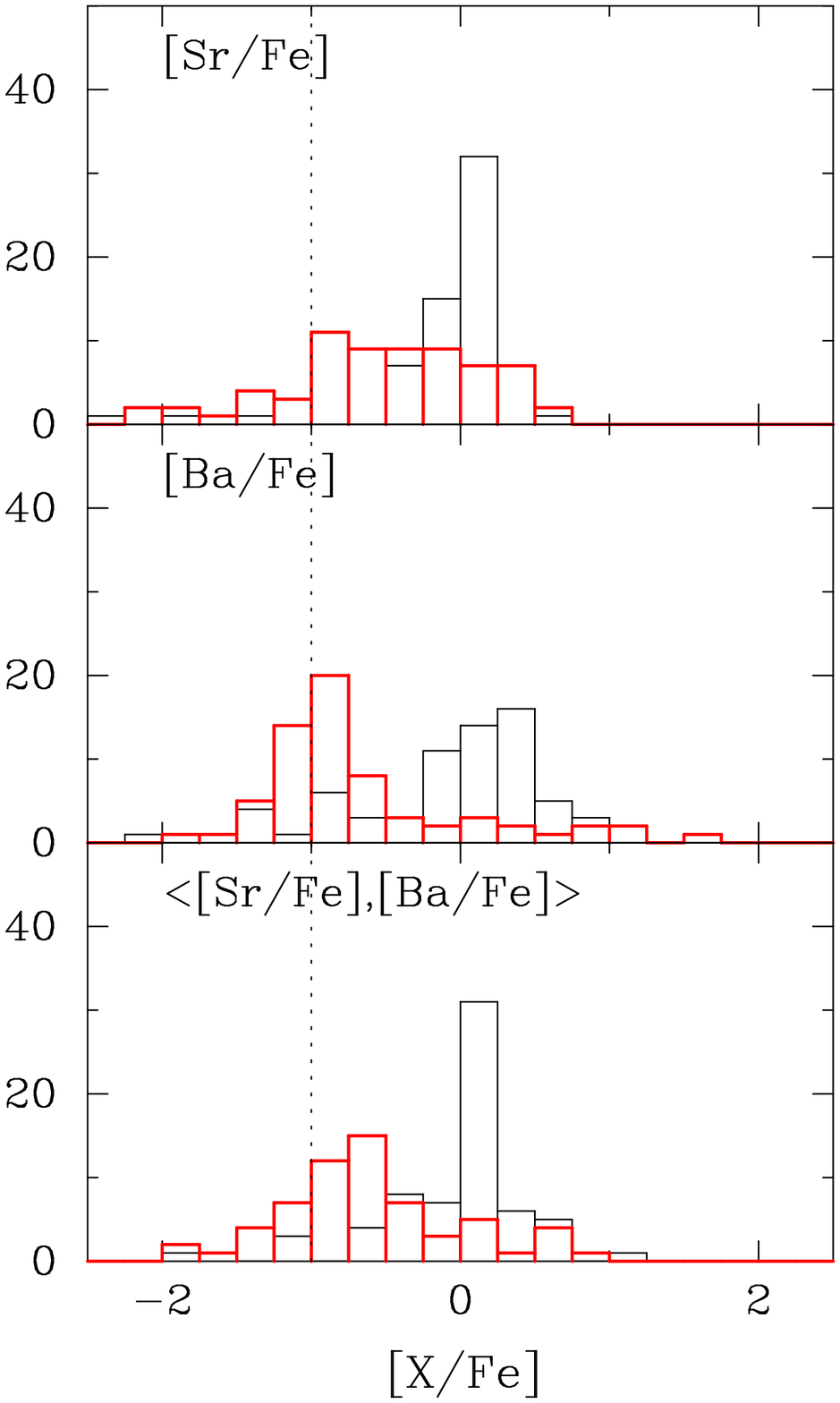}
\caption{Distributions of [Sr/Fe], [Ba/Fe] and their average ($<$[Sr/Fe], [Ba/Fe]$>$). 
Thin (black) and thick (red) lines mean those of stars with $-3.0<$[Fe/H]$<-2.5$ and [Fe/H]$<-3.0$, respectively. 
\label{fig:hist_srba}}
\end{figure*}

Figure~\ref{fig:hist_srba} shows histograms of the abundance ratios [Sr/Fe] and [Ba/Fe], as well as the average of the two values, 
of giants for the two metallicity ranges [Fe/H]$<-3.0$ and $-3.0<$[Fe/H]$<-2.5$. 
In the higher metallicity range, both [Sr/Fe] and [Ba/Fe] have peaks of the distributions at around 0.0. 
We note that the peak is much sharper in [Sr/Fe], and the distribution of [Ba/Fe] shows longer tail toward higher values. 
By contrast, the [Ba/Fe] distribution in [Fe/H]$<-3.0$ is significantly lower with a peak at around $-1.0$. 
Although majority of stars have [Sr/Fe]$>-1$ even in this metallicity range, the distribution shows a tail toward lower [Sr/Fe]. 
The average of [Sr/Fe] and [Ba/Fe] shows more smooth distributions having peaks at around 0.0 and $-0.7$ for $-3.0<$[Fe/H]$<-2.5$ and [Fe/H]$<-3.0$, respectively.

The difference of the distribution in [Fe/H]$<-3.0$ reflects the
difference of the origins of these two neutron-capture elements. 
Ba in the metallicity range $-3.0<$[Fe/H]$<-2.5$ would be mostly provided 
by the r-process \citep[e.g., ][]{Ishimaru2015ApJ} with possible small contribution of the s-process. 
Their contributions would be smaller to stars with lower metallicity (\FeHlt{-3.0}), reflecting relatively long timescale of the evolution of progenitors. 
Sr is also produced by the r-process, however, it should have another source in the early universe 
for which many models of the production process and site have been proposed \citep[e.g., ][ and references therein]{Cowan2021RvMP}, 
resulting in higher [Sr/Fe] ratios than [Ba/Fe] in \FeHlt{-3.0} as found in Figure~\ref{fig:hist_srba}. 

There are relatively small number of stars that have very low Ba
and/or Sr abundances. The numbers of stars having [Ba/Fe]$<-1.0$ and
[Sr/Fe]$<-1.0$ are 29 and 16, respectively. Among them 13 stars have
[Sr/Fe]$<-1.0$ and [Ba/Fe]$<-1.0$. All of them are red giants, and most of them are extremely metal-poor ([Fe/H]$\lesssim -3$). 
The frequency of these objects is, hence, higher than 10\% in this metallicity range. The low, but non-zero, 
abundances of neutron-capture elements in these stars suggest the existence of a kind of
neutron-capture reaction that operates as often as the process responsible for lighter elements \citep{Roederer2013AJ}, 
or small and ubiquitous pollution of r-process elements in the very early phase of chemical evolution. 

\section{Summary and concluding remarks}\label{sec:summary}

The joint project between LAMOST and Subaru has provided us with a homogeneous sample of 386 VMP stars with stellar parameters 
and abundances for more than 20 species that have been determined based on high-resolution spectra, 
and high-precision astrometry which is used to estimate the surface gravity for most objects. 
The sample covers a rather wide range in both metallicities and evolutionary status, 
and thus an unprecedentedly valuable sample to better constrain the Galactic chemical evolution models, as well as the production of elements in the early time.

Our abundance measurements mostly confirm the abundance trends and distributions found by previous studies, but provide some updates. 
We here summarize the results obtained by our abundance measurements. 

\begin{itemize}

\item The Li abundances of main-sequence turnoff stars (\tefft$>5500$~K) with [Fe/H]$>-2.5$ show a clear plateau with $A$(Li)=2.2, whereas those with lower metallicity have lower values ($A$(Li)$=2.0$ on average) showing possible decreasing trend with decreasing metallicity as found by previous studies. Red giants ( \tefft$<5500$~K) with $\log g \gtrsim 2$ have an almost constant value of $A$(Li)$=1.0$ regardless of metallicity. More evolved giants with $\log g \lesssim 1.5$ do not show Li absorption features, except for an anomalously Li-enhanced object, suggesting additional internal mixing in such highly evolved stars. Li is not detected in horizontal branch stars (\tefft$>5500$~K and \logg$<3.0$) in our sample.

\item The frequency of CEMP stars ([C/Fe]) estimated from the main-sequence turnoff stars is 20--30\%. The frequency of CEMP is much lower ($\sim$8\%) in giants. The frequency of CEMP-s stars is much lower in giants as expected from dilution of carbon-enhanced material provided by companion AGB stars by the first dredge-up. The frequency of CEMP-no in giants are also lower than that in turnoff stars, although the difference is not as significant as found for CEMP-s stars. The CEMP-no stars with $0.7<$[C/Fe]$<1.0$ are not well separated from other (carbon-normal) stars, suggesting a continuous distribution of [C/Fe] upto 1.0. This relatively small excess of carbon may disappear when they evolve to cool red giants by extra mixing.

\item Over-abundances of $\alpha$-elements (Mg, Si, Ca, and Ti) with respect to Fe in very metal-poor stars are confirmed. Among them, Mg (and Si) abundance ratios show a decreasing trend with increasing metallicity, which is not well reproduced by current Galactic chemical evolution models. Correlations of abundance ratios between Mg and Si and between Ca and Ti are found while the correlation between the two groups (Mg-Si and Ca-Ti) is weaker, suggesting the enrichment processes are different between them to a certain extent. We have identified 12 low-Mg stars ([Mg/Fe]$<0$) in a wide range of metallicity ($-3.8<$[Fe/H]$<-1.7$). The over-abundances of Ti and Sc are not reproduced at all by current chemical evolution models, and some essential revision of nucleosynthesis modeling would be required to solve the discrepancy.

\item The abundance ratios of iron-peak elements are approximately solar values, taking the possible NLTE effects into account for the Cr abundances derived for giants. An exception is Zn, which shows increasing trend with decreasing metallicity in [Fe/H]$<-2.5$. The high Zn abundance ratios, as well as the small over-abundances of Co, are not well reproduced by chemical evolution models. 

\item The frequency of r-process-enhanced stars (r-II stars) estimated from giants is about 5\%, which agrees with previous estimates. The r-II stars with the most significant excess of Eu are found at [Fe/H]$\sim -3$, but r-II stars are also identified at higher metallicity ([Fe/H]$>-2.5$). Sr and Ba abundances show large dispersion, but their distribution is significantly different in [Fe/H]$<-3$, suggesting the existence of extra sources to produce Sr at such low metallicity. 

\end{itemize}

This work provides the largest uniform sample of VMP stars with detailed chemical abundances, 
which should be a valuable data set for studies on the formation and evolution history of our Galaxy.
The full data file of parameters and abundances presented in Table~\ref{tab:stellar_param} and ~\ref{tab:abun_error} 
will be available online.
Combined with analysis of kinematic properties of this sample, which will be reported in the next paper of this series, 
we will be able to investigate the nature of these old stars and halo components of our Milky Way from a chemo-dynamical view, 
e.g., the relation between the accreted halo components and the decreasing trend with increasing metallicity for Mg (and Si) abundances. 

\acknowledgments

We thank the anonymous referee for the careful reading of the manuscript and useful comments.
This research is based on data collected at Subaru Telescope, which is operated by the National Astronomical Observatory of Japan. 
We are honored and grateful for the opportunity of observing the Universe from Maunakea, which has the cultural, historical and natural significance in Hawaii. 
Guoshoujing Telescope (the Large Sky Area Multi-Object Fiber Spectroscopic Telescope, LAMOST) 
is a National Major Scientific Project built by the Chinese Academy of Sciences. 
Funding for the project has been provided by the National Development and Reform Commission. 
It is operated and managed by the National Astronomical Observatories, Chinese Academy of Sciences.
This work was supported by the National Key R\&D Program of China No.2019YFA0405502, the JSPS - CAS Joint Research Program, 
the National Natural Science Foundation of China grant Nos. 11988101, 11973049, and 11625313, 
and the science research grants from the China Manned Space Project with No. CMSCSST-2021-B05. 
H.L. and Q.X. acknowledge support from the Strategic Priority Research Program of Chinese Academy of Sciences, Grant No. XDB34020205 
and the Youth Innovation Promotion Association of the CAS (id. Y202017 and 2020058). 
This work was also supported by a Grant-in-Aid for Scientific Research (KAKENHI) 
(JP16H02168, JP17K14249, JP19HP8019, JP20H05855, and JP20HP8012) from the Japan Society for the Promotion of Science.

%

\vspace{5mm}
\facilities{LAMOST, Subaru}





\bibliographystyle{apj} 
\bibliography{lamosubaru_paperII} 

\begin{thebibliography}{210}
\expandafter\ifx\csname natexlab\endcsname\relax\def\natexlab#1{#1}\fi

\bibitem[{{Aguado} {et~al.}(2017){Aguado}, {Gonz{\'a}lez Hern{\'a}ndez},
  {Allende Prieto}, \& {Rebolo}}]{Aguado2017AA}
{Aguado}, D.~S., {Gonz{\'a}lez Hern{\'a}ndez}, J.~I., {Allende Prieto}, C., \&
  {Rebolo}, R. 2017, \aap, 605, A40

\bibitem[{{Aguado} {et~al.}(2019){Aguado}, {Gonz{\'a}lez Hern{\'a}ndez},
  {Allende Prieto}, \& {Rebolo}}]{Aguado2019ApJ}
---. 2019, \apjl, 874, L21

\bibitem[{{Aguado} {et~al.}(2021){Aguado}, {Myeong}, {Belokurov}, {Evans},
  {Koposov}, {Allende Prieto}, {Lanfranchi}, {Matteucci}, {Shetrone},
  {Sbordone}, {Navarrete}, {Gonz{\'a}lez Hern{\'a}ndez}, {Chanam{\'e}},
  {Peralta de Arriba}, \& {Yuan}}]{Aguado2021MNRAS}
{Aguado}, D.~S., {et~al.} 2021, \mnras, 500, 889

\bibitem[{{Alonso} {et~al.}(1996){Alonso}, {Arribas}, \&
  {Martinez-Roger}}]{Alonso1996AAS}
{Alonso}, A., {Arribas}, S., \& {Martinez-Roger}, C. 1996, \aaps, 117, 227

\bibitem[{{Alonso} {et~al.}(1999){Alonso}, {Arribas}, \&
  {Mart{\'\i}nez-Roger}}]{Alonso1999AAS}
{Alonso}, A., {Arribas}, S., \& {Mart{\'\i}nez-Roger}, C. 1999, \aaps, 140, 261

\bibitem[{{Amarsi} {et~al.}(2016){Amarsi}, {Lind}, {Asplund}, {Barklem}, \&
  {Collet}}]{Amarsi2016MNRAS}
{Amarsi}, A.~M., {Lind}, K., {Asplund}, M., {Barklem}, P.~S., \& {Collet}, R.
  2016, \mnras, 463, 1518

\bibitem[{{Andrievsky} {et~al.}(2007){Andrievsky}, {Spite}, {Korotin}, {Spite},
  {Bonifacio}, {Cayrel}, {Hill}, \& {Fran{\c{c}}ois}}]{Andrievsky2007A&A}
{Andrievsky}, S.~M., {Spite}, M., {Korotin}, S.~A., {Spite}, F., {Bonifacio},
  P., {Cayrel}, R., {Hill}, V., \& {Fran{\c{c}}ois}, P. 2007, \aap, 464, 1081

\bibitem[{{Aoki} {et~al.}(2009{\natexlab{a}}){Aoki}, {Barklem}, {Beers},
  {Christlieb}, {Inoue}, {Garc{\'{\i}}a P{\'e}rez}, {Norris}, \&
  {Carollo}}]{Aoki2009ApJ}
{Aoki}, W., {Barklem}, P.~S., {Beers}, T.~C., {Christlieb}, N., {Inoue}, S.,
  {Garc{\'{\i}}a P{\'e}rez}, A.~E., {Norris}, J.~E., \& {Carollo}, D.
  2009{\natexlab{a}}, \apj, 698, 1803

\bibitem[{{Aoki} {et~al.}(2007){Aoki}, {Beers}, {Christlieb}, {Norris}, {Ryan},
  \& {Tsangarides}}]{Aoki2007ApJ}
{Aoki}, W., {Beers}, T.~C., {Christlieb}, N., {Norris}, J.~E., {Ryan}, S.~G.,
  \& {Tsangarides}, S. 2007, \apj, 655, 492

\bibitem[{{Aoki} {et~al.}(2018){Aoki}, {Matsuno}, {Honda}, {Ishigaki}, {Li},
  {Suda}, \& {Kumar}}]{Aoki2018PASJ}
{Aoki}, W., {Matsuno}, T., {Honda}, S., {Ishigaki}, M.~N., {Li}, H., {Suda},
  T., \& {Kumar}, Y.~B. 2018, \pasj, 70, 94

\bibitem[{{Aoki} {et~al.}(2002){Aoki}, {Norris}, {Ryan}, {Beers}, \&
  {Ando}}]{Aoki2002ApJ}
{Aoki}, W., {Norris}, J.~E., {Ryan}, S.~G., {Beers}, T.~C., \& {Ando}, H. 2002,
  \pasj, 54, 933

\bibitem[{{Aoki} {et~al.}(2013{\natexlab{a}}){Aoki}, {Suda}, {Boyd}, {Kajino},
  \& {Famiano}}]{Aoki2013ApJ}
{Aoki}, W., {Suda}, T., {Boyd}, R.~N., {Kajino}, T., \& {Famiano}, M.~A.
  2013{\natexlab{a}}, \apjl, 766, L13

\bibitem[{{Aoki} {et~al.}(2014){Aoki}, {Tominaga}, {Beers}, {Honda}, \&
  {Lee}}]{Aoki2014Science}
{Aoki}, W., {Tominaga}, N., {Beers}, T.~C., {Honda}, S., \& {Lee}, Y.~S. 2014,
  Science, 345, 912

\bibitem[{{Aoki} {et~al.}(2009{\natexlab{b}}){Aoki}, {Arimoto}, {Sadakane},
  {Tolstoy}, {Battaglia}, {Jablonka}, {Shetrone}, {Letarte}, {Irwin}, {Hill},
  {Francois}, {Venn}, {Primas}, {Helmi}, {Kaufer}, {Tafelmeyer}, {Szeifert}, \&
  {Babusiaux}}]{Aoki2009AA}
{Aoki}, W., {et~al.} 2009{\natexlab{b}}, \aap, 502, 569

\bibitem[{{Aoki} {et~al.}(2013{\natexlab{b}}){Aoki}, {Beers}, {Lee}, {Honda},
  {Ito}, {Takada-Hidai}, {Frebel}, {Suda}, {Fujimoto}, {Carollo}, \&
  {Sivarani}}]{Aoki2013AJ}
---. 2013{\natexlab{b}}, \aj, 145, 13

\bibitem[{{Asplund} {et~al.}(2009){Asplund}, {Grevesse}, {Sauval}, \&
  {Scott}}]{Asplund2009ARAA}
{Asplund}, M., {Grevesse}, N., {Sauval}, A.~J., \& {Scott}, P. 2009, \araa, 47,
  481

\bibitem[{{Asplund} {et~al.}(2006){Asplund}, {Lambert}, {Nissen}, {Primas}, \&
  {Smith}}]{Asplund2006ApJ}
{Asplund}, M., {Lambert}, D.~L., {Nissen}, P.~E., {Primas}, F., \& {Smith},
  V.~V. 2006, \apj, 644, 229

\bibitem[{{Barklem} {et~al.}(2005){Barklem}, {Christlieb}, {Beers}, {Hill},
  {Bessell}, {Holmberg}, {Marsteller}, {Rossi}, {Zickgraf}, \&
  {Reimers}}]{Barklem2005AA}
{Barklem}, P.~S., {et~al.} 2005, \aap, 439, 129

\bibitem[{{Beers} \& {Christlieb}(2005)}]{Beers&Christlieb2005ARAA}
{Beers}, T.~C., \& {Christlieb}, N. 2005, \araa, 43, 531

\bibitem[{{Beers} {et~al.}(1985){Beers}, {Preston}, \&
  {Shectman}}]{Beers1985AJ}
{Beers}, T.~C., {Preston}, G.~W., \& {Shectman}, S.~A. 1985, \aj, 90, 2089

\bibitem[{{Beers} {et~al.}(1992){Beers}, {Preston}, \&
  {Shectman}}]{Beers1992AJ}
---. 1992, \aj, 103, 1987

\bibitem[{{Bergemann} \& {Cescutti}(2010)}]{Bergeman2010AA}
{Bergemann}, M., \& {Cescutti}, G. 2010, \aap, 522, A9

\bibitem[{{Bonifacio} {et~al.}(2009){Bonifacio}, {Spite}, {Cayrel}, {Hill},
  {Spite}, {Fran{\c c}ois}, {Plez}, {Ludwig}, {Caffau}, {Molaro}, {Depagne},
  {Andersen}, {Barbuy}, {Beers}, {Nordstr{\"o}m}, \&
  {Primas}}]{Bonifacio2009AA}
{Bonifacio}, P., {et~al.} 2009, \aap, 501, 519

\bibitem[{{Bonifacio} {et~al.}(2018){Bonifacio}, {Caffau}, {Spite}, {Spite},
  {Sbordone}, {Monaco}, {Fran{\c{c}}ois}, {Plez}, {Molaro}, {Gallagher},
  {Cayrel}, {Christlieb}, {Klessen}, {Koch}, {Ludwig}, {Steffen}, {Zaggia}, \&
  {Abate}}]{Bonifacio2018AA}
---. 2018, \aap, 612, A65

\bibitem[{{Borghs} {et~al.}(1983){Borghs}, {de Bisschop}, {van Hove}, \&
  {Silverans}}]{Borghs1983HyInt}
{Borghs}, G., {de Bisschop}, P., {van Hove}, M., \& {Silverans}, R.~E. 1983,
  Hyperfine Interactions, 15, 177

\bibitem[{{Caffau} {et~al.}(2011){Caffau}, {Bonifacio}, {Fran{\c c}ois},
  {Sbordone}, {Monaco}, {Spite}, {Spite}, {Ludwig}, {Cayrel}, {Zaggia},
  {Hammer}, {Randich}, {Molaro}, \& {Hill}}]{Caffau2011Nature}
{Caffau}, E., {et~al.} 2011, \nat, 477, 67

\bibitem[{{Caffau} {et~al.}(2013){Caffau}, {Bonifacio}, {Fran{\c{c}}ois},
  {Sbordone}, {Spite}, {Monaco}, {Plez}, {Spite}, {Zaggia}, {Ludwig}, {Cayrel},
  {Molaro}, {Randich}, {Hammer}, \& {Hill}}]{Caffau2013AA}
---. 2013, \aap, 560, A15

\bibitem[{{Castelli} \& {Kurucz}(2003)}]{Castelli&Kurucz2003IAUS}
{Castelli}, F., \& {Kurucz}, R.~L. 2003, in IAU Symposium, Vol. 210, Modelling
  of Stellar Atmospheres, ed. N.~{Piskunov}, W.~W. {Weiss}, \& D.~F. {Gray},
  A20

\bibitem[{{Cayrel} {et~al.}(2004){Cayrel}, {Depagne}, {Spite}, {Hill}, {Spite},
  {Fran{\c c}ois}, {Plez}, {Beers}, {Primas}, {Andersen}, {Barbuy},
  {Bonifacio}, {Molaro}, \& {Nordstr{\"o}m}}]{Cayrel2004AA}
{Cayrel}, R., {et~al.} 2004, \aap, 416, 1117

\bibitem[{{Charbonnel} \& {Primas}(2005)}]{Charbonnel&Primas2005AA}
{Charbonnel}, C., \& {Primas}, F. 2005, \aap, 442, 961

\bibitem[{{Chiaki} {et~al.}(2017){Chiaki}, {Tominaga}, \&
  {Nozawa}}]{Chiaki2017MNRAS}
{Chiaki}, G., {Tominaga}, N., \& {Nozawa}, T. 2017, \mnras, 472, L115

\bibitem[{{Chieffi} {et~al.}(1998){Chieffi}, {Limongi}, \&
  {Straniero}}]{Chieffi1998ApJ}
{Chieffi}, A., {Limongi}, M., \& {Straniero}, O. 1998, \apj, 502, 737

\bibitem[{{Choplin} {et~al.}(2018){Choplin}, {Hirschi}, {Meynet},
  {Ekstr{\"o}m}, {Chiappini}, \& {Laird}}]{Choplin2018AA}
{Choplin}, A., {Hirschi}, R., {Meynet}, G., {Ekstr{\"o}m}, S., {Chiappini}, C.,
  \& {Laird}, A. 2018, \aap, 618, A133

\bibitem[{{Christlieb} {et~al.}(2004){Christlieb}, {Gustafsson}, {Korn},
  {Barklem}, {Beers}, {Bessell}, {Karlsson}, \&
  {Mizuno-Wiedner}}]{Christlieb2004ApJ}
{Christlieb}, N., {Gustafsson}, B., {Korn}, A.~J., {Barklem}, P.~S., {Beers},
  T.~C., {Bessell}, M.~S., {Karlsson}, T., \& {Mizuno-Wiedner}, M. 2004, \apj,
  603, 708

\bibitem[{{Christlieb} {et~al.}(2008){Christlieb}, {Sch{\"o}rck}, {Frebel},
  {Beers}, {Wisotzki}, \& {Reimers}}]{Christlieb2008AA}
{Christlieb}, N., {Sch{\"o}rck}, T., {Frebel}, A., {Beers}, T.~C., {Wisotzki},
  L., \& {Reimers}, D. 2008, \aap, 484, 721

\bibitem[{{Christlieb} {et~al.}(2002){Christlieb}, {Bessell}, {Beers},
  {Gustafsson}, {Korn}, {Barklem}, {Karlsson}, {Mizuno-Wiedner}, \&
  {Rossi}}]{Christlieb2002Nature}
{Christlieb}, N., {et~al.} 2002, \nat, 419, 904

\bibitem[{{Cohen} {et~al.}(2013){Cohen}, {Christlieb}, {Thompson}, {McWilliam},
  {Shectman}, {Reimers}, {Wisotzki}, \& {Kirby}}]{Cohen2013ApJ}
{Cohen}, J.~G., {Christlieb}, N., {Thompson}, I., {McWilliam}, A., {Shectman},
  S., {Reimers}, D., {Wisotzki}, L., \& {Kirby}, E. 2013, \apj, 778, 56

\bibitem[{{Collet} {et~al.}(2006){Collet}, {Asplund}, \&
  {Trampedach}}]{Collet2006ApJ}
{Collet}, R., {Asplund}, M., \& {Trampedach}, R. 2006, \apjl, 644, L121

\bibitem[{{Cowan} {et~al.}(2021){Cowan}, {Sneden}, {Lawler}, {Aprahamian},
  {Wiescher}, {Langanke}, {Mart{\'\i}nez-Pinedo}, \&
  {Thielemann}}]{Cowan2021RvMP}
{Cowan}, J.~J., {Sneden}, C., {Lawler}, J.~E., {Aprahamian}, A., {Wiescher},
  M., {Langanke}, K., {Mart{\'\i}nez-Pinedo}, G., \& {Thielemann}, F.-K. 2021,
  Reviews of Modern Physics, 93, 015002

\bibitem[{{Cristallo} {et~al.}(2015){Cristallo}, {Straniero}, {Piersanti}, \&
  {Gobrecht}}]{Cristallo2015ApJS}
{Cristallo}, S., {Straniero}, O., {Piersanti}, L., \& {Gobrecht}, D. 2015,
  \apjs, 219, 40

\bibitem[{{Cyburt} {et~al.}(2016){Cyburt}, {Fields}, {Olive}, \&
  {Yeh}}]{Cyburt2016a}
{Cyburt}, R.~H., {Fields}, B.~D., {Olive}, K.~A., \& {Yeh}, T.-H. 2016, Reviews
  of Modern Physics, 88, 015004

\bibitem[{{De Silva} {et~al.}(2015){De Silva}, {Freeman}, {Bland-Hawthorn},
  {Martell}, {de Boer}, {Asplund}, {Keller}, {Sharma}, {Zucker}, {Zwitter},
  {Anguiano}, {Bacigalupo}, {Bayliss}, {Beavis}, {Bergemann}, {Campbell},
  {Cannon}, {Carollo}, {Casagrande}, {Casey}, {Da Costa}, {D'Orazi}, {Dotter},
  {Duong}, {Heger}, {Ireland}, {Kafle}, {Kos}, {Lattanzio}, {Lewis}, {Lin},
  {Lind}, {Munari}, {Nataf}, {O'Toole}, {Parker}, {Reid}, {Schlesinger},
  {Sheinis}, {Simpson}, {Stello}, {Ting}, {Traven}, {Watson}, {Wittenmyer},
  {Yong}, \& {{\v{Z}}erjal}}]{DeSilva2015MNRAS}
{De Silva}, G.~M., {et~al.} 2015, \mnras, 449, 2604

\bibitem[{{Ekstr{\"o}m} {et~al.}(2008){Ekstr{\"o}m}, {Meynet}, {Chiappini},
  {Hirschi}, \& {Maeder}}]{Ekstrom2008AA}
{Ekstr{\"o}m}, S., {Meynet}, G., {Chiappini}, C., {Hirschi}, R., \& {Maeder},
  A. 2008, \aap, 489, 685

\bibitem[{{Ezzeddine} {et~al.}(2019){Ezzeddine}, {Frebel}, {Roederer},
  {Tominaga}, {Tumlinson}, {Ishigaki}, {Nomoto}, {Placco}, \&
  {Aoki}}]{Ezzeddine2019ApJ}
{Ezzeddine}, R., {et~al.} 2019, \apj, 876, 97

\bibitem[{{Ezzeddine} {et~al.}(2020){Ezzeddine}, {Rasmussen}, {Frebel},
  {Chiti}, {Hinojisa}, {Placco}, {Ji}, {Beers}, {Hansen}, {Roederer}, {Sakari},
  \& {Melendez}}]{Ezzeddine2020ApJ}
---. 2020, \apj, 898, 150

\bibitem[{{Fields} {et~al.}(2020){Fields}, {Olive}, {Yeh}, \&
  {Young}}]{Fields2019a}
{Fields}, B.~D., {Olive}, K.~A., {Yeh}, T.-H., \& {Young}, C. 2020, \jcap,
  2020, 010

\bibitem[{{Fishlock} {et~al.}(2014){Fishlock}, {Karakas}, {Lugaro}, \&
  {Yong}}]{Fishlock2014ApJ}
{Fishlock}, C.~K., {Karakas}, A.~I., {Lugaro}, M., \& {Yong}, D. 2014, \apj,
  797, 44

\bibitem[{{Fran{\c c}ois} {et~al.}(2007){Fran{\c c}ois}, {Depagne}, {Hill},
  {Spite}, {Spite}, {Plez}, {Beers}, {Andersen}, {James}, {Barbuy}, {Cayrel},
  {Bonifacio}, {Molaro}, {Nordstr{\"o}m}, \& {Primas}}]{Francois2007AA}
{Fran{\c c}ois}, P., {et~al.} 2007, \aap, 476, 935

\bibitem[{{Frebel} {et~al.}(2013){Frebel}, {Casey}, {Jacobson}, \&
  {Yu}}]{Frebel2013ApJ}
{Frebel}, A., {Casey}, A.~R., {Jacobson}, H.~R., \& {Yu}, Q. 2013, \apj, 769,
  57

\bibitem[{{Frebel} {et~al.}(2008){Frebel}, {Collet}, {Eriksson}, {Christlieb},
  \& {Aoki}}]{Frebel2008ApJ}
{Frebel}, A., {Collet}, R., {Eriksson}, K., {Christlieb}, N., \& {Aoki}, W.
  2008, \apj, 684, 588

\bibitem[{{Frebel} \& {Norris}(2015)}]{Frebel&Norris2015ARAA}
{Frebel}, A., \& {Norris}, J.~E. 2015, \araa, 53, 631

\bibitem[{{Frebel} {et~al.}(2005){Frebel}, {Aoki}, {Christlieb}, {Ando},
  {Asplund}, {Barklem}, {Beers}, {Eriksson}, {Fechner}, {Fujimoto}, {Honda},
  {Kajino}, {Minezaki}, {Nomoto}, {Norris}, {Ryan}, {Takada-Hidai},
  {Tsangarides}, \& {Yoshii}}]{Frebel2005Nature}
{Frebel}, A., {et~al.} 2005, \nat, 434, 871

\bibitem[{{Fu} {et~al.}(2015){Fu}, {Bressan}, {Molaro}, \& {Marigo}}]{Fu2015a}
{Fu}, X., {Bressan}, A., {Molaro}, P., \& {Marigo}, P. 2015, \mnras, 452, 3256

\bibitem[{{Gaia Collaboration} {et~al.}(2016){Gaia Collaboration}, {Prusti},
  {de Bruijne}, {Brown}, {Vallenari}, {Babusiaux}, {Bailer-Jones}, {Bastian},
  {Biermann}, {Evans}, {Eyer}, {Jansen}, {Jordi}, {Klioner}, {Lammers},
  {Lindegren}, {Luri}, {Mignard}, {Milligan}, {Panem}, {Poinsignon},
  {Pourbaix}, {Randich}, {Sarri}, {Sartoretti}, {Siddiqui}, {Soubiran},
  {Valette}, {van Leeuwen}, {Walton}, {Aerts}, {Arenou}, {Cropper}, {Drimmel},
  {H{\o}g}, {Katz}, {Lattanzi}, {O'Mullane}, {Grebel}, {Holland}, {Huc},
  {Passot}, {Bramante}, {Cacciari}, {Casta{\~n}eda}, {Chaoul}, {Cheek}, {De
  Angeli}, {Fabricius}, {Guerra}, {Hern{\'a}ndez}, {Jean-Antoine-Piccolo},
  {Masana}, {Messineo}, {Mowlavi}, {Nienartowicz}, {Ord{\'o}{\~n}ez-Blanco},
  {Panuzzo}, {Portell}, {Richards}, {Riello}, {Seabroke}, {Tanga},
  {Th{\'e}venin}, {Torra}, {Els}, {Gracia-Abril}, {Comoretto},
  {Garcia-Reinaldos}, {Lock}, {Mercier}, {Altmann}, {Andrae}, {Astraatmadja},
  {Bellas-Velidis}, {Benson}, {Berthier}, {Blomme}, {Busso}, {Carry},
  {Cellino}, {Clementini}, {Cowell}, {Creevey}, {Cuypers}, {Davidson}, {De
  Ridder}, {de Torres}, {Delchambre}, {Dell'Oro}, {Ducourant}, {Fr{\'e}mat},
  {Garc{\'\i}a-Torres}, {Gosset}, {Halbwachs}, {Hambly}, {Harrison}, {Hauser},
  {Hestroffer}, {Hodgkin}, {Huckle}, {Hutton}, {Jasniewicz}, {Jordan},
  {Kontizas}, {Korn}, {Lanzafame}, {Manteiga}, {Moitinho}, {Muinonen},
  {Osinde}, {Pancino}, {Pauwels}, {Petit}, {Recio-Blanco}, {Robin}, {Sarro},
  {Siopis}, {Smith}, {Smith}, {Sozzetti}, {Thuillot}, {van Reeven}, {Viala},
  {Abbas}, {Abreu Aramburu}, {Accart}, {Aguado}, {Allan}, {Allasia},
  {Altavilla}, {{\'A}lvarez}, {Alves}, {Anderson}, {Andrei}, {Anglada Varela},
  {Antiche}, {Antoja}, {Ant{\'o}n}, {Arcay}, {Atzei}, {Ayache}, {Bach},
  {Baker}, {Balaguer-N{\'u}{\~n}ez}, {Barache}, {Barata}, {Barbier}, {Barblan},
  {Baroni}, {Barrado y Navascu{\'e}s}, {Barros}, {Barstow}, {Becciani},
  {Bellazzini}, {Bellei}, {Bello Garc{\'\i}a}, {Belokurov}, {Bendjoya},
  {Berihuete}, {Bianchi}, {Bienaym{\'e}}, {Billebaud}, {Blagorodnova},
  {Blanco-Cuaresma}, {Boch}, {Bombrun}, {Borrachero}, {Bouquillon}, {Bourda},
  {Bouy}, {Bragaglia}, {Breddels}, {Brouillet}, {Br{\"u}semeister},
  {Bucciarelli}, {Budnik}, {Burgess}, {Burgon}, {Burlacu}, {Busonero}, {Buzzi},
  {Caffau}, {Cambras}, {Campbell}, {Cancelliere}, {Cantat-Gaudin}, {Carlucci},
  {Carrasco}, {Castellani}, {Charlot}, {Charnas}, {Charvet}, {Chassat},
  {Chiavassa}, {Clotet}, {Cocozza}, {Collins}, {Collins}, {Costigan}, {Crifo},
  {Cross}, {Crosta}, {Crowley}, {Dafonte}, {Damerdji}, {Dapergolas}, {David},
  {David}, {De Cat}, {de Felice}, {de Laverny}, {De Luise}, {De March}, {de
  Martino}, {de Souza}, {Debosscher}, {del Pozo}, {Delbo}, {Delgado},
  {Delgado}, {di Marco}, {Di Matteo}, {Diakite}, {Distefano}, {Dolding}, {Dos
  Anjos}, {Drazinos}, {Dur{\'a}n}, {Dzigan}, {Ecale}, {Edvardsson}, {Enke},
  {Erdmann}, {Escolar}, {Espina}, {Evans}, {Eynard Bontemps}, {Fabre},
  {Fabrizio}, {Faigler}, {Falc{\~a}o}, {Farr{\`a}s Casas}, {Faye}, {Federici},
  {Fedorets}, {Fern{\'a}ndez-Hern{\'a}ndez}, {Fernique}, {Fienga}, {Figueras},
  {Filippi}, {Findeisen}, {Fonti}, {Fouesneau}, {Fraile}, {Fraser}, {Fuchs},
  {Furnell}, {Gai}, {Galleti}, {Galluccio}, {Garabato}, {Garc{\'\i}a-Sedano},
  {Gar{\'e}}, {Garofalo}, {Garralda}, {Gavras}, {Gerssen}, {Geyer}, {Gilmore},
  {Girona}, {Giuffrida}, {Gomes}, {Gonz{\'a}lez-Marcos},
  {Gonz{\'a}lez-N{\'u}{\~n}ez}, {Gonz{\'a}lez-Vidal}, {Granvik}, {Guerrier},
  {Guillout}, {Guiraud}, {G{\'u}rpide}, {Guti{\'e}rrez-S{\'a}nchez}, {Guy},
  {Haigron}, {Hatzidimitriou}, {Haywood}, {Heiter}, {Helmi}, {Hobbs},
  {Hofmann}, {Holl}, {Holland}, {Hunt}, {Hypki}, {Icardi}, {Irwin}, {Jevardat
  de Fombelle}, {Jofr{\'e}}, {Jonker}, {Jorissen}, {Julbe}, {Karampelas},
  {Kochoska}, {Kohley}, {Kolenberg}, {Kontizas}, {Koposov}, {Kordopatis},
  {Koubsky}, {Kowalczyk}, {Krone-Martins}, {Kudryashova}, {Kull}, {Bachchan},
  {Lacoste-Seris}, {Lanza}, {Lavigne}, {Le Poncin-Lafitte}, {Lebreton},
  {Lebzelter}, {Leccia}, {Leclerc}, {Lecoeur-Taibi}, {Lemaitre}, {Lenhardt},
  {Leroux}, {Liao}, {Licata}, {Lindstr{\o}m}, {Lister}, {Livanou}, {Lobel},
  {L{\"o}ffler}, {L{\'o}pez}, {Lopez-Lozano}, {Lorenz}, {Loureiro},
  {MacDonald}, {Magalh{\~a}es Fernandes}, {Managau}, {Mann}, {Mantelet},
  {Marchal}, {Marchant}, {Marconi}, {Marie}, {Marinoni}, {Marrese},
  {Marschalk{\'o}}, {Marshall}, {Mart{\'\i}n-Fleitas}, {Martino}, {Mary},
  {Matijevi{\v{c}}}, {Mazeh}, {McMillan}, {Messina}, {Mestre}, {Michalik},
  {Millar}, {Miranda}, {Molina}, {Molinaro}, {Molinaro}, {Moln{\'a}r},
  {Moniez}, {Montegriffo}, {Monteiro}, {Mor}, {Mora}, {Morbidelli}, {Morel},
  {Morgenthaler}, {Morley}, {Morris}, {Mulone}, {Muraveva}, {Musella},
  {Narbonne}, {Nelemans}, {Nicastro}, {Noval}, {Ord{\'e}novic},
  {Ordieres-Mer{\'e}}, {Osborne}, {Pagani}, {Pagano}, {Pailler}, {Palacin},
  {Palaversa}, {Parsons}, {Paulsen}, {Pecoraro}, {Pedrosa}, {Pentik{\"a}inen},
  {Pereira}, {Pichon}, {Piersimoni}, {Pineau}, {Plachy}, {Plum}, {Poujoulet},
  {Pr{\v{s}}a}, {Pulone}, {Ragaini}, {Rago}, {Rambaux}, {Ramos-Lerate},
  {Ranalli}, {Rauw}, {Read}, {Regibo}, {Renk}, {Reyl{\'e}}, {Ribeiro},
  {Rimoldini}, {Ripepi}, {Riva}, {Rixon}, {Roelens}, {Romero-G{\'o}mez},
  {Rowell}, {Royer}, {Rudolph}, {Ruiz-Dern}, {Sadowski}, {Sagrist{\`a}
  Sell{\'e}s}, {Sahlmann}, {Salgado}, {Salguero}, {Sarasso}, {Savietto},
  {Schnorhk}, {Schultheis}, {Sciacca}, {Segol}, {Segovia}, {Segransan},
  {Serpell}, {Shih}, {Smareglia}, {Smart}, {Smith}, {Solano}, {Solitro},
  {Sordo}, {Soria Nieto}, {Souchay}, {Spagna}, {Spoto}, {Stampa}, {Steele},
  {Steidelm{\"u}ller}, {Stephenson}, {Stoev}, {Suess}, {S{\"u}veges}, {Surdej},
  {Szabados}, {Szegedi-Elek}, {Tapiador}, {Taris}, {Tauran}, {Taylor},
  {Teixeira}, {Terrett}, {Tingley}, {Trager}, {Turon}, {Ulla}, {Utrilla},
  {Valentini}, {van Elteren}, {Van Hemelryck}, {van Leeuwen}, {Varadi},
  {Vecchiato}, {Veljanoski}, {Via}, {Vicente}, {Vogt}, {Voss}, {Votruba},
  {Voutsinas}, {Walmsley}, {Weiler}, {Weingrill}, {Werner}, {Wevers},
  {Whitehead}, {Wyrzykowski}, {Yoldas}, {{\v{Z}}erjal}, {Zucker}, {Zurbach},
  {Zwitter}, {Alecu}, {Allen}, {Allende Prieto}, {Amorim},
  {Anglada-Escud{\'e}}, {Arsenijevic}, {Azaz}, {Balm}, {Beck}, {Bernstein},
  {Bigot}, {Bijaoui}, {Blasco}, {Bonfigli}, {Bono}, {Boudreault}, {Bressan},
  {Brown}, {Brunet}, {Bunclark}, {Buonanno}, {Butkevich}, {Carret}, {Carrion},
  {Chemin}, {Ch{\'e}reau}, {Corcione}, {Darmigny}, {de Boer}, {de Teodoro}, {de
  Zeeuw}, {Delle Luche}, {Domingues}, {Dubath}, {Fodor}, {Fr{\'e}zouls},
  {Fries}, {Fustes}, {Fyfe}, {Gallardo}, {Gallegos}, {Gardiol}, {Gebran},
  {Gomboc}, {G{\'o}mez}, {Grux}, {Gueguen}, {Heyrovsky}, {Hoar}, {Iannicola},
  {Isasi Parache}, {Janotto}, {Joliet}, {Jonckheere}, {Keil}, {Kim},
  {Klagyivik}, {Klar}, {Knude}, {Kochukhov}, {Kolka}, {Kos}, {Kutka}, {Lainey},
  {LeBouquin}, {Liu}, {Loreggia}, {Makarov}, {Marseille}, {Martayan},
  {Martinez-Rubi}, {Massart}, {Meynadier}, {Mignot}, {Munari}, {Nguyen},
  {Nordlander}, {Ocvirk}, {O'Flaherty}, {Olias Sanz}, {Ortiz}, {Osorio},
  {Oszkiewicz}, {Ouzounis}, {Palmer}, {Park}, {Pasquato}, {Peltzer}, {Peralta},
  {P{\'e}turaud}, {Pieniluoma}, {Pigozzi}, {Poels}, {Prat}, {Prod'homme},
  {Raison}, {Rebordao}, {Risquez}, {Rocca-Volmerange}, {Rosen}, {Ruiz-Fuertes},
  {Russo}, {Sembay}, {Serraller Vizcaino}, {Short}, {Siebert}, {Silva},
  {Sinachopoulos}, {Slezak}, {Soffel}, {Sosnowska}, {Strai{\v{z}}ys}, {ter
  Linden}, {Terrell}, {Theil}, {Tiede}, {Troisi}, {Tsalmantza}, {Tur},
  {Vaccari}, {Vachier}, {Valles}, {Van Hamme}, {Veltz}, {Virtanen}, {Wallut},
  {Wichmann}, {Wilkinson}, {Ziaeepour}, \& {Zschocke}}]{Gaia2016AA}
{Gaia Collaboration} {et~al.} 2016, \aap, 595, A1

\bibitem[{{Gaia Collaboration} {et~al.}(2018){Gaia Collaboration}, {Brown},
  {Vallenari}, {Prusti}, {de Bruijne}, {Babusiaux}, {Bailer-Jones}, {Biermann},
  {Evans}, {Eyer}, {Jansen}, {Jordi}, {Klioner}, {Lammers}, {Lindegren},
  {Luri}, {Mignard}, {Panem}, {Pourbaix}, {Randich}, {Sartoretti}, {Siddiqui},
  {Soubiran}, {van Leeuwen}, {Walton}, {Arenou}, {Bastian}, {Cropper},
  {Drimmel}, {Katz}, {Lattanzi}, {Bakker}, {Cacciari}, {Casta{\~n}eda},
  {Chaoul}, {Cheek}, {De Angeli}, {Fabricius}, {Guerra}, {Holl}, {Masana},
  {Messineo}, {Mowlavi}, {Nienartowicz}, {Panuzzo}, {Portell}, {Riello},
  {Seabroke}, {Tanga}, {Th{\'e}venin}, {Gracia-Abril}, {Comoretto},
  {Garcia-Reinaldos}, {Teyssier}, {Altmann}, {Andrae}, {Audard},
  {Bellas-Velidis}, {Benson}, {Berthier}, {Blomme}, {Burgess}, {Busso},
  {Carry}, {Cellino}, {Clementini}, {Clotet}, {Creevey}, {Davidson}, {De
  Ridder}, {Delchambre}, {Dell'Oro}, {Ducourant},
  {Fern{\'a}ndez-Hern{\'a}ndez}, {Fouesneau}, {Fr{\'e}mat}, {Galluccio},
  {Garc{\'\i}a-Torres}, {Gonz{\'a}lez-N{\'u}{\~n}ez}, {Gonz{\'a}lez-Vidal},
  {Gosset}, {Guy}, {Halbwachs}, {Hambly}, {Harrison}, {Hern{\'a}ndez},
  {Hestroffer}, {Hodgkin}, {Hutton}, {Jasniewicz}, {Jean-Antoine-Piccolo},
  {Jordan}, {Korn}, {Krone-Martins}, {Lanzafame}, {Lebzelter}, {L{\"o}ffler},
  {Manteiga}, {Marrese}, {Mart{\'\i}n-Fleitas}, {Moitinho}, {Mora}, {Muinonen},
  {Osinde}, {Pancino}, {Pauwels}, {Petit}, {Recio-Blanco}, {Richards},
  {Rimoldini}, {Robin}, {Sarro}, {Siopis}, {Smith}, {Sozzetti}, {S{\"u}veges},
  {Torra}, {van Reeven}, {Abbas}, {Abreu Aramburu}, {Accart}, {Aerts},
  {Altavilla}, {{\'A}lvarez}, {Alvarez}, {Alves}, {Anderson}, {Andrei},
  {Anglada Varela}, {Antiche}, {Antoja}, {Arcay}, {Astraatmadja}, {Bach},
  {Baker}, {Balaguer-N{\'u}{\~n}ez}, {Balm}, {Barache}, {Barata}, {Barbato},
  {Barblan}, {Barklem}, {Barrado}, {Barros}, {Barstow}, {Bartholom{\'e}
  Mu{\~n}oz}, {Bassilana}, {Becciani}, {Bellazzini}, {Berihuete}, {Bertone},
  {Bianchi}, {Bienaym{\'e}}, {Blanco-Cuaresma}, {Boch}, {Boeche}, {Bombrun},
  {Borrachero}, {Bossini}, {Bouquillon}, {Bourda}, {Bragaglia}, {Bramante},
  {Breddels}, {Bressan}, {Brouillet}, {Br{\"u}semeister}, {Brugaletta},
  {Bucciarelli}, {Burlacu}, {Busonero}, {Butkevich}, {Buzzi}, {Caffau},
  {Cancelliere}, {Cannizzaro}, {Cantat-Gaudin}, {Carballo}, {Carlucci},
  {Carrasco}, {Casamiquela}, {Castellani}, {Castro-Ginard}, {Charlot},
  {Chemin}, {Chiavassa}, {Cocozza}, {Costigan}, {Cowell}, {Crifo}, {Crosta},
  {Crowley}, {Cuypers}, {Dafonte}, {Damerdji}, {Dapergolas}, {David}, {David},
  {de Laverny}, {De Luise}, {De March}, {de Martino}, {de Souza}, {de Torres},
  {Debosscher}, {del Pozo}, {Delbo}, {Delgado}, {Delgado}, {Di Matteo},
  {Diakite}, {Diener}, {Distefano}, {Dolding}, {Drazinos}, {Dur{\'a}n},
  {Edvardsson}, {Enke}, {Eriksson}, {Esquej}, {Eynard Bontemps}, {Fabre},
  {Fabrizio}, {Faigler}, {Falc{\~a}o}, {Farr{\`a}s Casas}, {Federici},
  {Fedorets}, {Fernique}, {Figueras}, {Filippi}, {Findeisen}, {Fonti},
  {Fraile}, {Fraser}, {Fr{\'e}zouls}, {Gai}, {Galleti}, {Garabato},
  {Garc{\'\i}a-Sedano}, {Garofalo}, {Garralda}, {Gavel}, {Gavras}, {Gerssen},
  {Geyer}, {Giacobbe}, {Gilmore}, {Girona}, {Giuffrida}, {Glass}, {Gomes},
  {Granvik}, {Gueguen}, {Guerrier}, {Guiraud}, {Guti{\'e}rrez-S{\'a}nchez},
  {Haigron}, {Hatzidimitriou}, {Hauser}, {Haywood}, {Heiter}, {Helmi}, {Heu},
  {Hilger}, {Hobbs}, {Hofmann}, {Holland}, {Huckle}, {Hypki}, {Icardi},
  {Jan{\ss}en}, {Jevardat de Fombelle}, {Jonker}, {Juh{\'a}sz}, {Julbe},
  {Karampelas}, {Kewley}, {Klar}, {Kochoska}, {Kohley}, {Kolenberg},
  {Kontizas}, {Kontizas}, {Koposov}, {Kordopatis}, {Kostrzewa-Rutkowska},
  {Koubsky}, {Lambert}, {Lanza}, {Lasne}, {Lavigne}, {Le Fustec}, {Le
  Poncin-Lafitte}, {Lebreton}, {Leccia}, {Leclerc}, {Lecoeur-Taibi},
  {Lenhardt}, {Leroux}, {Liao}, {Licata}, {Lindstr{\o}m}, {Lister}, {Livanou},
  {Lobel}, {L{\'o}pez}, {Managau}, {Mann}, {Mantelet}, {Marchal}, {Marchant},
  {Marconi}, {Marinoni}, {Marschalk{\'o}}, {Marshall}, {Martino}, {Marton},
  {Mary}, {Massari}, {Matijevi{\v{c}}}, {Mazeh}, {McMillan}, {Messina},
  {Michalik}, {Millar}, {Molina}, {Molinaro}, {Moln{\'a}r}, {Montegriffo},
  {Mor}, {Morbidelli}, {Morel}, {Morris}, {Mulone}, {Muraveva}, {Musella},
  {Nelemans}, {Nicastro}, {Noval}, {O'Mullane}, {Ord{\'e}novic},
  {Ord{\'o}{\~n}ez-Blanco}, {Osborne}, {Pagani}, {Pagano}, {Pailler},
  {Palacin}, {Palaversa}, {Panahi}, {Pawlak}, {Piersimoni}, {Pineau}, {Plachy},
  {Plum}, {Poggio}, {Poujoulet}, {Pr{\v{s}}a}, {Pulone}, {Racero}, {Ragaini},
  {Rambaux}, {Ramos-Lerate}, {Regibo}, {Reyl{\'e}}, {Riclet}, {Ripepi}, {Riva},
  {Rivard}, {Rixon}, {Roegiers}, {Roelens}, {Romero-G{\'o}mez}, {Rowell},
  {Royer}, {Ruiz-Dern}, {Sadowski}, {Sagrist{\`a} Sell{\'e}s}, {Sahlmann},
  {Salgado}, {Salguero}, {Sanna}, {Santana-Ros}, {Sarasso}, {Savietto},
  {Schultheis}, {Sciacca}, {Segol}, {Segovia}, {S{\'e}gransan}, {Shih},
  {Siltala}, {Silva}, {Smart}, {Smith}, {Solano}, {Solitro}, {Sordo}, {Soria
  Nieto}, {Souchay}, {Spagna}, {Spoto}, {Stampa}, {Steele},
  {Steidelm{\"u}ller}, {Stephenson}, {Stoev}, {Suess}, {Surdej}, {Szabados},
  {Szegedi-Elek}, {Tapiador}, {Taris}, {Tauran}, {Taylor}, {Teixeira},
  {Terrett}, {Teyssandier}, {Thuillot}, {Titarenko}, {Torra Clotet}, {Turon},
  {Ulla}, {Utrilla}, {Uzzi}, {Vaillant}, {Valentini}, {Valette}, {van Elteren},
  {Van Hemelryck}, {van Leeuwen}, {Vaschetto}, {Vecchiato}, {Veljanoski},
  {Viala}, {Vicente}, {Vogt}, {von Essen}, {Voss}, {Votruba}, {Voutsinas},
  {Walmsley}, {Weiler}, {Wertz}, {Wevers}, {Wyrzykowski}, {Yoldas},
  {{\v{Z}}erjal}, {Ziaeepour}, {Zorec}, {Zschocke}, {Zucker}, {Zurbach}, \&
  {Zwitter}}]{Gaia2018AA}
---. 2018, \aap, 616, A1

\bibitem[{{Gaia Collaboration} {et~al.}(2021){Gaia Collaboration}, {Brown},
  {Vallenari}, {Prusti}, {de Bruijne}, {Babusiaux}, {Biermann}, {Creevey},
  {Evans}, {Eyer}, {Hutton}, {Jansen}, {Jordi}, {Klioner}, {Lammers},
  {Lindegren}, {Luri}, {Mignard}, {Panem}, {Pourbaix}, {Randich}, {Sartoretti},
  {Soubiran}, {Walton}, {Arenou}, {Bailer-Jones}, {Bastian}, {Cropper},
  {Drimmel}, {Katz}, {Lattanzi}, {van Leeuwen}, {Bakker}, {Cacciari},
  {Casta{\~n}eda}, {De Angeli}, {Ducourant}, {Fabricius}, {Fouesneau},
  {Fr{\'e}mat}, {Guerra}, {Guerrier}, {Guiraud}, {Jean-Antoine Piccolo},
  {Masana}, {Messineo}, {Mowlavi}, {Nicolas}, {Nienartowicz}, {Pailler},
  {Panuzzo}, {Riclet}, {Roux}, {Seabroke}, {Sordo}, {Tanga}, {Th{\'e}venin},
  {Gracia-Abril}, {Portell}, {Teyssier}, {Altmann}, {Andrae}, {Bellas-Velidis},
  {Benson}, {Berthier}, {Blomme}, {Brugaletta}, {Burgess}, {Busso}, {Carry},
  {Cellino}, {Cheek}, {Clementini}, {Damerdji}, {Davidson}, {Delchambre},
  {Dell'Oro}, {Fern{\'a}ndez-Hern{\'a}ndez}, {Galluccio}, {Garc{\'\i}a-Lario},
  {Garcia-Reinaldos}, {Gonz{\'a}lez-N{\'u}{\~n}ez}, {Gosset}, {Haigron},
  {Halbwachs}, {Hambly}, {Harrison}, {Hatzidimitriou}, {Heiter},
  {Hern{\'a}ndez}, {Hestroffer}, {Hodgkin}, {Holl}, {Jan{\ss}en}, {Jevardat de
  Fombelle}, {Jordan}, {Krone-Martins}, {Lanzafame}, {L{\"o}ffler}, {Lorca},
  {Manteiga}, {Marchal}, {Marrese}, {Moitinho}, {Mora}, {Muinonen}, {Osborne},
  {Pancino}, {Pauwels}, {Petit}, {Recio-Blanco}, {Richards}, {Riello},
  {Rimoldini}, {Robin}, {Roegiers}, {Rybizki}, {Sarro}, {Siopis}, {Smith},
  {Sozzetti}, {Ulla}, {Utrilla}, {van Leeuwen}, {van Reeven}, {Abbas}, {Abreu
  Aramburu}, {Accart}, {Aerts}, {Aguado}, {Ajaj}, {Altavilla}, {{\'A}lvarez},
  {{\'A}lvarez Cid-Fuentes}, {Alves}, {Anderson}, {Anglada Varela}, {Antoja},
  {Audard}, {Baines}, {Baker}, {Balaguer-N{\'u}{\~n}ez}, {Balbinot}, {Balog},
  {Barache}, {Barbato}, {Barros}, {Barstow}, {Bartolom{\'e}}, {Bassilana},
  {Bauchet}, {Baudesson-Stella}, {Becciani}, {Bellazzini}, {Bernet}, {Bertone},
  {Bianchi}, {Blanco-Cuaresma}, {Boch}, {Bombrun}, {Bossini}, {Bouquillon},
  {Bragaglia}, {Bramante}, {Breedt}, {Bressan}, {Brouillet}, {Bucciarelli},
  {Burlacu}, {Busonero}, {Butkevich}, {Buzzi}, {Caffau}, {Cancelliere},
  {C{\'a}novas}, {Cantat-Gaudin}, {Carballo}, {Carlucci}, {Carnerero},
  {Carrasco}, {Casamiquela}, {Castellani}, {Castro-Ginard}, {Castro Sampol},
  {Chaoul}, {Charlot}, {Chemin}, {Chiavassa}, {Cioni}, {Comoretto}, {Cooper},
  {Cornez}, {Cowell}, {Crifo}, {Crosta}, {Crowley}, {Dafonte}, {Dapergolas},
  {David}, {David}, {de Laverny}, {De Luise}, {De March}, {De Ridder}, {de
  Souza}, {de Teodoro}, {de Torres}, {del Peloso}, {del Pozo}, {Delbo},
  {Delgado}, {Delgado}, {Delisle}, {Di Matteo}, {Diakite}, {Diener},
  {Distefano}, {Dolding}, {Eappachen}, {Edvardsson}, {Enke}, {Esquej}, {Fabre},
  {Fabrizio}, {Faigler}, {Fedorets}, {Fernique}, {Fienga}, {Figueras},
  {Fouron}, {Fragkoudi}, {Fraile}, {Franke}, {Gai}, {Garabato},
  {Garcia-Gutierrez}, {Garc{\'\i}a-Torres}, {Garofalo}, {Gavras}, {Gerlach},
  {Geyer}, {Giacobbe}, {Gilmore}, {Girona}, {Giuffrida}, {Gomel}, {Gomez},
  {Gonzalez-Santamaria}, {Gonz{\'a}lez-Vidal}, {Granvik},
  {Guti{\'e}rrez-S{\'a}nchez}, {Guy}, {Hauser}, {Haywood}, {Helmi}, {Hidalgo},
  {Hilger}, {H{\l}adczuk}, {Hobbs}, {Holland}, {Huckle}, {Jasniewicz},
  {Jonker}, {Juaristi Campillo}, {Julbe}, {Karbevska}, {Kervella}, {Khanna},
  {Kochoska}, {Kontizas}, {Kordopatis}, {Korn}, {Kostrzewa-Rutkowska},
  {Kruszy{\'n}ska}, {Lambert}, {Lanza}, {Lasne}, {Le Campion}, {Le Fustec},
  {Lebreton}, {Lebzelter}, {Leccia}, {Leclerc}, {Lecoeur-Taibi}, {Liao},
  {Licata}, {Lindstr{\o}m}, {Lister}, {Livanou}, {Lobel}, {Madrero Pardo},
  {Managau}, {Mann}, {Marchant}, {Marconi}, {Marcos Santos}, {Marinoni},
  {Marocco}, {Marshall}, {Martin Polo}, {Mart{\'\i}n-Fleitas}, {Masip},
  {Massari}, {Mastrobuono-Battisti}, {Mazeh}, {McMillan}, {Messina},
  {Michalik}, {Millar}, {Mints}, {Molina}, {Molinaro}, {Moln{\'a}r},
  {Montegriffo}, {Mor}, {Morbidelli}, {Morel}, {Morris}, {Mulone}, {Munoz},
  {Muraveva}, {Murphy}, {Musella}, {Noval}, {Ord{\'e}novic}, {Orr{\`u}},
  {Osinde}, {Pagani}, {Pagano}, {Palaversa}, {Palicio}, {Panahi}, {Pawlak},
  {Pe{\~n}alosa Esteller}, {Penttil{\"a}}, {Piersimoni}, {Pineau}, {Plachy},
  {Plum}, {Poggio}, {Poretti}, {Poujoulet}, {Pr{\v{s}}a}, {Pulone}, {Racero},
  {Ragaini}, {Rainer}, {Raiteri}, {Rambaux}, {Ramos}, {Ramos-Lerate}, {Re
  Fiorentin}, {Regibo}, {Reyl{\'e}}, {Ripepi}, {Riva}, {Rixon}, {Robichon},
  {Robin}, {Roelens}, {Rohrbasser}, {Romero-G{\'o}mez}, {Rowell}, {Royer},
  {Rybicki}, {Sadowski}, {Sagrist{\`a} Sell{\'e}s}, {Sahlmann}, {Salgado},
  {Salguero}, {Samaras}, {Sanchez Gimenez}, {Sanna}, {Santove{\~n}a},
  {Sarasso}, {Schultheis}, {Sciacca}, {Segol}, {Segovia}, {S{\'e}gransan},
  {Semeux}, {Shahaf}, {Siddiqui}, {Siebert}, {Siltala}, {Slezak}, {Smart},
  {Solano}, {Solitro}, {Souami}, {Souchay}, {Spagna}, {Spoto}, {Steele},
  {Steidelm{\"u}ller}, {Stephenson}, {S{\"u}veges}, {Szabados}, {Szegedi-Elek},
  {Taris}, {Tauran}, {Taylor}, {Teixeira}, {Thuillot}, {Tonello}, {Torra},
  {Torra}, {Turon}, {Unger}, {Vaillant}, {van Dillen}, {Vanel}, {Vecchiato},
  {Viala}, {Vicente}, {Voutsinas}, {Weiler}, {Wevers}, {Wyrzykowski}, {Yoldas},
  {Yvard}, {Zhao}, {Zorec}, {Zucker}, {Zurbach}, \&
  {Zwitter}}]{GaiaEDR3_2021AA}
---. 2021, \aap, 649, A1

\bibitem[{{Gallagher} {et~al.}(2017){Gallagher}, {Caffau}, {Bonifacio},
  {Ludwig}, {Steffen}, {Homeier}, \& {Plez}}]{Gallagher2017}
{Gallagher}, A.~J., {Caffau}, E., {Bonifacio}, P., {Ludwig}, H.~G., {Steffen},
  M., {Homeier}, D., \& {Plez}, B. 2017, \aap, 598, L10

\bibitem[{{Gao} {et~al.}(2020){Gao}, {Lind}, {Amarsi}, {Buder},
  {Bland-Hawthorn}, {Campbell}, {Asplund}, {Casey}, {de Silva}, {Freeman},
  {Hayden}, {Lewis}, {Martell}, {Simpson}, {Sharma}, {Zucker}, {Zwitter},
  {Horner}, {Munari}, {Nordlander}, {Stello}, {Ting}, {Traven}, {Wittenmyer},
  \& {GALAH Collaboration}}]{Gao2020a}
{Gao}, X., {et~al.} 2020, \mnras, 497, L30

\bibitem[{{Gilmore} {et~al.}(2012){Gilmore}, {Randich}, {Asplund}, {Binney},
  {Bonifacio}, {Drew}, {Feltzing}, {Ferguson}, {Jeffries}, {Micela},
  {Negueruela}, {Prusti}, {Rix}, {Vallenari}, {Alfaro}, {Allende-Prieto},
  {Babusiaux}, {Bensby}, {Blomme}, {Bragaglia}, {Flaccomio}, {Fran{\c{c}}ois},
  {Irwin}, {Koposov}, {Korn}, {Lanzafame}, {Pancino}, {Paunzen},
  {Recio-Blanco}, {Sacco}, {Smiljanic}, {Van Eck}, {Walton}, {Aden}, {Aerts},
  {Affer}, {Alcala}, {Altavilla}, {Alves}, {Antoja}, {Arenou}, {Argiroffi},
  {Asensio Ramos}, {Bailer-Jones}, {Balaguer-Nunez}, {Bayo}, {Barbuy},
  {Barisevicius}, {Barrado y Navascues}, {Battistini}, {Bellas Velidis},
  {Bellazzini}, {Belokurov}, {Bergemann}, {Bertelli}, {Biazzo}, {Bienayme},
  {Bland-Hawthorn}, {Boeche}, {Bonito}, {Boudreault}, {Bouvier}, {Brandao},
  {Brown}, {de Bruijne}, {Burleigh}, {Caballero}, {Caffau}, {Calura},
  {Capuzzo-Dolcetta}, {Caramazza}, {Carraro}, {Casagrande}, {Casewell},
  {Chapman}, {Chiappini}, {Chorniy}, {Christlieb}, {Cignoni}, {Cocozza},
  {Colless}, {Collet}, {Collins}, {Correnti}, {Covino}, {Crnojevic}, {Cropper},
  {Cunha}, {Damiani}, {David}, {Delgado}, {Duffau}, {Edvardsson}, {Eldridge},
  {Enke}, {Eriksson}, {Evans}, {Eyer}, {Famaey}, {Fellhauer}, {Ferreras},
  {Figueras}, {Fiorentino}, {Flynn}, {Folha}, {Franciosini}, {Frasca},
  {Freeman}, {Fremat}, {Friel}, {Gaensicke}, {Gameiro}, {Garzon}, {Geier},
  {Geisler}, {Gerhard}, {Gibson}, {Gomboc}, {Gomez}, {Gonzalez-Fernandez},
  {Gonzalez Hernandez}, {Gosset}, {Grebel}, {Greimel}, {Groenewegen},
  {Grundahl}, {Guarcello}, {Gustafsson}, {Hadrava}, {Hatzidimitriou}, {Hambly},
  {Hammersley}, {Hansen}, {Haywood}, {Heber}, {Heiter}, {Held}, {Helmi},
  {Hensler}, {Herrero}, {Hill}, {Hodgkin}, {Huelamo}, {Huxor}, {Ibata},
  {Jackson}, {de Jong}, {Jonker}, {Jordan}, {Jordi}, {Jorissen}, {Katz},
  {Kawata}, {Keller}, {Kharchenko}, {Klement}, {Klutsch}, {Knude}, {Koch},
  {Kochukhov}, {Kontizas}, {Koubsky}, {Lallement}, {de Laverny}, {van Leeuwen},
  {Lemasle}, {Lewis}, {Lind}, {Lindstrom}, {Lobel}, {Lopez Santiago}, {Lucas},
  {Ludwig}, {Lueftinger}, {Magrini}, {Maiz Apellaniz}, {Maldonado}, {Marconi},
  {Marino}, {Martayan}, {Martinez-Valpuesta}, {Matijevic}, {McMahon},
  {Messina}, {Meyer}, {Miglio}, {Mikolaitis}, {Minchev}, {Minniti}, {Moitinho},
  {Momany}, {Monaco}, {Montalto}, {Monteiro}, {Monier}, {Montes}, {Mora},
  {Moraux}, {Morel}, {Mowlavi}, {Mucciarelli}, {Munari}, {Napiwotzki},
  {Nardetto}, {Naylor}, {Naze}, {Nelemans}, {Okamoto}, {Ortolani}, {Pace},
  {Palla}, {Palous}, {Parker}, {Penarrubia}, {Pillitteri}, {Piotto}, {Posbic},
  {Prisinzano}, {Puzeras}, {Quirrenbach}, {Ragaini}, {Read}, {Read}, {Reyle},
  {De Ridder}, {Robichon}, {Robin}, {Roeser}, {Romano}, {Royer}, {Ruchti},
  {Ruzicka}, {Ryan}, {Ryde}, {Santos}, {Sanz Forcada}, {Sarro Baro},
  {Sbordone}, {Schilbach}, {Schmeja}, {Schnurr}, {Schoenrich}, {Scholz},
  {Seabroke}, {Sharma}, {De Silva}, {Smith}, {Solano}, {Sordo}, {Soubiran},
  {Sousa}, {Spagna}, {Steffen}, {Steinmetz}, {Stelzer}, {Stempels},
  {Tabernero}, {Tautvaisiene}, {Thevenin}, {Torra}, {Tosi}, {Tolstoy}, {Turon},
  {Walker}, {Wambsganss}, {Worley}, {Venn}, {Vink}, {Wyse}, {Zaggia},
  {Zeilinger}, {Zoccali}, {Zorec}, {Zucker}, {Zwitter}, \& {Gaia-ESO Survey
  Team}}]{Gilmore2012Msngr}
{Gilmore}, G., {et~al.} 2012, The Messenger, 147, 25

\bibitem[{{Goswami} \& {Prantzos}(2000)}]{Goswami2000A&A}
{Goswami}, A., \& {Prantzos}, N. 2000, \aap, 359, 191

\bibitem[{{Gratton} {et~al.}(2000){Gratton}, {Sneden}, {Carretta}, \&
  {Bragaglia}}]{Gratton2000AA}
{Gratton}, R.~G., {Sneden}, C., {Carretta}, E., \& {Bragaglia}, A. 2000, \aap,
  354, 169

\bibitem[{{Hansen} {et~al.}(2016{\natexlab{a}}){Hansen}, {Andersen},
  {Nordstr{\"o}m}, {Beers}, {Placco}, {Yoon}, \& {Buchhave}}]{Hansen2016AAa}
{Hansen}, T.~T., {Andersen}, J., {Nordstr{\"o}m}, B., {Beers}, T.~C., {Placco},
  V.~M., {Yoon}, J., \& {Buchhave}, L.~A. 2016{\natexlab{a}}, \aap, 586, A160

\bibitem[{{Hansen} {et~al.}(2016{\natexlab{b}}){Hansen}, {Andersen},
  {Nordstr{\"o}m}, {Beers}, {Placco}, {Yoon}, \& {Buchhave}}]{Hansen2016AAb}
---. 2016{\natexlab{b}}, \aap, 588, A3

\bibitem[{{Hansen} {et~al.}(2020){Hansen}, {Marshall}, {Simon}, {Li},
  {Bernstein}, {Pace}, {Ferguson}, {Nagasawa}, {Kuehn}, {Carollo}, {Geha},
  {James}, {Walker}, {Diehl}, {Aguena}, {Allam}, {Avila}, {Bertin}, {Brooks},
  {Buckley-Geer}, {Burke}, {Rosell}, {Kind}, {Carretero}, {Costanzi}, {Da
  Costa}, {Desai}, {De Vicente}, {Doel}, {Eckert}, {Eifler}, {Everett},
  {Ferrero}, {Frieman}, {Garc{\'\i}a-Bellido}, {Gaztanaga}, {Gerdes}, {Gruen},
  {Gruendl}, {Gschwend}, {Gutierrez}, {Hinton}, {Hollowood}, {Honscheid},
  {Kuropatkin}, {Maia}, {March}, {Miquel}, {Palmese}, {Paz-Chinch{\'o}n},
  {Plazas}, {Sanchez}, {Santiago}, {Scarpine}, {Serrano}, {Smith},
  {Soares-Santos}, {Suchyta}, {Swanson}, {Tarle}, {Varga}, {Wilkinson}, \& {DES
  Collaboration}}]{Hansen2020ApJ}
{Hansen}, T.~T., {et~al.} 2020, \apj, 897, 183

\bibitem[{{Hartwig} {et~al.}(2019){Hartwig}, {Ishigaki}, {Klessen}, \&
  {Yoshida}}]{Hartwig2019MNRAS}
{Hartwig}, T., {Ishigaki}, M.~N., {Klessen}, R.~S., \& {Yoshida}, N. 2019,
  \mnras, 482, 1204

\bibitem[{{Hartwig} \& {Yoshida}(2019)}]{Hartwig2019ApJ}
{Hartwig}, T., \& {Yoshida}, N. 2019, \apjl, 870, L3

\bibitem[{{Heger} \& {Woosley}(2010)}]{Heger&Woosley2010ApJ}
{Heger}, A., \& {Woosley}, S.~E. 2010, \apj, 724, 341

\bibitem[{{Helmi} {et~al.}(2018){Helmi}, {Babusiaux}, {Koppelman}, {Massari},
  {Veljanoski}, \& {Brown}}]{Helmi2018Nature}
{Helmi}, A., {Babusiaux}, C., {Koppelman}, H.~H., {Massari}, D., {Veljanoski},
  J., \& {Brown}, A. G.~A. 2018, \nat, 563, 85

\bibitem[{{Helmi} {et~al.}(1999){Helmi}, {White}, {de Zeeuw}, \&
  {Zhao}}]{Helmi1999Nature}
{Helmi}, A., {White}, S. D.~M., {de Zeeuw}, P.~T., \& {Zhao}, H. 1999, \nat,
  402, 53

\bibitem[{{Hirai} {et~al.}(2015){Hirai}, {Ishimaru}, {Saitoh}, {Fujii},
  {Hidaka}, \& {Kajino}}]{Hirai2015ApJ}
{Hirai}, Y., {Ishimaru}, Y., {Saitoh}, T.~R., {Fujii}, M.~S., {Hidaka}, J., \&
  {Kajino}, T. 2015, \apj, 814, 41

\bibitem[{{Hirai} {et~al.}(2018){Hirai}, {Saitoh}, {Ishimaru}, \&
  {Wanajo}}]{Hirai2018ApJ}
{Hirai}, Y., {Saitoh}, T.~R., {Ishimaru}, Y., \& {Wanajo}, S. 2018, \apj, 855,
  63

\bibitem[{{Hirschi}(2007)}]{Hirschi2007AA}
{Hirschi}, R. 2007, \aap, 461, 571

\bibitem[{{Hollek} {et~al.}(2011){Hollek}, {Frebel}, {Roederer}, {Sneden},
  {Shetrone}, {Beers}, {Kang}, \& {Thom}}]{Hollek2011ApJ}
{Hollek}, J.~K., {Frebel}, A., {Roederer}, I.~U., {Sneden}, C., {Shetrone}, M.,
  {Beers}, T.~C., {Kang}, S.-j., \& {Thom}, C. 2011, \apj, 742, 54

\bibitem[{{Holmbeck} {et~al.}(2020){Holmbeck}, {Hansen}, {Beers}, {Placco},
  {Whitten}, {Rasmussen}, {Roederer}, {Ezzeddine}, {Sakari}, {Frebel}, {Drout},
  {Simon}, {Thompson}, {Bland-Hawthorn}, {Gibson}, {Grebel}, {Kordopatis},
  {Kunder}, {Mel{\'e}ndez}, {Navarro}, {Reid}, {Seabroke}, {Steinmetz},
  {Watson}, \& {Wyse}}]{Holmbeck2020ApJS}
{Holmbeck}, E.~M., {et~al.} 2020, \apjs, 249, 30

\bibitem[{{Honda} {et~al.}(2004{\natexlab{a}}){Honda}, {Aoki}, {Kajino},
  {Ando}, {Beers}, {Izumiura}, {Sadakane}, \& {Takada-Hidai}}]{Honda2004ApJ}
{Honda}, S., {Aoki}, W., {Kajino}, T., {Ando}, H., {Beers}, T.~C., {Izumiura},
  H., {Sadakane}, K., \& {Takada-Hidai}, M. 2004{\natexlab{a}}, \apj, 607, 474

\bibitem[{{Honda} {et~al.}(2004{\natexlab{b}}){Honda}, {Aoki}, {Ando},
  {Izumiura}, {Kajino}, {Kambe}, {Kawanomoto}, {Noguchi}, {Okita}, {Sadakane},
  {Sato}, {Takada-Hidai}, {Takeda}, {Watanabe}, {Beers}, {Norris}, \&
  {Ryan}}]{Honda2004ApJS}
{Honda}, S., {et~al.} 2004{\natexlab{b}}, \apjs, 152, 113

\bibitem[{{Ibata} {et~al.}(1994){Ibata}, {Gilmore}, \&
  {Irwin}}]{Ibata1994Nature}
{Ibata}, R.~A., {Gilmore}, G., \& {Irwin}, M.~J. 1994, \nat, 370, 194

\bibitem[{{Iben}(1967)}]{Iben1967ApJ}
{Iben}, Icko, J. 1967, \apj, 147, 624

\bibitem[{{Ishimaru} {et~al.}(2015){Ishimaru}, {Wanajo}, \&
  {Prantzos}}]{Ishimaru2015ApJ}
{Ishimaru}, Y., {Wanajo}, S., \& {Prantzos}, N. 2015, \apjl, 804, L35

\bibitem[{{Ivans} {et~al.}(2003){Ivans}, {Sneden}, {James}, {Preston},
  {Fulbright}, {H{\"o}flich}, {Carney}, \& {Wheeler}}]{Ivans2003ApJ}
{Ivans}, I.~I., {Sneden}, C., {James}, C.~R., {Preston}, G.~W., {Fulbright},
  J.~P., {H{\"o}flich}, P.~A., {Carney}, B.~W., \& {Wheeler}, J.~C. 2003, \apj,
  592, 906

\bibitem[{{Iwamoto} {et~al.}(1999){Iwamoto}, {Brachwitz}, {Nomoto},
  {Kishimoto}, {Umeda}, {Hix}, \& {Thielemann}}]{Iwamoto1999ApJS}
{Iwamoto}, K., {Brachwitz}, F., {Nomoto}, K., {Kishimoto}, N., {Umeda}, H.,
  {Hix}, W.~R., \& {Thielemann}, F.-K. 1999, \apjs, 125, 439

\bibitem[{{Jacobson} {et~al.}(2015){Jacobson}, {Keller}, {Frebel}, {Casey},
  {Asplund}, {Bessell}, {Da Costa}, {Lind}, {Marino}, {Norris}, {Pe{\~n}a},
  {Schmidt}, {Tisserand}, {Walsh}, {Yong}, \& {Yu}}]{Jacobson2015ApJ}
{Jacobson}, H.~R., {et~al.} 2015, \apj, 807, 171

\bibitem[{{Ji} {et~al.}(2016){Ji}, {Frebel}, {Chiti}, \&
  {Simon}}]{Ji2016Nature}
{Ji}, A.~P., {Frebel}, A., {Chiti}, A., \& {Simon}, J.~D. 2016, \nat, 531, 610

\bibitem[{{Jones} {et~al.}(2019){Jones}, {C{\^o}t{\'e}}, {R{\"o}pke}, \&
  {Wanajo}}]{Jones2019ApJ}
{Jones}, S., {C{\^o}t{\'e}}, B., {R{\"o}pke}, F.~K., \& {Wanajo}, S. 2019,
  \apj, 882, 170

\bibitem[{{Kang} \& {Lee}(2012)}]{Kang&Lee2012MNRAS}
{Kang}, W., \& {Lee}, S.-G. 2012, \mnras, 425, 3162

\bibitem[{{Karakas}(2010)}]{Karakas2010MNRAS}
{Karakas}, A.~I. 2010, \mnras, 403, 1413

\bibitem[{{Karakas} {et~al.}(2018){Karakas}, {Lugaro}, {Carlos}, {Cseh},
  {Kamath}, \& {Garc{\'\i}a-Hern{\'a}ndez}}]{Karakas2018MNRAS}
{Karakas}, A.~I., {Lugaro}, M., {Carlos}, M., {Cseh}, B., {Kamath}, D., \&
  {Garc{\'\i}a-Hern{\'a}ndez}, D.~A. 2018, \mnras, 477, 421

\bibitem[{{Keller} {et~al.}(2007){Keller}, {Schmidt}, {Bessell}, {Conroy},
  {Francis}, {Granlund}, {Kowald}, {Oates}, {Martin-Jones}, {Preston},
  {Tisserand}, {Vaccarella}, \& {Waterson}}]{Keller2007PASA}
{Keller}, S.~C., {et~al.} 2007, \pasa, 24, 1

\bibitem[{{Keller} {et~al.}(2014){Keller}, {Bessell}, {Frebel}, {Casey},
  {Asplund}, {Jacobson}, {Lind}, {Norris}, {Yong}, {Heger}, {Magic}, {da
  Costa}, {Schmidt}, \& {Tisserand}}]{Keller2014Nature}
---. 2014, \nat, 506, 463

\bibitem[{{Kobayashi} {et~al.}(2014){Kobayashi}, {Ishigaki}, {Tominaga}, \&
  {Nomoto}}]{Kobayashi2014ApJ}
{Kobayashi}, C., {Ishigaki}, M.~N., {Tominaga}, N., \& {Nomoto}, K. 2014,
  \apjl, 785, L5

\bibitem[{{Kobayashi} {et~al.}(2020){Kobayashi}, {Karakas}, \&
  {Lugaro}}]{Kobayashi2020ApJ}
{Kobayashi}, C., {Karakas}, A.~I., \& {Lugaro}, M. 2020, \apj, 900, 179

\bibitem[{{Kobayashi} {et~al.}(2011){Kobayashi}, {Karakas}, \&
  {Umeda}}]{Kobayashi2011MNRAS}
{Kobayashi}, C., {Karakas}, A.~I., \& {Umeda}, H. 2011, \mnras, 414, 3231

\bibitem[{{Kobayashi} \& {Nomoto}(2009)}]{Kobayashi&Nomoto2009ApJ}
{Kobayashi}, C., \& {Nomoto}, K. 2009, \apj, 707, 1466

\bibitem[{{Kobayashi} {et~al.}(2015){Kobayashi}, {Nomoto}, \&
  {Hachisu}}]{Kobayashi2015ApJ}
{Kobayashi}, C., {Nomoto}, K., \& {Hachisu}, I. 2015, \apjl, 804, L24

\bibitem[{{Kobayashi} {et~al.}(1998){Kobayashi}, {Tsujimoto}, {Nomoto},
  {Hachisu}, \& {Kato}}]{Kobayashi1998ApJ}
{Kobayashi}, C., {Tsujimoto}, T., {Nomoto}, K., {Hachisu}, I., \& {Kato}, M.
  1998, \apjl, 503, L155

\bibitem[{{Kobayashi} {et~al.}(2006){Kobayashi}, {Umeda}, {Nomoto}, {Tominaga},
  \& {Ohkubo}}]{Kobayashi2006ApJ}
{Kobayashi}, C., {Umeda}, H., {Nomoto}, K., {Tominaga}, N., \& {Ohkubo}, T.
  2006, \apj, 653, 1145

\bibitem[{{Koch} {et~al.}(2008){Koch}, {McWilliam}, {Grebel}, {Zucker}, \&
  {Belokurov}}]{Koch2008ApJ}
{Koch}, A., {McWilliam}, A., {Grebel}, E.~K., {Zucker}, D.~B., \& {Belokurov},
  V. 2008, \apjl, 688, L13

\bibitem[{{Koppelman} {et~al.}(2019){Koppelman}, {Helmi}, {Massari},
  {Price-Whelan}, \& {Starkenburg}}]{Koppelman2019AA}
{Koppelman}, H.~H., {Helmi}, A., {Massari}, D., {Price-Whelan}, A.~M., \&
  {Starkenburg}, T.~K. 2019, \aap, 631, L9

\bibitem[{{Korn} {et~al.}(2006){Korn}, {Grundahl}, {Richard}, {Barklem},
  {Mashonkina}, {Collet}, {Piskunov}, \& {Gustafsson}}]{Korn2006a}
{Korn}, A.~J., {Grundahl}, F., {Richard}, O., {Barklem}, P.~S., {Mashonkina},
  L., {Collet}, R., {Piskunov}, N., \& {Gustafsson}, B. 2006, \nat, 442, 657

\bibitem[{{Kroupa}(2002)}]{Kroupa2002Sci}
{Kroupa}, P. 2002, Science, 295, 82

\bibitem[{{Kroupa}(2008)}]{Kroupa2008ASPC}
{Kroupa}, P. 2008, in Astronomical Society of the Pacific Conference Series,
  Vol. 390, Pathways Through an Eclectic Universe, ed. J.~H. {Knapen}, T.~J.
  {Mahoney}, \& A.~{Vazdekis}, 3

\bibitem[{{Kubryk} {et~al.}(2015){Kubryk}, {Prantzos}, \&
  {Athanassoula}}]{Kubryk2015A&A}
{Kubryk}, M., {Prantzos}, N., \& {Athanassoula}, E. 2015, \aap, 580, A127

\bibitem[{{Kumar} {et~al.}(2020){Kumar}, {Reddy}, {Campbell}, {Maben}, {Zhao},
  \& {Ting}}]{Kumar2020NatAs}
{Kumar}, Y.~B., {Reddy}, B.~E., {Campbell}, S.~W., {Maben}, S., {Zhao}, G., \&
  {Ting}, Y.-S. 2020, Nature Astronomy, 4, 1059

\bibitem[{{Lai} {et~al.}(2008){Lai}, {Bolte}, {Johnson}, {Lucatello}, {Heger},
  \& {Woosley}}]{Lai2008ApJ}
{Lai}, D.~K., {Bolte}, M., {Johnson}, J.~A., {Lucatello}, S., {Heger}, A., \&
  {Woosley}, S.~E. 2008, \apj, 681, 1524

\bibitem[{{Lawler} {et~al.}(2001{\natexlab{a}}){Lawler}, {Bonvallet}, \&
  {Sneden}}]{Lawler2001ApJ_La}
{Lawler}, J.~E., {Bonvallet}, G., \& {Sneden}, C. 2001{\natexlab{a}}, \apj,
  556, 452

\bibitem[{{Lawler} {et~al.}(2001{\natexlab{b}}){Lawler}, {Wickliffe}, {den
  Hartog}, \& {Sneden}}]{Lawler2001ApJ_Eu}
{Lawler}, J.~E., {Wickliffe}, M.~E., {den Hartog}, E.~A., \& {Sneden}, C.
  2001{\natexlab{b}}, \apj, 563, 1075

\bibitem[{{Lee} {et~al.}(2017){Lee}, {Beers}, {Kim}, {Placco}, {Yoon},
  {Carollo}, {Masseron}, \& {Jung}}]{Lee2017ApJ}
{Lee}, Y.~S., {Beers}, T.~C., {Kim}, Y.~K., {Placco}, V., {Yoon}, J.,
  {Carollo}, D., {Masseron}, T., \& {Jung}, J. 2017, \apj, 836, 91

\bibitem[{{Li} {et~al.}(2018{\natexlab{a}}){Li}, {Aoki}, {Matsuno}, {Bharat
  Kumar}, {Shi}, {Suda}, \& {Zhao}}]{Li2018ApJL}
{Li}, H., {Aoki}, W., {Matsuno}, T., {Bharat Kumar}, Y., {Shi}, J., {Suda}, T.,
  \& {Zhao}, G. 2018{\natexlab{a}}, \apjl, 852, L31

\bibitem[{{Li} {et~al.}(2015{\natexlab{a}}){Li}, {Aoki}, {Zhao}, {Honda},
  {Christlieb}, \& {Suda}}]{Li2015PASJ}
{Li}, H., {Aoki}, W., {Zhao}, G., {Honda}, S., {Christlieb}, N., \& {Suda}, T.
  2015{\natexlab{a}}, \pasj, 67, 84

\bibitem[{{Li} {et~al.}(2018{\natexlab{b}}){Li}, {Tan}, \& {Zhao}}]{Li2018ApJS}
{Li}, H., {Tan}, K., \& {Zhao}, G. 2018{\natexlab{b}}, \apjs, 238, 16

\bibitem[{{Li} {et~al.}(2015{\natexlab{b}}){Li}, {Aoki}, {Honda}, {Zhao},
  {Christlieb}, \& {Suda}}]{Li2015RAA}
{Li}, H.-N., {Aoki}, W., {Honda}, S., {Zhao}, G., {Christlieb}, N., \& {Suda},
  T. 2015{\natexlab{b}}, Research in Astronomy and Astrophysics, 15, 1264

\bibitem[{{Li} {et~al.}(2015{\natexlab{c}}){Li}, {Zhao}, {Christlieb}, {Wang},
  {Wang}, {Zhang}, {Hou}, \& {Yuan}}]{Li2015ApJ}
{Li}, H.-N., {Zhao}, G., {Christlieb}, N., {Wang}, L., {Wang}, W., {Zhang}, Y.,
  {Hou}, Y., \& {Yuan}, H. 2015{\natexlab{c}}, \apj, 798, 110

\bibitem[{{Limongi} \& {Chieffi}(2012)}]{Limongi2012ApJS}
{Limongi}, M., \& {Chieffi}, A. 2012, \apjs, 199, 38

\bibitem[{{Limongi} \& {Chieffi}(2018)}]{Limongi2018ApJS}
---. 2018, \apjs, 237, 13

\bibitem[{{Lind} {et~al.}(2011){Lind}, {Asplund}, {Barklem}, \&
  {Belyaev}}]{Lind2011AA}
{Lind}, K., {Asplund}, M., {Barklem}, P.~S., \& {Belyaev}, A.~K. 2011, \aap,
  528, A103

\bibitem[{{Lind} {et~al.}(2012){Lind}, {Bergemann}, \&
  {Asplund}}]{Lind2012MNRAS}
{Lind}, K., {Bergemann}, M., \& {Asplund}, M. 2012, \mnras, 427, 50

\bibitem[{{Lind} {et~al.}(2009){Lind}, {Primas}, {Charbonnel}, {Grundahl}, \&
  {Asplund}}]{Lind2009AA}
{Lind}, K., {Primas}, F., {Charbonnel}, C., {Grundahl}, F., \& {Asplund}, M.
  2009, \aap, 503, 545

\bibitem[{{Liu} {et~al.}(2015){Liu}, {Zhao}, \& {Hou}}]{Liu2015RAA}
{Liu}, X.-W., {Zhao}, G., \& {Hou}, J.-L. 2015, Research in Astronomy and
  Astrophysics, 15, 1089

\bibitem[{{Lugaro} {et~al.}(2012){Lugaro}, {Karakas}, {Stancliffe}, \&
  {Rijs}}]{Lugaro2012ApJ}
{Lugaro}, M., {Karakas}, A.~I., {Stancliffe}, R.~J., \& {Rijs}, C. 2012, \apj,
  747, 2

\bibitem[{{Maeder} {et~al.}(2015){Maeder}, {Meynet}, \&
  {Chiappini}}]{Maeder2015AA}
{Maeder}, A., {Meynet}, G., \& {Chiappini}, C. 2015, \aap, 576, A56

\bibitem[{{Mardini} {et~al.}(2020){Mardini}, {Placco}, {Meiron}, {Ishchenko},
  {Avramov}, {Mazzarini}, {Berczik}, {Arca Sedda}, {Beers}, {Frebel}, {Taani},
  {Donnari}, {Al-Wardat}, \& {Zhao}}]{Mohammad2020ApJ}
{Mardini}, M.~K., {et~al.} 2020, \apj, 903, 88

\bibitem[{{Mashonkina} {et~al.}(2010){Mashonkina}, {Christlieb}, {Barklem},
  {Hill}, {Beers}, \& {Velichko}}]{Mashonkina2010AA}
{Mashonkina}, L., {Christlieb}, N., {Barklem}, P.~S., {Hill}, V., {Beers},
  T.~C., \& {Velichko}, A. 2010, \aap, 516, A46

\bibitem[{{Masseron} {et~al.}(2012){Masseron}, {Johnson}, {Lucatello},
  {Karakas}, {Plez}, {Beers}, \& {Christlieb}}]{Masseron2012ApJ}
{Masseron}, T., {Johnson}, J.~A., {Lucatello}, S., {Karakas}, A., {Plez}, B.,
  {Beers}, T.~C., \& {Christlieb}, N. 2012, \apj, 751, 14

\bibitem[{{Matsuno} {et~al.}(2017{\natexlab{a}}){Matsuno}, {Aoki}, {Beers},
  {Lee}, \& {Honda}}]{Matsuno2017AJ}
{Matsuno}, T., {Aoki}, W., {Beers}, T.~C., {Lee}, Y.~S., \& {Honda}, S.
  2017{\natexlab{a}}, \aj, 154, 52

\bibitem[{{Matsuno} {et~al.}(2019){Matsuno}, {Aoki}, \&
  {Suda}}]{Matsuno2019ApJL}
{Matsuno}, T., {Aoki}, W., \& {Suda}, T. 2019, \apjl, 874, L35

\bibitem[{{Matsuno} {et~al.}(2017{\natexlab{b}}){Matsuno}, {Aoki}, {Suda}, \&
  {Li}}]{Matsuno2017PASJ}
{Matsuno}, T., {Aoki}, W., {Suda}, T., \& {Li}, H. 2017{\natexlab{b}}, \pasj,
  69, 24

\bibitem[{{McWilliam} {et~al.}(1995){McWilliam}, {Preston}, {Sneden}, \&
  {Searle}}]{McWilliam1995AJ}
{McWilliam}, A., {Preston}, G.~W., {Sneden}, C., \& {Searle}, L. 1995, \aj,
  109, 2757

\bibitem[{{Mel{\'e}ndez} {et~al.}(2010){Mel{\'e}ndez}, {Casagrande},
  {Ram{\'{\i}}rez}, {Asplund}, \& {Schuster}}]{Melendez2010AA}
{Mel{\'e}ndez}, J., {Casagrande}, L., {Ram{\'{\i}}rez}, I., {Asplund}, M., \&
  {Schuster}, W.~J. 2010, \aap, 515, L3

\bibitem[{{Meynet} {et~al.}(2006){Meynet}, {Ekstr{\"o}m}, \&
  {Maeder}}]{Meynet2006AA}
{Meynet}, G., {Ekstr{\"o}m}, S., \& {Maeder}, A. 2006, \aap, 447, 623

\bibitem[{{Mucciarelli} {et~al.}(2012){Mucciarelli}, {Salaris}, \&
  {Bonifacio}}]{Mucciarelli2012MNRAS}
{Mucciarelli}, A., {Salaris}, M., \& {Bonifacio}, P. 2012, \mnras, 419, 2195

\bibitem[{{Mucciarelli} {et~al.}(2018){Mucciarelli}, {Salaris}, {Monaco},
  {Bonifacio}, {Fu}, \& {Villanova}}]{Mucciarelli2018AA}
{Mucciarelli}, A., {Salaris}, M., {Monaco}, L., {Bonifacio}, P., {Fu}, X., \&
  {Villanova}, S. 2018, \aap, 618, A134

\bibitem[{{Myeong} {et~al.}(2018){Myeong}, {Evans}, {Belokurov}, {Sanders}, \&
  {Koposov}}]{Myeong2018ApJ}
{Myeong}, G.~C., {Evans}, N.~W., {Belokurov}, V., {Sanders}, J.~L., \&
  {Koposov}, S.~E. 2018, \apjl, 856, L26

\bibitem[{{Naidu} {et~al.}(2020){Naidu}, {Conroy}, {Bonaca}, {Johnson}, {Ting},
  {Caldwell}, {Zaritsky}, \& {Cargile}}]{Naidu2020ApJ}
{Naidu}, R.~P., {Conroy}, C., {Bonaca}, A., {Johnson}, B.~D., {Ting}, Y.-S.,
  {Caldwell}, N., {Zaritsky}, D., \& {Cargile}, P.~A. 2020, \apj, 901, 48

\bibitem[{{Nissen} \& {Schuster}(2010)}]{Nissen&Schuster2010AA}
{Nissen}, P.~E., \& {Schuster}, W.~J. 2010, \aap, 511, L10

\bibitem[{{Nomoto} {et~al.}(2013){Nomoto}, {Kobayashi}, \&
  {Tominaga}}]{Nomoto2013ARAA}
{Nomoto}, K., {Kobayashi}, C., \& {Tominaga}, N. 2013, \araa, 51, 457

\bibitem[{{Norris} {et~al.}(2002){Norris}, {Ryan}, {Beers}, {Aoki}, \&
  {Ando}}]{Norris2002ApJ}
{Norris}, J.~E., {Ryan}, S.~G., {Beers}, T.~C., {Aoki}, W., \& {Ando}, H. 2002,
  \apjl, 569, L107

\bibitem[{{Norris} {et~al.}(1994){Norris}, {Ryan}, \&
  {Stringfellow}}]{Norris1994ApJ}
{Norris}, J.~E., {Ryan}, S.~G., \& {Stringfellow}, G.~S. 1994, \apj, 423, 386

\bibitem[{{Norris} \& {Yong}(2019)}]{Norris&Yong2019ApJ}
{Norris}, J.~E., \& {Yong}, D. 2019, \apj, 879, 37

\bibitem[{{Norris} {et~al.}(2013){Norris}, {Bessell}, {Yong}, {Christlieb},
  {Barklem}, {Asplund}, {Murphy}, {Beers}, {Frebel}, \&
  {Ryan}}]{Norris2013ApJa}
{Norris}, J.~E., {et~al.} 2013, \apj, 762, 25

\bibitem[{{Osorio} \& {Barklem}(2016)}]{Osorio&Barklem2016AA}
{Osorio}, Y., \& {Barklem}, P.~S. 2016, \aap, 586, A120

\bibitem[{{Placco} {et~al.}(2014{\natexlab{a}}){Placco}, {Frebel}, {Beers},
  {Christlieb}, {Lee}, {Kennedy}, {Rossi}, \& {Santucci}}]{Placco2014ApJa}
{Placco}, V.~M., {Frebel}, A., {Beers}, T.~C., {Christlieb}, N., {Lee}, Y.~S.,
  {Kennedy}, C.~R., {Rossi}, S., \& {Santucci}, R.~M. 2014{\natexlab{a}}, \apj,
  781, 40

\bibitem[{{Placco} {et~al.}(2014{\natexlab{b}}){Placco}, {Frebel}, {Beers}, \&
  {Stancliffe}}]{Placco2014ApJb}
{Placco}, V.~M., {Frebel}, A., {Beers}, T.~C., \& {Stancliffe}, R.~J.
  2014{\natexlab{b}}, \apj, 797, 21

\bibitem[{{Prantzos} {et~al.}(2018){Prantzos}, {Abia}, {Limongi}, {Chieffi}, \&
  {Cristallo}}]{Prantzos2018MNRAS}
{Prantzos}, N., {Abia}, C., {Limongi}, M., {Chieffi}, A., \& {Cristallo}, S.
  2018, \mnras, 476, 3432

\bibitem[{{Reggiani} {et~al.}(2017){Reggiani}, {Mel{\'e}ndez}, {Kobayashi},
  {Karakas}, \& {Placco}}]{Reggiani2017AA}
{Reggiani}, H., {Mel{\'e}ndez}, J., {Kobayashi}, C., {Karakas}, A., \&
  {Placco}, V. 2017, \aap, 608, A46

\bibitem[{{Richard}(2012)}]{Richard2012MSAIS}
{Richard}, O. 2012, Memorie della Societa Astronomica Italiana Supplementi, 22,
  211

\bibitem[{{Richard} {et~al.}(2005){Richard}, {Michaud}, \&
  {Richer}}]{Richard2005ApJ}
{Richard}, O., {Michaud}, G., \& {Richer}, J. 2005, \apj, 619, 538

\bibitem[{{Roederer}(2013)}]{Roederer2013AJ}
{Roederer}, I.~U. 2013, \aj, 145, 26

\bibitem[{{Roederer} \& {Barklem}(2018)}]{RoedererBarklem2018ApJ}
{Roederer}, I.~U., \& {Barklem}, P.~S. 2018, \apj, 857, 2

\bibitem[{{Roederer} \& {Gnedin}(2019)}]{Roederer2019ApJ}
{Roederer}, I.~U., \& {Gnedin}, O.~Y. 2019, \apj, 883, 84

\bibitem[{{Roederer} {et~al.}(2014){Roederer}, {Preston}, {Thompson},
  {Shectman}, {Sneden}, {Burley}, \& {Kelson}}]{Roederer2014AJ}
{Roederer}, I.~U., {Preston}, G.~W., {Thompson}, I.~B., {Shectman}, S.~A.,
  {Sneden}, C., {Burley}, G.~S., \& {Kelson}, D.~D. 2014, \aj, 147, 136

\bibitem[{{Roederer} {et~al.}(2018){Roederer}, {Sakari}, {Placco}, {Beers},
  {Ezzeddine}, {Frebel}, \& {Hansen}}]{Roederer2018ApJ}
{Roederer}, I.~U., {Sakari}, C.~M., {Placco}, V.~M., {Beers}, T.~C.,
  {Ezzeddine}, R., {Frebel}, A., \& {Hansen}, T.~T. 2018, \apj, 865, 129

\bibitem[{{Roederer} {et~al.}(2012){Roederer}, {Lawler}, {Cowan}, {Beers},
  {Frebel}, {Ivans}, {Schatz}, {Sobeck}, \& {Sneden}}]{Roederer2012ApJ}
{Roederer}, I.~U., {et~al.} 2012, \apjl, 747, L8

\bibitem[{{Roederer} {et~al.}(2016){Roederer}, {Mateo}, {Bailey}, {Song},
  {Bell}, {Crane}, {Loebman}, {Nidever}, {Olszewski}, {Shectman}, {Thompson},
  {Valluri}, \& {Walker}}]{Roederer2016AJ}
---. 2016, \aj, 151, 82

\bibitem[{{Romano} {et~al.}(2010){Romano}, {Karakas}, {Tosi}, \&
  {Matteucci}}]{Romano2010A&A}
{Romano}, D., {Karakas}, A.~I., {Tosi}, M., \& {Matteucci}, F. 2010, \aap, 522,
  A32

\bibitem[{{Ruiz-Lara} {et~al.}(2022){Ruiz-Lara}, {Matsuno}, {Sofie L{\"o}vdal},
  {Helmi}, {Dodd}, \& {Koppelman}}]{Ruiz-Lara2022arXiv}
{Ruiz-Lara}, T., {Matsuno}, T., {Sofie L{\"o}vdal}, S., {Helmi}, A., {Dodd},
  E., \& {Koppelman}, H.~H. 2022, arXiv e-prints, arXiv:2201.02405

\bibitem[{{Ryan} {et~al.}(1999){Ryan}, {Norris}, \& {Beers}}]{Ryan1999ApJ}
{Ryan}, S.~G., {Norris}, J.~E., \& {Beers}, T.~C. 1999, \apj, 523, 654

\bibitem[{{Sakari} {et~al.}(2018){Sakari}, {Placco}, {Farrell}, {Roederer},
  {Wallerstein}, {Beers}, {Ezzeddine}, {Frebel}, {Hansen}, {Holmbeck},
  {Sneden}, {Cowan}, {Venn}, {Davis}, {Matijevi{\v{c}}}, {Wyse},
  {Bland-Hawthorn}, {Chiappini}, {Freeman}, {Gibson}, {Grebel}, {Helmi},
  {Kordopatis}, {Kunder}, {Navarro}, {Reid}, {Seabroke}, {Steinmetz}, \&
  {Watson}}]{Sakari2018ApJ}
{Sakari}, C.~M., {et~al.} 2018, \apj, 868, 110

\bibitem[{{Sakari} {et~al.}(2019){Sakari}, {Roederer}, {Placco}, {Beers},
  {Ezzeddine}, {Frebel}, {Hansen}, {Sneden}, {Cowan}, {Wallerstein}, {Farrell},
  {Venn}, {Matijevi{\v{c}}}, {Wyse}, {Bland-Hawthorn}, {Chiappini}, {Freeman},
  {Gibson}, {Grebel}, {Helmi}, {Kordopatis}, {Kunder}, {Navarro}, {Reid},
  {Seabroke}, {Steinmetz}, \& {Watson}}]{Sakari2019ApJ}
---. 2019, \apj, 874, 148

\bibitem[{{Salvadori} {et~al.}(2019){Salvadori}, {Bonifacio}, {Caffau},
  {Korotin}, {Andreevsky}, {Spite}, \&
  {Sk{\'u}lad{\'o}ttir}}]{Salvadori2019MNRAS}
{Salvadori}, S., {Bonifacio}, P., {Caffau}, E., {Korotin}, S., {Andreevsky},
  S., {Spite}, M., \& {Sk{\'u}lad{\'o}ttir}, {\'A}. 2019, \mnras, 487, 4261

\bibitem[{{Salvadori} {et~al.}(2015){Salvadori}, {Sk{\'u}lad{\'o}ttir}, \&
  {Tolstoy}}]{Salvadori2015MNRAS}
{Salvadori}, S., {Sk{\'u}lad{\'o}ttir}, {\'A}., \& {Tolstoy}, E. 2015, \mnras,
  454, 1320

\bibitem[{{Sbordone} {et~al.}(2010){Sbordone}, {Bonifacio}, {Caffau}, {Ludwig},
  {Behara}, {Gonz{\'a}lez Hern{\'a}ndez}, {Steffen}, {Cayrel}, {Freytag},
  {van't Veer}, {Molaro}, {Plez}, {Sivarani}, {Spite}, {Spite}, {Beers},
  {Christlieb}, {Fran{\c c}ois}, \& {Hill}}]{Sbordone2010AA}
{Sbordone}, L., {et~al.} 2010, \aap, 522, A26

\bibitem[{{Sestito} {et~al.}(2019){Sestito}, {Longeard}, {Martin},
  {Starkenburg}, {Fouesneau}, {Gonz{\'a}lez Hern{\'a}ndez}, {Arentsen},
  {Ibata}, {Aguado}, {Carlberg}, {Jablonka}, {Navarro}, {Tolstoy}, \&
  {Venn}}]{Sestito2019MNRAS}
{Sestito}, F., {et~al.} 2019, \mnras, 484, 2166

\bibitem[{{Sestito} {et~al.}(2020){Sestito}, {Martin}, {Starkenburg},
  {Arentsen}, {Ibata}, {Longeard}, {Kielty}, {Youakim}, {Venn}, {Aguado},
  {Carlberg}, {Gonz{\'a}lez Hern{\'a}ndez}, {Hill}, {Jablonka}, {Kordopatis},
  {Malhan}, {Navarro}, {S{\'a}nchez-Janssen}, {Thomas}, {Tolstoy}, {Wilson},
  {Palicio}, {Bialek}, {Garcia-Dias}, {Lucchesi}, {North}, {Osorio}, {Patrick},
  \& {Peralta de Arriba}}]{Sestito2020MNRAS}
---. 2020, \mnras, 497, L7

\bibitem[{{Sestito} {et~al.}(2021){Sestito}, {Buck}, {Starkenburg}, {Martin},
  {Navarro}, {Venn}, {Obreja}, {Jablonka}, \& {Macci{\`o}}}]{Sestito2021MNRAS}
---. 2021, \mnras, 500, 3750

\bibitem[{{Shetrone} {et~al.}(2003){Shetrone}, {Venn}, {Tolstoy}, {Primas},
  {Hill}, \& {Kaufer}}]{Shetrone2003AJ}
{Shetrone}, M., {Venn}, K.~A., {Tolstoy}, E., {Primas}, F., {Hill}, V., \&
  {Kaufer}, A. 2003, \aj, 125, 684

\bibitem[{{Shetrone} {et~al.}(2019){Shetrone}, {Tayar}, {Johnson}, {Somers},
  {Pinsonneault}, {Holtzman}, {Hasselquist}, {Masseron}, {M{\'e}sz{\'a}ros},
  {J{\"o}nsson}, {Hawkins}, {Sobeck}, {Zamora}, \&
  {Garc{\'\i}a-Hern{\'a}ndez}}]{Shetrone2019ApJ}
{Shetrone}, M., {et~al.} 2019, \apj, 872, 137

\bibitem[{{Simmerer} {et~al.}(2004){Simmerer}, {Sneden}, {Cowan}, {Collier},
  {Woolf}, \& {Lawler}}]{Simmerer2004ApJ}
{Simmerer}, J., {Sneden}, C., {Cowan}, J.~J., {Collier}, J., {Woolf}, V.~M., \&
  {Lawler}, J.~E. 2004, \apj, 617, 1091

\bibitem[{{Sitnova} {et~al.}(2016){Sitnova}, {Mashonkina}, \&
  {Ryabchikova}}]{Sitnova2016MNRAS}
{Sitnova}, T.~M., {Mashonkina}, L.~I., \& {Ryabchikova}, T.~A. 2016, \mnras,
  461, 1000

\bibitem[{{Sneden}(1973)}]{Sneden1973ApJ}
{Sneden}, C. 1973, \apj, 184, 839

\bibitem[{{Sneden} {et~al.}(2003){Sneden}, {Cowan}, {Lawler}, {Ivans},
  {Burles}, {Beers}, {Primas}, {Hill}, {Truran}, {Fuller}, {Pfeiffer}, \&
  {Kratz}}]{Sneden2003ApJ}
{Sneden}, C., {et~al.} 2003, \apj, 591, 936

\bibitem[{{Sobeck} {et~al.}(2007){Sobeck}, {Lawler}, \&
  {Sneden}}]{Sobeck2007ApJ}
{Sobeck}, J.~S., {Lawler}, J.~E., \& {Sneden}, C. 2007, \apj, 667, 1267

\bibitem[{{Sobeck} {et~al.}(2011){Sobeck}, {Kraft}, {Sneden}, {Preston},
  {Cowan}, {Smith}, {Thompson}, {Shectman}, \& {Burley}}]{Sobeck2011AJ}
{Sobeck}, J.~S., {et~al.} 2011, \aj, 141, 175

\bibitem[{{Spite} \& {Spite}(1982)}]{Spite1982AA}
{Spite}, F., \& {Spite}, M. 1982, \aap, 115, 357

\bibitem[{{Spite} {et~al.}(2005){Spite}, {Cayrel}, {Plez}, {Hill}, {Spite},
  {Depagne}, {Fran{\c c}ois}, {Bonifacio}, {Barbuy}, {Beers}, {Andersen},
  {Molaro}, {Nordstr{\"o}m}, \& {Primas}}]{Spite2005AA}
{Spite}, M., {et~al.} 2005, \aap, 430, 655

\bibitem[{{Spite} {et~al.}(2006){Spite}, {Cayrel}, {Hill}, {Spite},
  {Fran{\c{c}}ois}, {Plez}, {Bonifacio}, {Molaro}, {Depagne}, {Andersen},
  {Barbuy}, {Beers}, {Nordstr{\"o}m}, \& {Primas}}]{Spite2006AA}
---. 2006, \aap, 455, 291

\bibitem[{{Stancliffe} {et~al.}(2007){Stancliffe}, {Glebbeek}, {Izzard}, \&
  {Pols}}]{Stancliffe2007AA}
{Stancliffe}, R.~J., {Glebbeek}, E., {Izzard}, R.~G., \& {Pols}, O.~R. 2007,
  \aap, 464, L57

\bibitem[{{Starkenburg} {et~al.}(2017){Starkenburg}, {Martin}, {Youakim},
  {Aguado}, {Allende Prieto}, {Arentsen}, {Bernard}, {Bonifacio}, {Caffau},
  {Carlberg}, {C{\^o}t{\'e}}, {Fouesneau}, {Fran{\c c}ois}, {Franke},
  {Gonz{\'a}lez Hern{\'a}ndez}, {Gwyn}, {Hill}, {Ibata}, {Jablonka},
  {Longeard}, {McConnachie}, {Navarro}, {S{\'a}nchez-Janssen}, {Tolstoy}, \&
  {Venn}}]{Starkenburg2017MNRAS}
{Starkenburg}, E., {et~al.} 2017, \mnras, 471, 2587

\bibitem[{{Suda} {et~al.}(2004){Suda}, {Aikawa}, {Machida}, {Fujimoto}, \&
  {Iben}}]{Suda2004ApJ}
{Suda}, T., {Aikawa}, M., {Machida}, M.~N., {Fujimoto}, M.~Y., \& {Iben}, Jr.,
  I. 2004, \apj, 611, 476

\bibitem[{{Suda} {et~al.}(2011){Suda}, {Yamada}, {Katsuta}, {Komiya},
  {Ishizuka}, {Aoki}, \& {Fujimoto}}]{Suda2011MNRAS}
{Suda}, T., {Yamada}, S., {Katsuta}, Y., {Komiya}, Y., {Ishizuka}, C., {Aoki},
  W., \& {Fujimoto}, M.~Y. 2011, \mnras, 412, 843

\bibitem[{{Suda} {et~al.}(2008){Suda}, {Katsuta}, {Yamada}, {Suwa}, {Ishizuka},
  {Komiya}, {Sorai}, {Aikawa}, \& {Fujimoto}}]{Suda2008PASJ}
{Suda}, T., {et~al.} 2008, \pasj, 60, 1159

\bibitem[{{Suda} {et~al.}(2017){Suda}, {Hidaka}, {Aoki}, {Katsuta}, {Yamada},
  {Fujimoto}, {Ohtani}, {Masuyama}, {Noda}, \& {Wada}}]{Suda2017PASJ}
---. 2017, \pasj, 69, 76

\bibitem[{{Takahashi} {et~al.}(2018){Takahashi}, {Yoshida}, \&
  {Umeda}}]{Takahashi2018ApJ}
{Takahashi}, K., {Yoshida}, T., \& {Umeda}, H. 2018, \apj, 857, 111

\bibitem[{{Takeda}(2019)}]{Takeda2019AA}
{Takeda}, Y. 2019, \aap, 622, A107

\bibitem[{{Thorburn}(1994)}]{Thorburn1994ApJ}
{Thorburn}, J.~A. 1994, \apj, 421, 318

\bibitem[{{Tolstoy} {et~al.}(2009){Tolstoy}, {Hill}, \&
  {Tosi}}]{Tolstoy2009ARAA}
{Tolstoy}, E., {Hill}, V., \& {Tosi}, M. 2009, \araa, 47, 371

\bibitem[{{Tominaga}(2009)}]{Tominaga2009ApJ}
{Tominaga}, N. 2009, \apj, 690, 526

\bibitem[{{Tominaga} {et~al.}(2014){Tominaga}, {Iwamoto}, \&
  {Nomoto}}]{Tominaga2014ApJ}
{Tominaga}, N., {Iwamoto}, N., \& {Nomoto}, K. 2014, \apj, 785, 98

\bibitem[{{Tominaga} {et~al.}(2007){Tominaga}, {Umeda}, \&
  {Nomoto}}]{Nozomu2007ApJ}
{Tominaga}, N., {Umeda}, H., \& {Nomoto}, K. 2007, \apj, 660, 516

\bibitem[{{Travaglio} {et~al.}(2004){Travaglio}, {Gallino}, {Arnone}, {Cowan},
  {Jordan}, \& {Sneden}}]{Travaglio2004ApJ}
{Travaglio}, C., {Gallino}, R., {Arnone}, E., {Cowan}, J., {Jordan}, F., \&
  {Sneden}, C. 2004, \apj, 601, 864

\bibitem[{{Umeda} \& {Nomoto}(2002)}]{Umeda&Nomoto2002ApJ}
{Umeda}, H., \& {Nomoto}, K. 2002, \apj, 565, 385

\bibitem[{{Umeda} \& {Nomoto}(2005)}]{Umeda&Nomoto2005ApJ}
---. 2005, \apj, 619, 427

\bibitem[{{Venn} {et~al.}(2012){Venn}, {Shetrone}, {Irwin}, {Hill}, {Jablonka},
  {Tolstoy}, {Lemasle}, {Divell}, {Starkenburg}, {Letarte}, {Baldner},
  {Battaglia}, {Helmi}, {Kaufer}, \& {Primas}}]{Venn2012ApJ}
{Venn}, K.~A., {et~al.} 2012, \apj, 751, 102

\bibitem[{{Venn} {et~al.}(2020){Venn}, {Kielty}, {Sestito}, {Starkenburg},
  {Martin}, {Aguado}, {Arentsen}, {Bonifacio}, {Caffau}, {Hill}, {Jablonka},
  {Lardo}, {Mashonkina}, {Navarro}, {Sneden}, {Thomas}, {Youakim},
  {Gonz{\'a}lez-Hern{\'a}ndez}, {S{\'a}nchez Janssen}, {Carlberg}, \&
  {Malhan}}]{Venn2020MNRAS}
---. 2020, \mnras, 492, 3241

\bibitem[{{Wanajo}(2018)}]{Wanajo2018ApJ}
{Wanajo}, S. 2018, \apj, 868, 65

\bibitem[{{Wanajo} \& {Ishimaru}(2006)}]{Wanajo2006NuPhA}
{Wanajo}, S., \& {Ishimaru}, Y. 2006, \nphysa, 777, 676

\bibitem[{{Woosley} \& {Weaver}(1995)}]{Woosley&Weaver1995ApJS}
{Woosley}, S.~E., \& {Weaver}, T.~A. 1995, \apjs, 101, 181

\bibitem[{{Xing} {et~al.}(2019){Xing}, {Zhao}, {Aoki}, {Honda}, {Li},
  {Ishigaki}, \& {Matsuno}}]{Xing2019NatAs}
{Xing}, Q.-F., {Zhao}, G., {Aoki}, W., {Honda}, S., {Li}, H.-N., {Ishigaki},
  M.~N., \& {Matsuno}, T. 2019, Nature Astronomy, 3, 631

\bibitem[{{Yamada} {et~al.}(2013){Yamada}, {Suda}, {Komiya}, {Aoki}, \&
  {Fujimoto}}]{Yamada2013MNRAS}
{Yamada}, S., {Suda}, T., {Komiya}, Y., {Aoki}, W., \& {Fujimoto}, M.~Y. 2013,
  \mnras, 436, 1362

\bibitem[{{Yan} {et~al.}(2021){Yan}, {Zhou}, {Zhang}, {Li}, {Gao}, {Shi},
  {Zhao}, {Aoki}, {Matsuno}, {Li}, {Xu}, {Li}, {Wu}, {Jin}, {Mosser}, {Bi},
  {Fu}, {Pan}, {Suda}, {Liu}, {Zhao}, \& {Liang}}]{Yan2021NatAs}
{Yan}, H.-L., {et~al.} 2021, Nature Astronomy, 5, 86

\bibitem[{{Yanny} {et~al.}(2009){Yanny}, {Rockosi}, {Newberg}, {Knapp},
  {Adelman-McCarthy}, {Alcorn}, {Allam}, {Allende Prieto}, {An}, {Anderson},
  {Anderson}, {Bailer-Jones}, {Bastian}, {Beers}, {Bell}, {Belokurov},
  {Bizyaev}, {Blythe}, {Bochanski}, {Boroski}, {Brinchmann}, {Brinkmann},
  {Brewington}, {Carey}, {Cudworth}, {Evans}, {Evans}, {Gates}, {G{\"a}nsicke},
  {Gillespie}, {Gilmore}, {Gomez-Moran}, {Grebel}, {Greenwell}, {Gunn},
  {Jordan}, {Jordan}, {Harding}, {Harris}, {Hendry}, {Holder}, {Ivans},
  {Ivezi{\v c}}, {Jester}, {Johnson}, {Kent}, {Kleinman}, {Kniazev},
  {Krzesinski}, {Kron}, {Kuropatkin}, {Lebedeva}, {Lee}, {Leger}, {L{\'e}pine},
  {Levine}, {Lin}, {Long}, {Loomis}, {Lupton}, {Malanushenko}, {Malanushenko},
  {Margon}, {Martinez-Delgado}, {McGehee}, {Monet}, {Morrison}, {Munn},
  {Neilsen}, {Nitta}, {Norris}, {Oravetz}, {Owen}, {Padmanabhan}, {Pan},
  {Peterson}, {Pier}, {Platson}, {Fiorentin}, {Richards}, {Rix}, {Schlegel},
  {Schneider}, {Schreiber}, {Schwope}, {Sibley}, {Simmons}, {Snedden}, {Smith},
  {Stark}, {Stauffer}, {Steinmetz}, {Stoughton}, {Subba Rao}, {Szalay},
  {Szkody}, {Thakar}, {Thirupathi}, {Tucker}, {Uomoto}, {Vanden Berk},
  {Vidrih}, {Wadadekar}, {Watters}, {Wilhelm}, {Wyse}, {Yarger}, \&
  {Zucker}}]{Yanny2009AJ}
{Yanny}, B., {et~al.} 2009, \aj, 137, 4377

\bibitem[{{Yong} {et~al.}(2013){Yong}, {Norris}, {Bessell}, {Christlieb},
  {Asplund}, {Beers}, {Barklem}, {Frebel}, \& {Ryan}}]{Yong2013ApJa}
{Yong}, D., {et~al.} 2013, \apj, 762, 26

\bibitem[{{Yong} {et~al.}(2021{\natexlab{a}}){Yong}, {Da Costa}, {Bessell},
  {Chiti}, {Frebel}, {Gao}, {Lind}, {Mackey}, {Marino}, {Murphy}, {Nordlander},
  {Asplund}, {Casey}, {Kobayashi}, {Norris}, \& {Schmidt}}]{Yong2021MNRAS}
---. 2021{\natexlab{a}}, \mnras, 507, 4102

\bibitem[{{Yong} {et~al.}(2021{\natexlab{b}}){Yong}, {Kobayashi}, {Da Costa},
  {Bessell}, {Chiti}, {Frebel}, {Lind}, {Mackey}, {Nordlander}, {Asplund},
  {Casey}, {Marino}, {Murphy}, \& {Schmidt}}]{Yong2021Nature}
---. 2021{\natexlab{b}}, \nat, 595, 223

\bibitem[{{Yoon} {et~al.}(2016){Yoon}, {Beers}, {Placco}, {Rasmussen},
  {Carollo}, {He}, {Hansen}, {Roederer}, \& {Zeanah}}]{Yoon2016ApJ}
{Yoon}, J., {et~al.} 2016, \apj, 833, 20

\bibitem[{{Yuan} {et~al.}(2020){Yuan}, {Myeong}, {Beers}, {Evans}, {Lee},
  {Banerjee}, {Gudin}, {Hattori}, {Li}, {Matsuno}, {Placco}, {Smith},
  {Whitten}, \& {Zhao}}]{Yuan2020ApJ}
{Yuan}, Z., {et~al.} 2020, \apj, 891, 39

\bibitem[{{Zhang} {et~al.}(2019){Zhang}, {Li}, {Zhao}, {Aoki}, \&
  {Matsuno}}]{Zhang2019PASJ}
{Zhang}, S., {Li}, H., {Zhao}, G., {Aoki}, W., \& {Matsuno}, T. 2019, \pasj,
  71, 89

\bibitem[{{Zhao} \& {Chen}(2021)}]{Zhao&Chen2021SCPMA}
{Zhao}, G., \& {Chen}, Y. 2021, Science China Physics, Mechanics, and
  Astronomy, 64, 239562

\bibitem[{{Zhao} {et~al.}(2006){Zhao}, {Chen}, {Shi}, {Liang}, {Hou}, {Chen},
  {Zhang}, \& {Li}}]{Zhao2006ChJAA}
{Zhao}, G., {Chen}, Y., {Shi}, J., {Liang}, Y., {Hou}, J., {Chen}, L., {Zhang},
  H., \& {Li}, A. 2006, Chinese Journal of Astronomy and Astrophysics, 6, 265

\bibitem[{{Zhao} \& {Magain}(1991)}]{Zhao&Magain1991AA}
{Zhao}, G., \& {Magain}, P. 1991, \aap, 244, 425

\bibitem[{{Zhao} {et~al.}(2012){Zhao}, {Zhao}, {Chu}, {Jing}, \&
  {Deng}}]{Zhao2012RAA}
{Zhao}, G., {Zhao}, Y.-H., {Chu}, Y.-Q., {Jing}, Y.-P., \& {Deng}, L.-C. 2012,
  Research in Astronomy and Astrophysics, 12, 723

\end{thebibliography}



\appendix

\section{Validation of stellar parameters derived from LAMOST}

\begin{figure*}
\epsscale{1.1}
\plotone{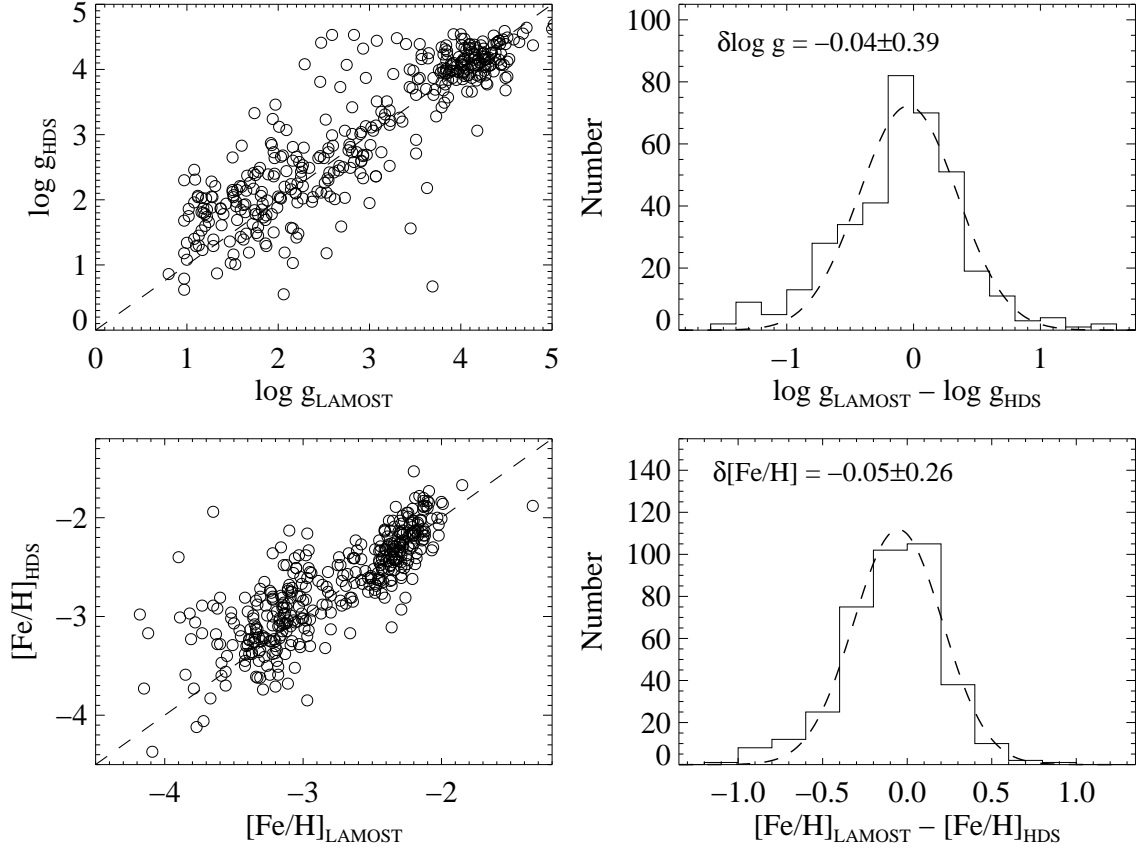}
\caption{Top left: Comparison between {\logg} derived from the LAMOST low-resolution spectra
and that from Subaru/HDS spectra as adopted in this paper, 
i.e. including parallax-derived {\logg} for 315 objects 
and corrected spectroscopic {\logg} for 70 program stars.
The one-to-one line is also plotted for reference.
Top right: Distribution of the {\logg} difference, together with a best-fit Gaussian
to indicate the measurement error.
The bottom panels are similar, but for comparison of the metallicity.
\label{fig:FeH_logg_comparison}}
\end{figure*}

Comparisons of stellar parameters derived from Subaru/HDS and LAMOST low-resolution spectra 
can be helpful to understand the efficiency of VMP selection from the LAMOST database, 
which has been shown in Figure~\ref{fig:Teff_comparison} for the effective temperature. 
Here the comparison for {\logg} and {\FeH} are shown in Figure~\ref{fig:FeH_logg_comparison}.
As can be seen in the bottom panels, rather good agreement can be found for {\FeH}
with a small offset and reasonably low measurement uncertainty of $-0.05\pm0.26$ dex.
Rather good consistency can be found even at the extremely low metallicity region.
Agreement of the surface gravity is also good in general, showing a very small offset of $-0.04$, 
which is mostly because that the parallax-derived surface gravity has been adopted 
as the initial guess for most of the program stars to derive stellar parameters from LAMOST spectra. 
The relatively large scatter of the uncertainty for {\logg} ($\sim 0.39$) is not unexpected, 
since even with high-resolution spectroscopy,
the uncertainty in estimating surface gravity is relatively large compared to {\Tefft} and {\FeH}, 
for objects that have no parallax measurements. 
However, the surface gravity estimated based on LAMOST spectra is generally consistent with
the high-resolution analyses, which is sufficient to separate stars located in
different evolutionary stages. 
Involving parallaxes obtained from further {\it Gaia} data releases 
will help improve the accuracy of {\logg} estimates based on LAMOST low-resolution spectra.
To sum, the stellar parameters derived from LAMOST low-resolution spectra
are generally robust and reliable for a large sample of VMP and EMP stars.

\section{Updating the surface gravity with Gaia EDR3}

\begin{figure*}
\epsscale{1.1}
\plotone{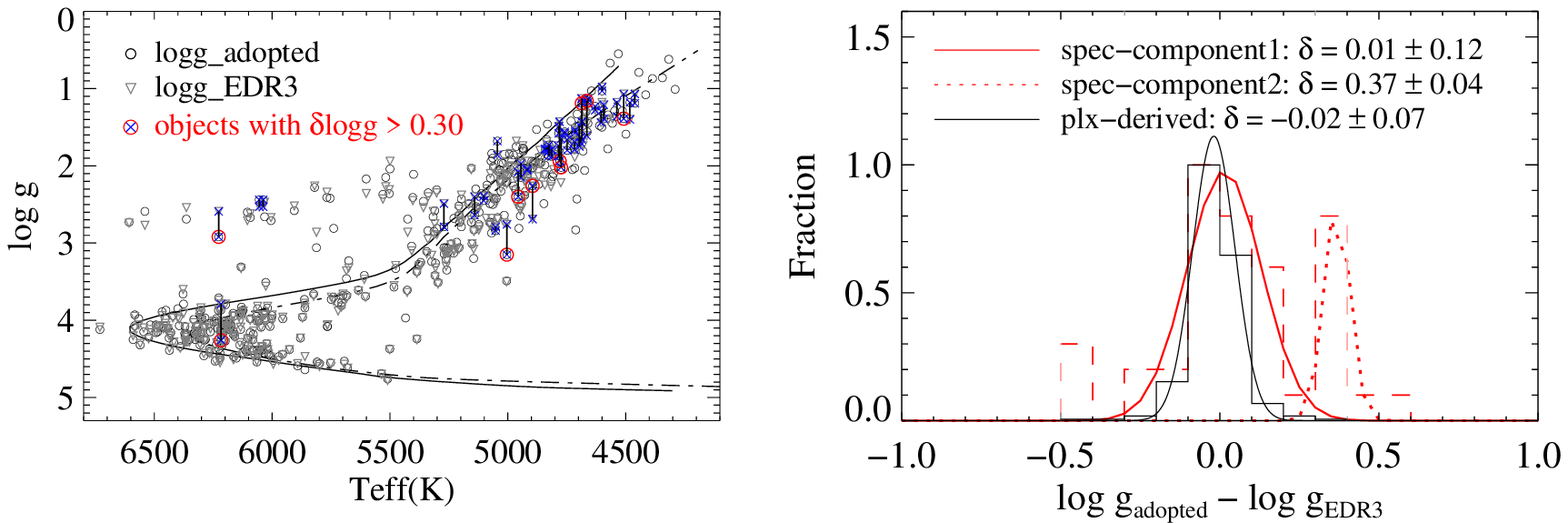}
\caption{Left: comparison of the adopted {\logg} (circle) and those derived from the parallax from {\it Gaia} EDR3 (triangle). 
Open symbols represent objects for which {\logg} is directly derived from the parallax, 
and crosses stand for those whose {\logg} is corrected from the spectroscopic result. 
Two values of {\logg} are connected for the 43 objects which have reliable parallax measurements in EDR3 but not in DR2. 
Ten program stars whose EDR3-based {\logg} differ from the DR2-based values with more than 0.3 has been marked out with a large red circle.
Right: difference of the two sets of {\logg}. Solid histogram refer to the distribution of the parallax-derived sub-sample, 
and the dashed histogram refers to that of spectroscopic-corrected {\logg} sub-sample. 
Note that the latter can be fit with two Gaussian distributions (dashed for component 1 and dotted for component 2), with the typical offset and dispersion shown on the top of the panel. 
\label{fig:logg_EDR3}}
\end{figure*}

{\it Gaia} has just published its latest data release, EDR3 \citep{GaiaEDR3_2021AA}, 
which provides more accurate measurements on the parallax and proper motion for Galactic stars. 
Although the main purpose of this paper is exploring the chemical abundances, 
it is still necessary to understand how much the results would be affected (mainly on {\logg}) 
if the new parallax measurement is adopted. 
Therefore, we have cross-matched our sample with Gaia EDR3, 
and found that there could be 43 objects (mostly giants) whose {\logg} have been derived 
based on the correction described in Section~\ref{subsec:stellar_parameter} 
now have reliable parallax measurement from Gaia EDR3. 
Comparisons of the two sets of {\logg} based on Gaia DR2 and EDR3 are shown in Figure~\ref{fig:logg_EDR3}.

As for objects whose {\logg} have been derived from parallax, DR2 and EDR3 are in good agreement, 
e.g., with an average difference of $0.01\pm0.07$ as shown in the right panel of Figure~\ref{fig:logg_EDR3}. 
Larger difference can be seen for objects for which the {\logg} correction has been adopted. The DR2-based-corrected and EDR3-parallax-derived {\logg} are in general consistent 
for the majority of the 43 objects with reliable parallax measurements from EDR3, 
i.e., the dashed distribution in red in the right panel of Figure~\ref{fig:logg_EDR3}. 
There are ten program stars showing a difference larger than 0.3 
between the adopted (Gaia DR2-based) and the EDR3-based {\logg}, which 
are marked in red in the left panel, 
and all of them are more distant than 6\,kpc (based on the EDR3 parallax). 

As can be seen from the right panel of Figure~\ref{fig:logg_EDR3}, 
It will only be the ten program stars (``spec-component2'') that could have noticeably different {\logg} 
if the EDR3 parallax is adopted. While for the majority of our sample, 
including the 315 objects that have parallax-derived {\logg} (``plx-derived'') and 59 objects with corrected spectroscopic {\logg} (``spec-component1''), 
the stellar parameters would not vary much, e.g., with a change smaller or comparable to the uncertainty for {\logg}. 
According to the error budget estimation in \S~\ref{subsec:abun_error}, 
even for objects whose uncertainty in {\logg} is as large as 0.30, 
the resulted variation in abundances is smaller than the total uncertainty of abundances. 
We therefore decide to keep the {\logg} estimation based on DR2 parallax measurement for current work. 
For the next paper of this series, which is going to study the kinematics of the sample, 
the latest version of Gaia measurement shall be used.




\end{document}